\documentclass[twocolumn, iop, apj, twocolappendix, numberedappendix, appendixfloats, linenumbers]{emulateapj}
\usepackage{CJKutf8}
\usepackage{threeparttable}
\usepackage{multirow}
\usepackage{times}
\usepackage{enumitem}
\usepackage{url} 
\usepackage{color}
\usepackage{amsmath,bm}
\usepackage{subfigure}
\usepackage{natbib}
\usepackage{graphicx}
\usepackage{graphics}
\usepackage{hyperref}                             
\usepackage{amsmath}
\usepackage{bm}
\usepackage{subfigure}
\usepackage{array}
\usepackage{booktabs}
\usepackage{lineno}

\shorttitle{Kinematics of the Galactic disk as revealed by the LAMOST Red clump stars}
\shortauthors{Sun et al.}
\submitted{Submitted to ApJ;\,\,\,\,Accepted October 23, 2023}

\begin{document}
%\linenumbers

\title{Mapping the Galactic disk with the LAMOST and Gaia Red clump sample:\\ VIII: Mapping the kinematics of the Galactic disk using mono-age and mono-abundance stellar populations}

\author{ Weixiang Sun\textsuperscript{1}}
\author{ Yang Huang\textsuperscript{2,3,9}}
\author{ Han Shen\textsuperscript{4}}
\author{ Chun Wang\textsuperscript{5}}
\author{ Huawei Zhang\textsuperscript{6,7}}
\author{ Zhijia Tian\textsuperscript{8}}
\author{ Xiaowei Liu\textsuperscript{4,9}}
\author{ Biwei Jiang\textsuperscript{1}}

\altaffiltext{1}{Department of Astronomy, Beijing Normal University, Beijing 100875, People's Republic of China; {\it sunweixiang@bnu.edu.cn}.}
\altaffiltext{2}{University of Chinese Academy of Sciences, Beijing 100049, People's Republic of China; {\it huangyang@ucas.ac.cn}.}
\altaffiltext{3}{National Astronomical Observatories, Chinese Academy of Sciences, Beijing 100012, People’s Republic of China.}
\altaffiltext{4}{South-Western Institute for Astronomy Research, Yunnan University, Kunming 650500, People's Republic of China; {\it x.liu@ynu.edu.cn}.}
\altaffiltext{5}{Tianjin Astrophysics Center, Tianjin Normal University, Tianjin 300387, People's Republic of China.}
\altaffiltext{6}{Department of Astronomy, Peking University, Beijing 100871, People's Republic of China.}
\altaffiltext{7}{Kavli Institute for Astronomy and Astrophysics, Peking University, Beijing 100871, People's Republic of China}
\altaffiltext{8}{Department of Astronomy, Yunnan University, Kunming 650500, People's Republic of China.}
\altaffiltext{9}{Corresponding authors}

\begin{abstract}

We present a comprehensive study of the kinematic properties of the different Galactic disk populations, as defined by the chemical abundance ratios and stellar ages, across a large disk volume (4.5 $\leq$ R $\leq$ 15.0\,kpc and $|Z|$ $\leq$ 3.0\,kpc), by using the LAMOST-Gaia red clump sample stars.
We determine the median velocities for various spatial and population bins, finding large-scale bulk motions, such as the wave-like behavior in radial velocity, the north-south discrepancy in azimuthal velocity and the warp signal in vertical velocity, and the amplitudes and spatial-dependences of those bulk motions show significant variations for different mono-age and mono-abundance populations.
The global spatial behaviors of the velocity dispersions clearly show a signal of spiral arms and, a signal of the disk perturbation event within 4\,Gyr, as well as the disk flaring in the outer region (i.e., $R \ge 12$\,kpc) mostly for young or alpha-poor stellar populations.
Our detailed measurements of age/[$\alpha$/Fe]-velocity dispersion relations for different disk volumes indicate that young/$\alpha$-poor populations are likely originated from dynamically heated by both giant molecular clouds and spiral arms, while old/$\alpha$-enhanced populations require an obvious contribution from other heating mechanisms such as merger and accretion, or born in the chaotic mergers of gas-rich systems and/or turbulent interstellar medium.

\end{abstract}


\section{Introduction}

As a benchmark galaxy, the Milky Way provides an excellent laboratory of astrophysics for understanding the formation and evolution of galaxies by analysis of a large sample of individual stars in exquisite details that are almost impossible for extragalactic galaxies \citep[e.g.,][]{Rix2013, Bensby2014, Hayden2020}.
The assembly history of the Milky Way is imprinted not only in the present-day stellar positions but also in its stellar kinematics.
As an example, by studying the stellar mean velocity and the velocity dispersion as a function of stellar positions and ages/abundances, we can investigate the perturbation and heating history of the Galactic disk and their related mechanisms \citep[e.g.,][]{Stromberg1946, Roman1950a, Roman1950b, Lee2011, Williams2013, Mackereth2019b, Han2020}.
Therefore, mapping the stellar kinematics in mono-age or mono-abundance phases across a large volume of the Galactic disk can help to draw a full picture of the disk and to lead a deep understanding of the dynamical history of our Galaxy \citep[e.g.,][]{Rix2013, Quillen2001, Holmberg2009, Minchev2013, Sun2015, Sharma2021}.
With the advents of the massive large-scale surveys, such as the SDSS \citep{York2000} for photometry, the LAMOST \citep[e.g.,][]{Deng2012, Cui2012, Liu2014, Yuan2015} for spectroscopy and, the Gaia \citep{Perryman2001, Gaia Collaboration2016} for astrometry, we are entering a golden era of Galactic studies with precise multi-dimensional information (e.g., 3D position, 3D velocity, chemical element abundance ratios and stellar age) now available for a huge sample of stars in the Galaxy.

In the era prior to the {\it Gaia} mission, studies on the Galactic kinematics are largely limited to the measurements of the kinematic properties in the solar neighborhood due to the lack of accurate measurements of proper motions \citep[e.g.,][]{Quillen2001, Aumer2009, Holmberg2009, Casagrande2011, Lee2011, Williams2013, Minchev2014, Sun2015}.
The results in the solar neighborhood indicate that velocity dispersions display a clear power-law-like increasing trend with stellar age \citep[e.g.,][]{Stromberg1946, Quillen2001, Aumer2009, Casagrande2011, Yu2018, Hayden2020, Mackereth2019b, Sharma2021}, which imply that the disk stars have experienced the heating from the Giant Molecular Clouds (GMCs) or transient spiral arms \citep[e.g.,][]{Spitzer1951, Spitzer1953, Barbanis1967, Jenkins1992}, or even some violent heating processes, such as the minor mergers of dwarf galaxies \citep[e.g.,][]{Quinn1993, Villalobos2008, Abadi2003, Minchev2014, Grand2016} and, infall of misaligned gas \citep[e.g.,][]{Roskar2010, Sharma2012, Aumer2013}.
However, more data, especially of larger volume, is required to understand the roles of different heating mechanisms for disk stars.
For the mean velocity of the disk, Widrow et al. ({\color{blue}{2012}}) measured the vertical velocity as a function of the height from the disk plane, and found significant vertical bulk motions.
Williams et al. ({\color{blue}{2013}}) present a detailed analysis of the local disk velocity field by nearly ten thousand red clump stars selected from the RAVE survey.
They find both clear radial and vertical bulk motions across the disk. 
Moreover, compression and rarefaction patterns are detected in the vertical bulk motions.
Those interesting results are later independently confirmed by Carlin et al. ({\color{blue}{2013}}) by over ten thousand turn-off stars selected from the LAMOST surveys.
On the shoulders of previous studies, Sun et al. ({\color{blue}{2015}}) investigate the bulk motions of 0.57 million Galactic disk stars as far as 2.0\,kpc for different populations,
and present a much clear picture of the local velocity field, by detecting significant large-scale bending- and breathing perturbations across the disk, as well as some breaks and ripples on the small-scale.
However, kinematics with improved measurements of proper motions are required to confirm the above findings and to compare them to comprehensive numerical simulations to reveal the origins of those perturbations of different scales.

With the recent releases of the Gaia surveys \citep[e.g.,][]{Gaia Collaboration2018, Gaia Collaboration2020}, precise astrometric information is now available for billion stars, and thus the accuracy of stellar kinematics has been much improved \citep[e.g.,][]{Gaia Collaboration2018, Antoja2018, Trick2019, Mackereth2019b}.
As one of the most representative works of the Gaia surveys, Gaia Collaboration, Katz et al. ({\color{blue}{2018}}) used DR2 data to derive the 3D kinematics of the Galactic disk with unprecedented accuracy and spatial resolution over a wide range of Galactocentric radii for the first time, and provide a detailed portrait of the disk on different phase spaces.
As an example, the results of their mapping kinematics indicate that the disk bulk motions reported by previous studies \citep[e.g.,][]{Siebert2011, Williams2013, Carlin2013, Carlin2014, Sun2015} are part of oscillation(s) on a kilo-parsec scale.
The results of Gaia Collaboration, Katz et al. ({\color{blue}{2018}}) also find the large-scale non-axisymmetric features of the Galactic disk, including the warp, flare, bending, as well as rich substructures.
These new results are further characterized and studied by later studies \citep[e.g.,][]{Antoja2018, Trick2019, Hunt2018, Wangc2019, Mackereth2019b, Li2020}.
Although the big advances achieved in disk kinematics, especially with the Gaia data released most recently, it is too early to say that the assembly history of the disk is fully explored given the lack of studies on mono-age/mono-abundance space, which is the key to understanding the evolution of Galactic disk.

At present, based on the advantages of the LAMOST and Gaia surveys, a large sample of red clump stars with accurate 3D position, 3D velocity, element abundance ratios and stellar ages has been constructed.
On top of this sample, one can dissect the disk kinematics in mono-age or abundance spaces for a significant volume and thus the evolved behaviors of the disk can be probed.

This paper is organized as follows: In Section\,2, we describe the data adopted in our analysis, and in Section\,3, we present the main results from measurements of the kinematics in mono-age and mono-abundance spaces, and discuss our main results in Section\,4.
Finally, our main conclusions are summarized in Section 5.

\begin{figure}[t]
\begin{center}
\includegraphics[scale=0.53,angle=0]{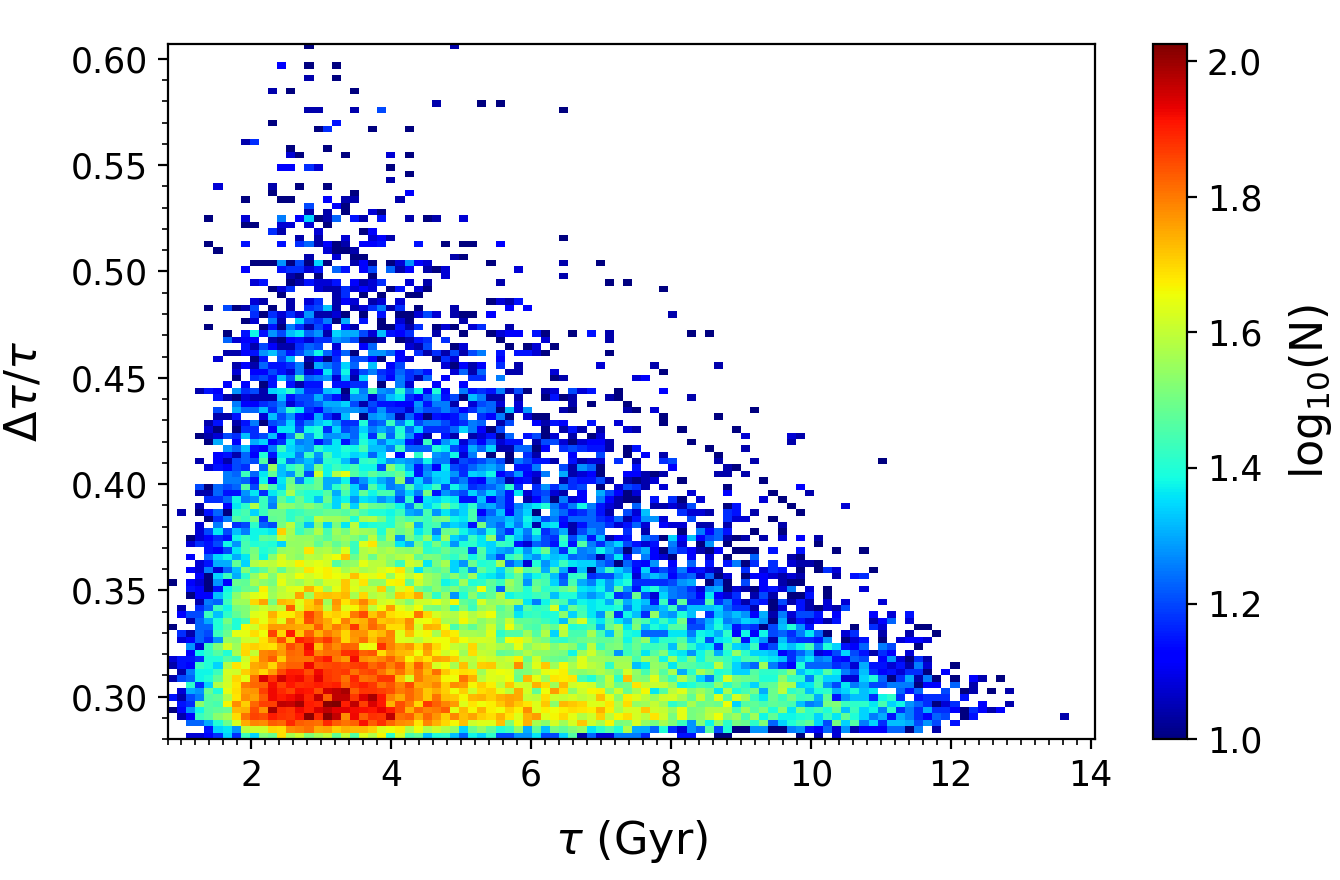}
\caption{The relative age error ($\Delta \tau/\tau$, defined as the ratio of the age error to age) as a function of age ($\tau$), color-coded by stellar densities on a logarithmic scale.
The horizontal and vertical bin sizes are 0.13\,Gyr and 0.003, respectively.
There are no less than 10 stars in each bin.
}
\end{center}
\end{figure}
%%\label{Fig. 1}

\begin{figure}[t]
\begin{center}
\includegraphics[scale=0.395,angle=0]{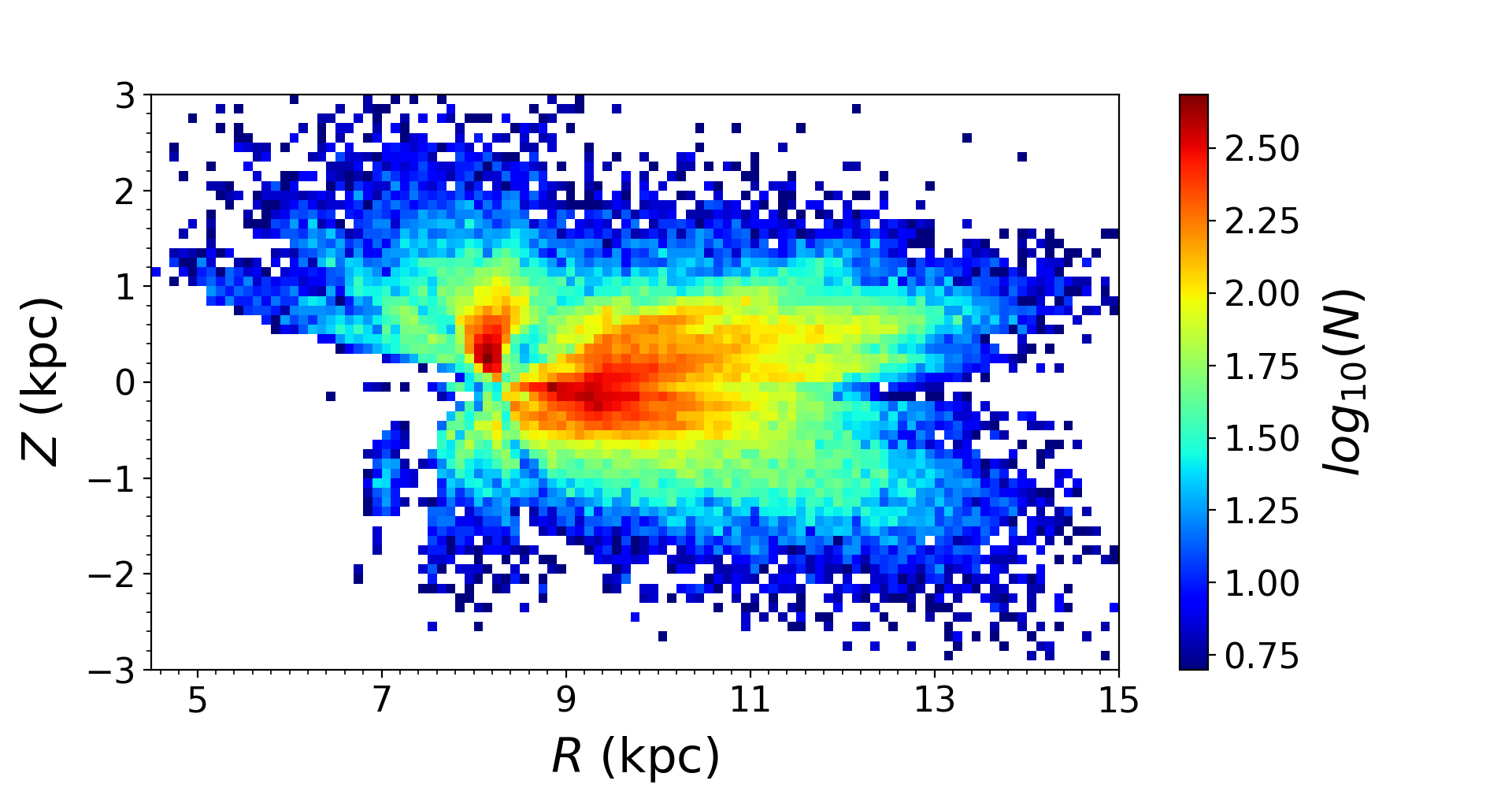}

\includegraphics[scale=0.26,angle=0]{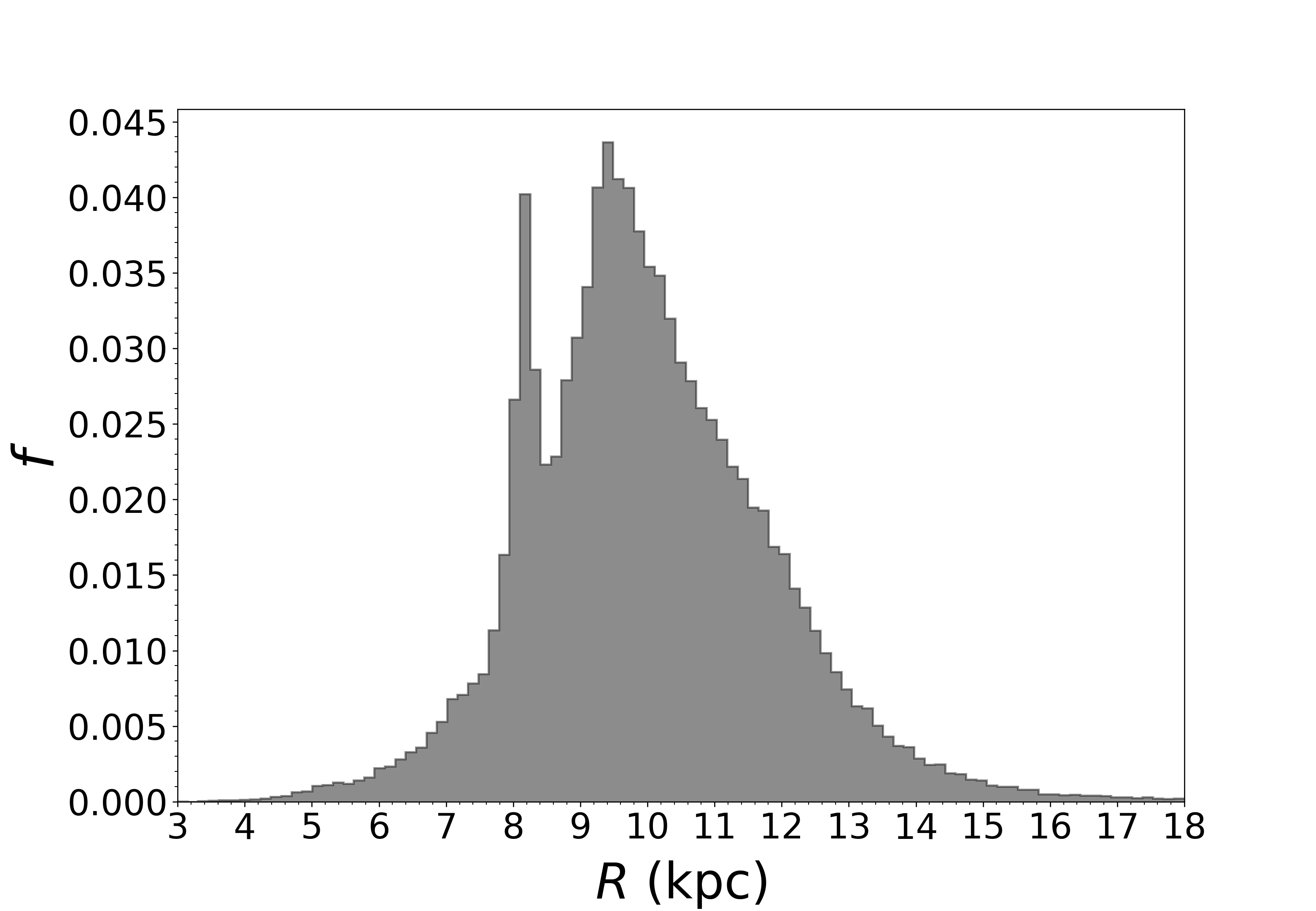}

\caption{{\bfseries Upper panel:} Spatial distribution of the LAMOST RC sample stars in the $R$ - $Z$ plane, color-coded by stellar number densities on a logarithmic scale.
Both axes are spaced by 0.1\,kpc, with a minimum of 5 stars per bin.
{\bfseries Bottom panel:} Histograms of fraction number density ($f = N_i/N_{\rm tot}$) distributions of the whole sample in the radial directions.
Here, $N_{i}$ is the number of stars in the individual radial bins and $N_{\rm tot}$ is the total number of stars of the whole sample.
}
\end{center}
\end{figure}
%%\label{Fig. 2}

\begin{figure*}[t]
\centering
\subfigure{
\includegraphics[width=8.82cm]{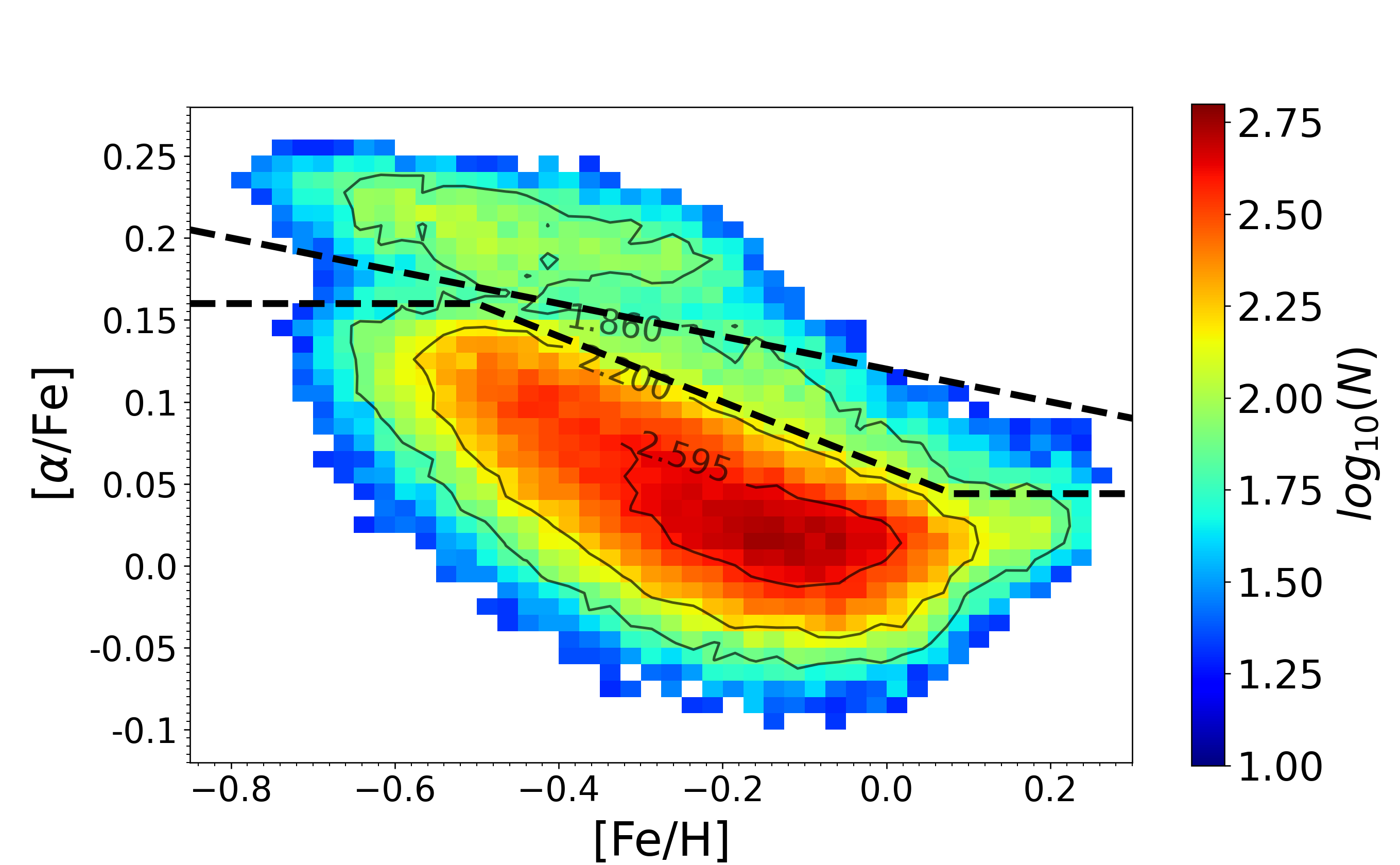}
}
\subfigure{
\includegraphics[width=8.82cm]{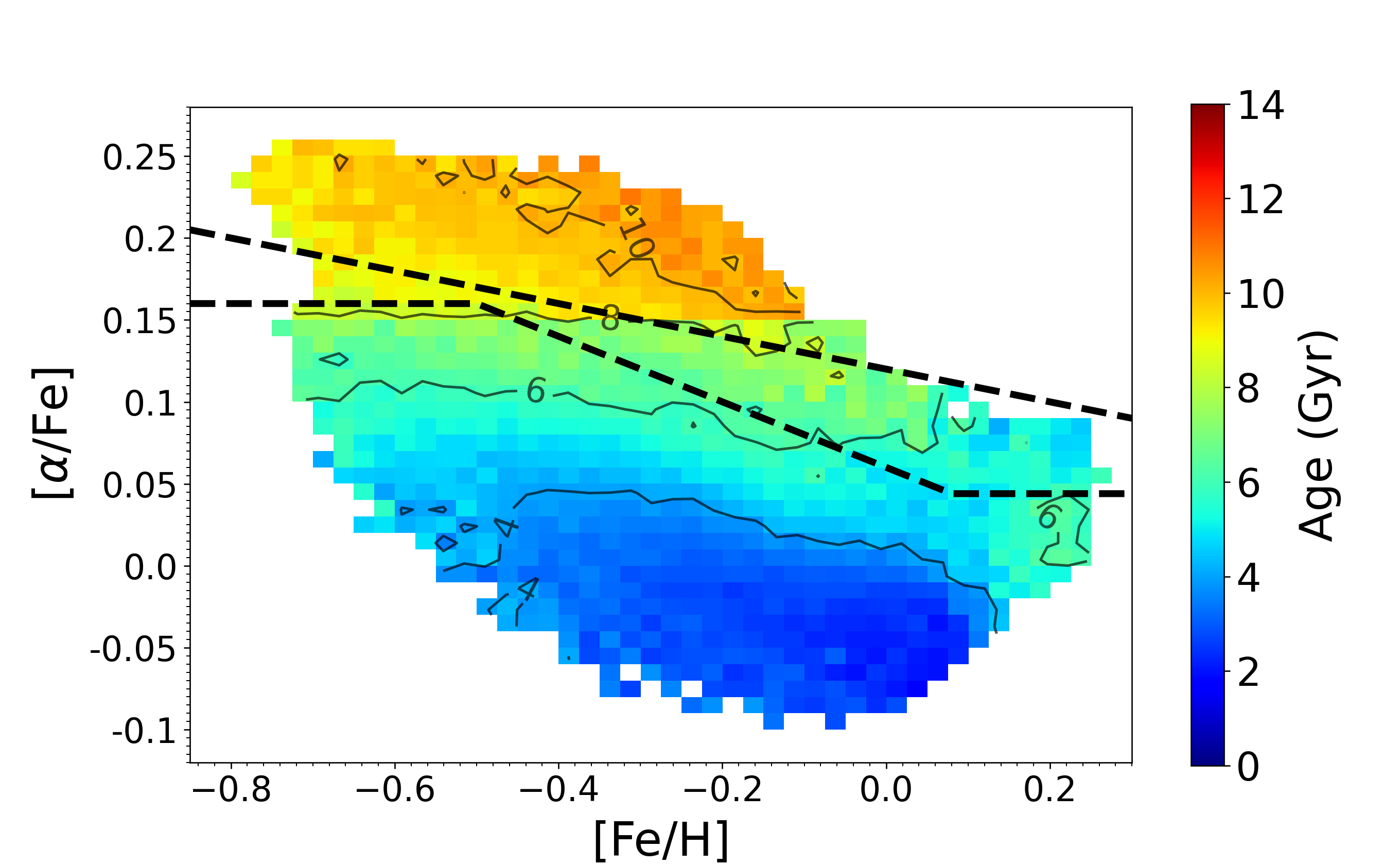}
}

\caption{{\bfseries Left panel}: The [Fe/H]--[$\alpha$/Fe] of the RC sample stars, color-coded by logarithmic stellar number densities.
The horizontal and vertical bin sizes are 0.025\,dex and 0.02\,dex, respectively.
There are no less than 20 stars in each bin.
The thin (with low-[$\alpha$/Fe]) and thick (with high-[$\alpha$/Fe]) disk stars are separated by the two dashed lines.
{\bfseries Right panel}: Similar to the left but color-coded by the mean stellar ages.
}
\end{figure*}
%%\label{Fig.3}

\section{Data}
\subsection{The LAMOST primary RC stars}

In this work, the sample data is adopted from Huang et al. ({\color{blue}{2020}}, hereafter Paper I), which provide nearly 140,000 primary red clump (RC) stars with the typical uncertainties of 5.0 km s$^{-1}$, 100 K, 0.10 dex, 0.10$-$0.15 dex and 0.03$-$0.05 dex respectively in line-of-sight velocity $V_{r}$, effective temperature $T_{\rm eff}$, surface gravity log\,$g$, metallicity [Fe/H] and [$\alpha$/Fe] abundance ratio \citep{Huang2018a}.
Stellar masses and ages are determined from all the features of the optical spectrum of each star from LAMOST, by using a machine-learning method trained with thousands of RC stars with accurate asteroseismic ages in the LAMOST-Kepler fields.
The typical uncertainties in masses and ages are, respectively, 15\% and 30\%, and the age uncertainties as a function of age are shown in Fig.\,1.
Taking advantage of the nature of the standard candle of RCs, the distances have been estimated with uncertainties better than 5\% $-$ 10\%, even better than the measurements from Gaia parallaxes of stars beyond $\sim$ 5.0\,kpc.
A full description of this sample data is presented in Paper I \citep{Huang2020}.
We note that the proper motions of this sample have been updated to the more accurate measurements from the Gaia EDR3 \citep[e.g.,][]{Gaia Collaboration2020}.

\subsection{Coordinate systems and sample selections}

In this paper, the standard Galactocentric cylindrical Coordinate ($R$, $\phi$, $Z$) has been used for the following analysis.
$R$ is the projected Galactocentric distance, increasing radial outwards, $\phi$ is in the direction of Galactic rotation and $Z$ towards the North Galactic Pole.
The three velocity components ($V_{R}$, $V_{\phi}$, $V_{z}$) are determined by the methods from Johnson \& Soderblom ({\color{blue}{1987}}) and Williams et al. ({\color{blue}{2013}}) by assuming that the values of the Galactocentric distance of the sun, the solar motions and local circular velocity have values $R_{\odot}$ = 8.34 kpc \citep{Reid2014}, ($U_{\odot}$, $V_{\odot}$, $W_{\odot}$) $=$ $(13.00, 12.24, 7.24)$ km s$^{-1}$ \citep{Schonrich2018} and $V_{c,0}$ = 238 km s$^{-1}$ \citep[e.g.,][]{Reid2004, Schonrich2010, Schonrich2012, Reid2014, Huang2015, Huang2016, Bland-Hawthorn2016}, respectively.

The velocity dispersion $\sigma$ of each bin is estimated by a 3$\sigma$-clipping procedure that removes outliers of extreme values.
The uncertainty of $\sigma$ is estimated by the classical method, $\Delta\sigma$ = $\sqrt{1/[2(N-1)]}$ $\sigma$ \citep{Huang2016}, where $N$ is the number of stars in each bin.
To ensure the accuracy of the calculated 3D velocities, we have further applied the following cuts to the sample stars:

\begin{itemize}[leftmargin=*]
\item {Age uncertainty $\leq$ 50\%  \ \& \ stellar age $\leq$ 14.0 Gyr;}

\item  {Distance uncertainty $\leq$ 10\%;}

\item {Vertical velocity\,$\ |V_{z}|$ $\leq$ 120 km s$^{-1}$;}

\item  {Metallicity [Fe/H] $\geq -1.0$ dex.}

\end{itemize}

The first two cuts combined with the signal-to-noise ratio cut (SNRs $>$ 20) that already applied in Paper I, ensure that the uncertainties of the measured 3D velocities are typically within 5.0 km s$^{-1}$ and mostly smaller than 15.0 km s$^{-1}$ even for stars of distances beyond 4.0$-$6.0 kpc.
The last two cuts are used to avoid the contaminations of the halo stars \citep{Huang2018b, Hayden2020}.
We also excluded the so-called ``young" [$\alpha$/Fe]-enhanced stars with age$\leq$ 6.0 Gyr and [$\alpha$/Fe] $\geq$ 0.15 dex since their true ages have been underestimated caused by the present large masses resulted from the binary merger or mass transfer \citep{Sun2020}.
With the above cuts, 128,537 RC stars are finally selected.
The spatial distribution of those selected stars in the $R$-$Z$ plane is shown in the upper panel of Fig.\,2, and the histogram of radial distribution of the whole sample as shown in the bottom panel of Fig.\,2.
This sample covers a significant volume of the Galactic disk, with $6.0 \leq R \leq 15.0$\,kpc, $|Z| \leq$ 3.0$-$4.0\,kpc.
It is clear that one bin is very much centred in the solar circle (as it is the case for stars towards the Kepler field) and another over-density is more around 9$-$10\,kpc from the Galactic center.

\subsection{The thin and thick disk stars of the RC sample}

The [Fe/H]--[$\alpha$/Fe] relation of the RC sample stars, is shown in the left panel of Fig.\,3.
The plot shows two obvious over-density sequences, which are respectively the chemically thin and thick disk populations.
To get pure thin/thick disk stars, inspired from previous studies \citep[e.g.,][]{Bensby2005, Lee2011, Brook2012, Haywood2013, Recio-Blanco2014, Nidever2014, Guiglion2015, Hayden2015, Queiroz2020, Sun2023}, two empirical cuts (see Fig.\,3) are adopted to separate the chemical thin disk (104,691 stars below the lower cut) and thick disk (14,444 stars above the upper cut) populations.

The [Fe/H]--[$\alpha$/Fe] relation, color-coded by the mean stellar ages of the individual bin, is shown in the right panel of Fig.\,3.
The thin disk population has a wide range of ages from the youngest at 1.0$-$2.0\,Gyr up to 6.0$-$8.0\,Gyr, whereas the thick disk population are generally older than 7.0$-$8.0\,Gyr, and peak around 8.0$-$9.0\,Gyr, which is well consistent with recent results \citep[e.g.,][]{Haywood2013, Bensby2014, Bergemann2014}, and slightly smaller than some early results with a small sample of stars \citep[e.g.,][]{Fuhrmann2011}.
We just note that the [Fe/H]--[$\alpha$/Fe] diagrams revealed by our RC sample contain lots of information on the chemical evolution history of the Galactic disk.
This is not the topic of this study and will be detailedly discussed in our further studies.

\begin{figure}[t]
\begin{center}
\includegraphics[scale=0.373,angle=0]{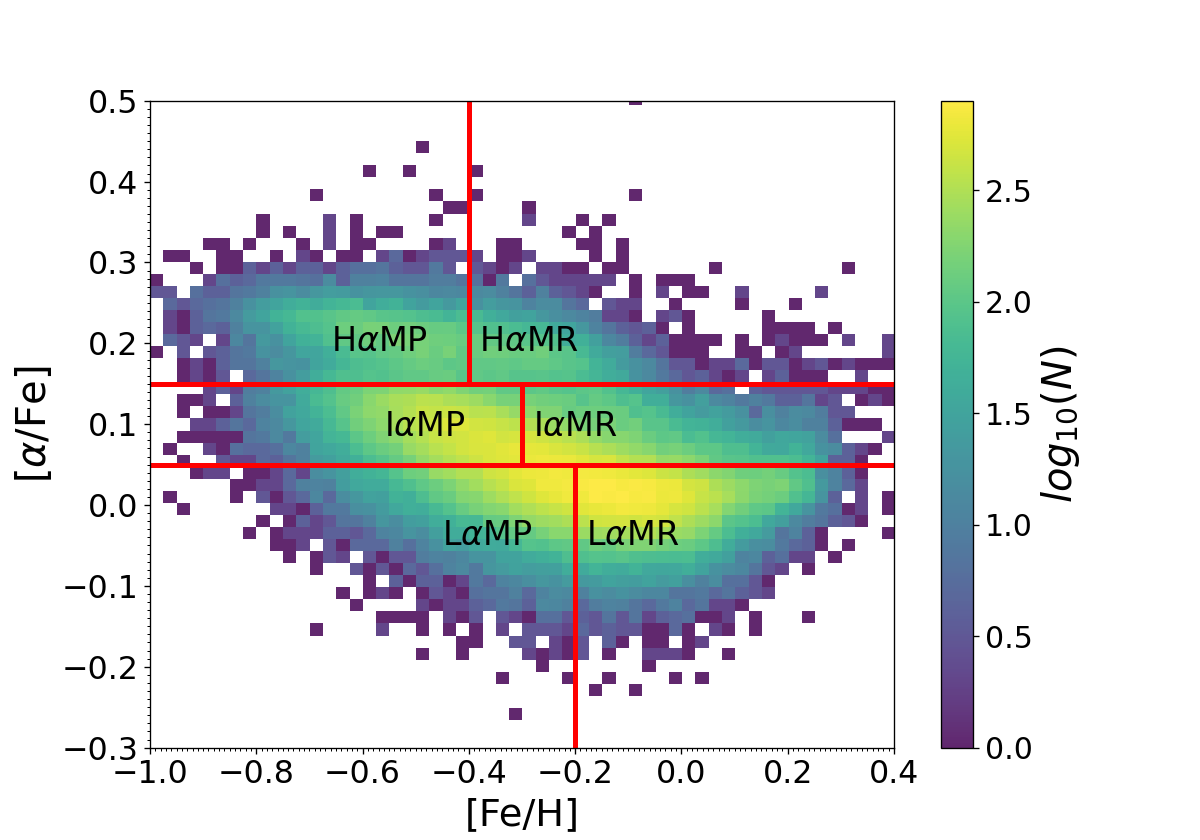}
\caption{Same as Fig.\,3, but with no less than 1 star in each bin.
The red lines define different mono-[$\alpha$/Fe]-[Fe/H] populations, and the names of various populations are marked at corresponding positions in the diagram.}
\end{center}
\end{figure}
%%\label{fig4}

\begin{table*}
\caption{The properties of sub-populations}

\centering
\setlength{\tabcolsep}{3mm}{
\begin{tabular}{lllllllll}
\hline
\hline
\specialrule{0em}{5pt}{0pt}
Name                                         &           [$\alpha$/Fe]            &        [Fe/H]      &     $\left\langle R \right\rangle$   &          $\left\langle V_{\phi} \right\rangle$        &  $\sigma_{\phi}$  &     Number     &    Fraction \\
                                             &                (dex)               &        (dex)       &     (kpc)     &        (km\,s$^{-1}$)           &    (km\,s$^{-1}$)    &           &               \\
\specialrule{0em}{5pt}{0pt}
\hline
\specialrule{0em}{3pt}{0pt}
Thin disk                                    &                 -                  &         -          &     10.14     &            225.68               &          21.10       &    104,691 &      81.45\%\\ [0.2cm]
Thick disk                                   &                 -                  &         -          &     8.54      &            182.43               &          46.11       &   14,444  &      11.23\%   \\
\specialrule{0em}{3pt}{0pt}
\hline 
\specialrule{0em}{3pt}{0pt}
0.0 $<$ age $\leq$ 4.0\,Gyr (YAGE)           &                -                   &         -          &     10.05     &            227.02               &          19.39       &   52,734  &      41.03\%  \\[0.2cm]
4.0 $<$ age $\leq$ 8.0\,Gyr (IAGE)           &                -                   &         -          &     10.10     &            223.35               &          23.58       &   50,134  &      39.00\%  \\[0.2cm]
8.0 $<$ age $\leq$ 14.0\,Gyr (OAGE)          &                -                   &         -          &     9.38      &            204.21               &          36.11       &   25,669  &      19.97\%  \\
\specialrule{0em}{3pt}{0pt}
\hline
\specialrule{0em}{3pt}{0pt}
Low [$\alpha$/Fe] metal-poor (L$\alpha$MP)   & $-$0.3 $\leq$ [$\alpha$/Fe] $<$ 0.05 & [Fe/H] $<$ $-$0.2    &     10.60     &            228.28               &          19.39       &   23,540  &      18.31\%  \\[0.2cm]
Low [$\alpha$/Fe] metal-rich (L$\alpha$MR)   & $-$0.3 $\leq$ [$\alpha$/Fe] $<$ 0.05 & [Fe/H] $\geq$ $-$0.2 &     9.51      &            223.22               &          20.52       &   41,233  &      32.07\%  \\[0.2cm]
Intermediate [$\alpha$/Fe] metal-poor (I$\alpha$MP)& 0.05 $\leq$ [$\alpha$/Fe] $<$ 0.15 & [Fe/H] $<$ $-$0.3    &     10.91     &            228.23               &          21.57       &   28,103  &      21.86\%  \\[0.2cm]
Intermediate [$\alpha$/Fe] metal-poor (I$\alpha$MR)& 0.05 $\leq$ [$\alpha$/Fe] $<$ 0.15 & [Fe/H] $\geq$ $-$0.3    &     9.60      &            218.34               &          26.22       &   20,240  &      15.74\%  \\[0.2cm]
High [$\alpha$/Fe] metal-poor (H$\alpha$MP)  & 0.15 $\leq$ [$\alpha$/Fe] $<$ 0.5  & [Fe/H] $<$ $-$0.4    &     8.92      &            179.05               &          51.88       &   9,844   &      7.65\%   \\[0.2cm]
High [$\alpha$/Fe] metal-poor (H$\alpha$MR)  & 0.15 $\leq$ [$\alpha$/Fe] $<$ 0.5  & [Fe/H] $\geq$ $-$0.4    &     8.52      &            191.90               &          37.90       &   5,588   &      4.34\%   \\
\specialrule{0em}{3pt}{0pt}
\hline
\specialrule{0em}{3pt}{0pt}
\end{tabular}}
\label{tab:datasets}
\end{table*}

\subsection{The definitions of the mono-age and mono-abundance populations of the RC sample} 
To make a more detailed analysis of the disk kinematics, the mono-age and mono-abundance populations are defined in this study for the following analysis.
For the definition of the mono-age populations, we simply divide the RC sample stars into 3 mono-age populations, i.e., the young age population with age smaller than 4\,Gyr (hereafter the YAGE population), the intermediate age population with 4.0 $<$ age $\le$ 8.0\,Gyr (hereafter the IAGE population) and the old age population with age greater than 8.0\,Gyr (hereafter the OAGE population).
The choice of the three age bins as the definition of mono-age populations is based on the typical age uncertainty and the age distribution on [Fe/H]--[$\alpha$/Fe] plane, along with the number of stars in each sub-sample.
The number of stars in each sub-population is present in Table\,1.
For the mono-[$\alpha$/Fe]-[Fe/H] populations, we divided the whole sample into 3 mono-[$\alpha$/Fe] series similar to the age series based on the predicted age--[$\alpha$/Fe] relation from the classical Galactic chemical evolution (GCE) models \citep[e.g.,][]{Matteucci2001, Matteucci2012, Pagel2009, Chiappini2009}.
Therefore, we divide the RC sample stars into low [$\alpha$/Fe] population with $-$0.3 $\leq$ [$\alpha$/Fe] $<$ 0.05\,dex, intermediate [$\alpha$/Fe] population with $-$0.3 $\leq$ [$\alpha$/Fe] $<$ 0.05\,dex, and high [$\alpha$/Fe] population with $-$0.3 $\leq$ [$\alpha$/Fe] $<$ 0.05\,dex.
Since previous studies suggested that the metal-poor and metal-rich populations exhibit different behaviors in kinematics even with the same [$\alpha$/Fe] abundances \citep[e.g.,][]{Hayden2020, Sun2023}, and therefore, we further separate each [$\alpha$/Fe] population into the metal-poor population and metal-rich population by considering the almost comparable number of stars in each sub-population.
Finally, we define 6 mono-abundance populations, that is shown in Fig.\,4 and are detailedly presented in Table 1.

\section{Results}
\subsection{Mapping the mean velocity field of disk for various populations}

\subsubsection{Radial velocity}

The radial velocity $V_{R}$, as a function of $R$ and $Z$, for various populations as defined in the above section as shown in the upper panels of Figs.\,5--6 for whole, thin and thick stars and Appendix Figs.\,A1--A2 for different populations characterized by stellar ages and chemical abundance ratios.

Generally, $V_{R}$ seems to present oscillations, with two local maximums at $R$ around 5.5\,kpc and 12.0\,kpc and a local minimum at $R$ around 9.0\,kpc.
The amplitude is about 15 km\,s$^{-1}$.
The radial gradient of $V_{R}$ displays negative values at $R$ ranging from $\sim$ 5.5\,kpc to $\sim$ 9.0\,kpc and, shows positive value at $R$ ranging from $\sim$ 9.0\,kpc to $\sim$ 12.0\,kpc, then shows again negative outwards.
In the outskirt of the disk ($R > 9.0$\,kpc), the $V_{R}$ shows a vertical gradient, with stars in the upper disk region (north) moving outward faster than the lower disk region (south), this behavior is more obvious for young populations (see YAGE and LaMR/MP in Appendix Fig.\,A1).

\begin{figure*}[t]
\centering
\subfigure{
\includegraphics[width=16.5cm]{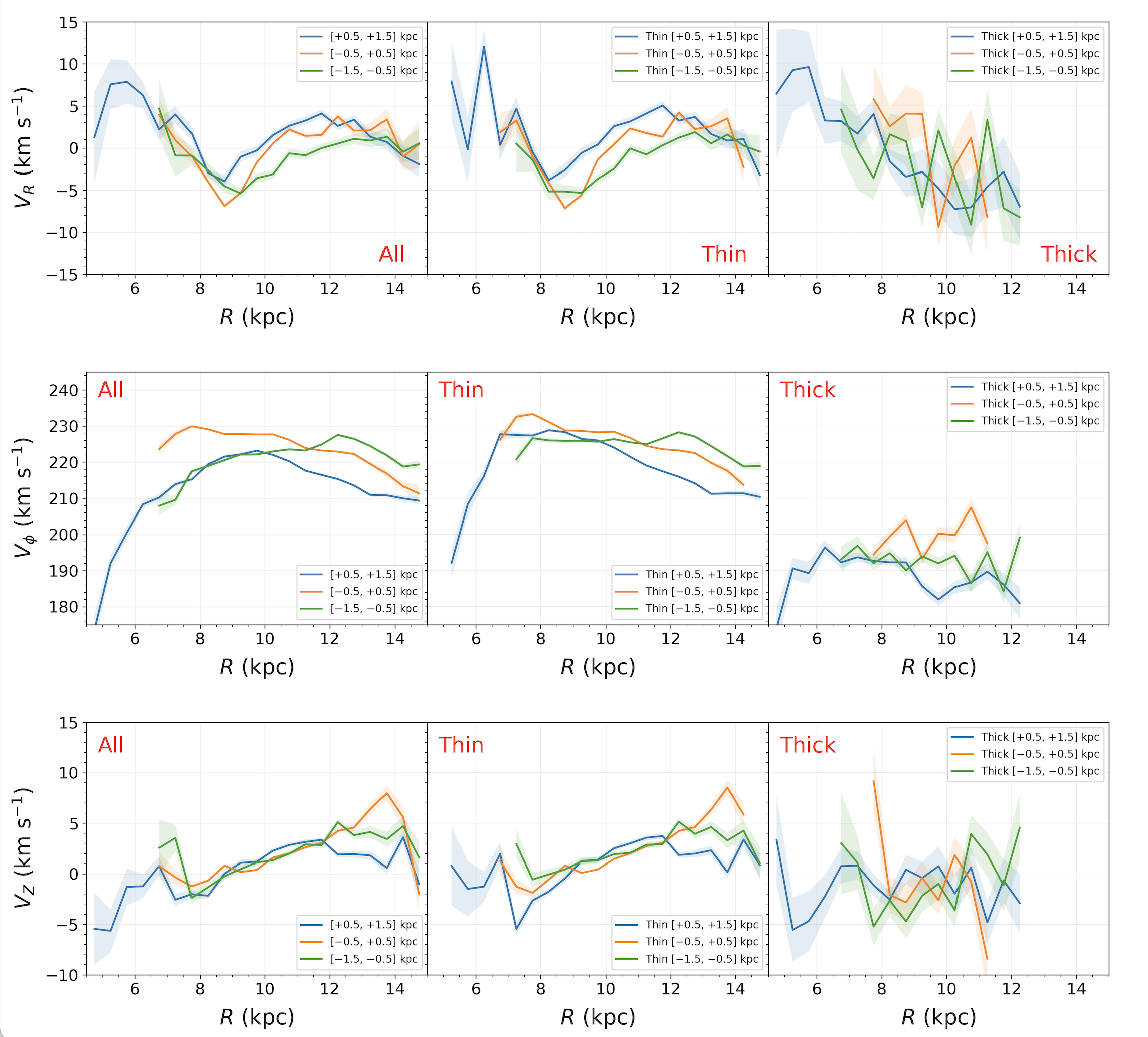}
}

\caption{Radial velocity (upper), azimuthal velocity (middle) and vertical velocity (bottom) of the whole RC sample stars (first column) and thin/thick disk populations (second and third columns) as a function of $R$ for different disk heights (lines marked by different colors).
The number of stars in each bin is required to be greater than 50 and the width of the bin is set to 0.5\,kpc.}
\end{figure*}
%%\label{Fig.5}

Using RAVE data, Siebert et al. ({\color{blue}{2011}}) found a negative radial gradient of $V_{R}$ at $R$ ranging from $\sim$ 6.0\,kpc to $\sim$ 9.0\,kpc, and this behavior has been further identified by Williams et al. ({\color{blue}{2013}}), who also report the north-south differences in velocities and, a similar result also are found in LAMOST observations \citep[e.g.,][]{Carlin2013, Carlin2014}.
Huang et al. ({\color{blue}{2016}}), Tian et al. ({\color{blue}{2017}}) and Liu et al. ({\color{blue}{2017}}) have found a positive gradient of median line-of-sight velocity with $R$ in the direction of anti-center.
These gradients are also well detected in Gaia DR2 data  \citep[e.g.,][]{Gaia Collaboration2018}.
The north-south differences in $V_{R}$ are also clearly seen in Gaia DR2 data, the results indicate that $V_{R}$ at $R$ ranging from $\sim$ 8.5\,kpc to $\sim$ 9.5\,kpc is negative for stars with $|Z|$ $\leq$ 0.6 $\sim$ 0.8\,kpc and, positive for stars above the disk plane.
Using the LAMOST-Gaia DR2 sample, Wang et al. ({\color{blue}{2019}}) further confirmed that the north-south differences in velocities show variations for various populations.

In the upper panels of Fig.\,5, the thin disk has a strong signal of the oscillation, which is consistent with the results of APOGEE \citep{Mackereth2019b}.
For the thick disk, such oscillation does not exist, and the $V_{R}$ shows a global decreasing trend from inner to outer, while with high-frequency fluctuations.
This is an interesting trend not reported yet.
We encourage further observational and simulation efforts to confirm it and understand its origin.

Our results in more detailed populations indicate that the oscillation and the north-south differences in $V_{R}$, can also be found in the populations of the thin disk, YAGE, IAGE, L$\alpha$MP, L$\alpha$MR, I$\alpha$MR and I$\alpha$MR, whereas other populations show likely weak or no signal of such behaviors (see the upper panels of Figs.\,5--6 and Appendix Figs.\,A1--A2).
In particular, the thick disk population shows no oscillation and no north-south differences in $V_{R}$.

\begin{figure*}[t]
\centering
\subfigure{
\includegraphics[width=16.5cm]{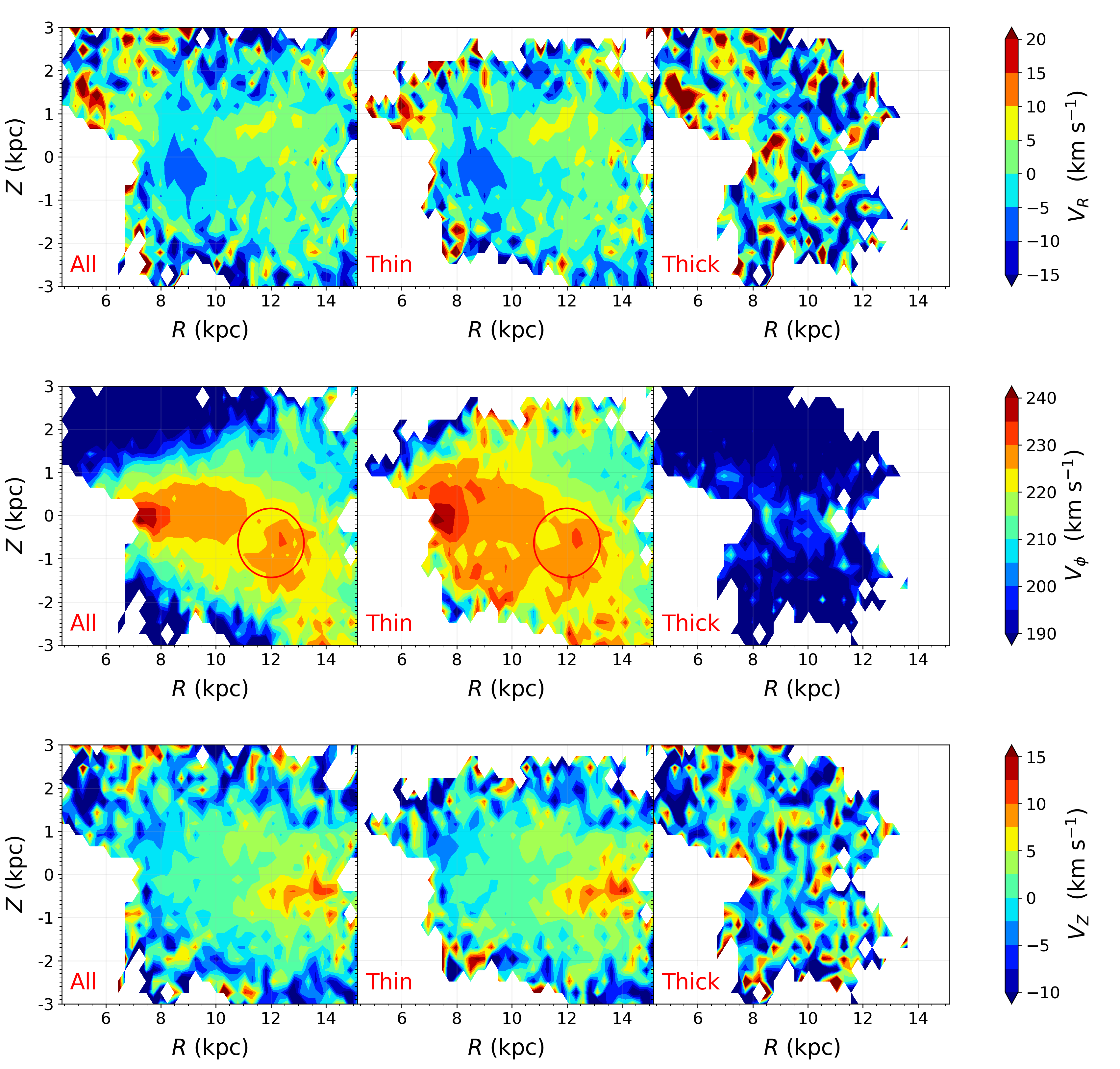}
}

\caption{Radial velocity (upper), azimuthal velocity (middle) and vertical velocity (bottom) of the whole RC sample stars (first column) and thin/thick disk populations (second and third columns) in $R$-$Z$ plane.
The color bars on the right of each panel represent the values of the velocities of different components.
There are at least 5 stars in each bin, and both $R$ and $Z$ bins have a width of 0.2\,kpc.
The red circles in the middle panels mark a possible hotter substructure, which is discussed in section 4.1.}
\end{figure*}
%%\label{Fig.6}

\subsubsection{Azimuthal velocity}

The middle panels of Fig.\,5 and Appendix Fig.\,A3 show the azimuthal velocity $V_{\phi}$, as a function of $R$ and $Z$, for various populations, and its distributions in the $R$-$Z$ plane are shown in the middle panels of Fig.\,6 and Appendix Fig.\,A4.

For the whole sample, $V_{\phi}$--$R$ of all Galactic $|Z|$ bins presents a relatively smooth inverted U$-$shape, with maximum values of $Z$ = [$-$1.5, $-$0.5]\,kpc, $Z$ = [$-$0.5, 0.5]\,kpc, and $Z$ = [0.5, 1.5]\,kpc are respectively $\sim$ 228\,km\,s$^{-1}$ at $R$ $\sim$ 12.5\,kpc, 230\,km\,s$^{-1}$ at $R$ $\sim$ 7.5\,kpc and 223\,km\,s$^{-1}$ at $R$ $\sim$ 9.5\,kpc.
At $R$ $\leq$ 11.0\,kpc, $V_{\phi}$ is largest at $|Z|$ $\leq$ 0.5\,kpc, and decreases as $|Z|$, which is consistent with the expectations \citep[e.g.,][]{Binney1987, Binney2008}.

Interestingly, $V_{\phi}$ shows clear north-south symmetry at $R > 10$\,kpc, with stars in the south plane rotating faster than stars in the north plane.
Moreover, this asymmetry is found in nearly all populations.

Using SEGUE data, Lee et al. ({\color{blue}{2011}}) detect negative gradients at $V_{\phi}$--$R$ plane for both thin and thick disk populations at $R$ between $\sim$7.0\,kpc and $\sim$10.0\,kpc, which have been further identified and characterized by further studies \citep[e.g.,][]{Williams2013, Guiglion2015, Yan2019, Han2020}.
In our result, $V_{\phi}$--$R$ for thin disk population, shows a positive gradient at $R$ less than $\sim$ 7.0\,kpc, and displays a negative gradient at $R$ larger than $\sim$ 7.0\,kpc.
While the thick disk population shows weak or no radial gradient in $V_{\phi}$.
Our results show that the gradient at $V_{\phi}$--$R$ plane is evolved, with negative gradients for thin or Y/I-AGE (see Appendix Fig.\,A3) populations as already found by previous studies and nearly zero or even positive gradients for thick or OAGE population.

In the middle panels of Fig\,6, an obvious break can be found at $R$ $\sim$ 11.0\,kpc.
This break may be the main reason of the dip that has been observed in the past \citep[e.g.,][]{Sikivie2003, Duffy2008, Sofue2009, de Boer2011, Huang2016}.
In our results, we suggest this break feature may be linked to the younger stars since break can also be clearly seen in the thin disk, YAGE and IAGE populations (Appendix Fig.\,A4).

\subsubsection{Vertical velocity}

The bottom panels of Figs.\,5--6 and Appendix Figs.\,A5--A6 display vertical velocity $V_{Z}$ of various populations, as a function of $R$ and $Z$, as well as the distributions in $R$--$Z$ plane.
A global increasing trend is detected in $V_{Z}$ for the whole sample stars from inner disk to outer disk, but with complex vertical dependencies.
The increasing trend of $V_{Z}$ with $R$ reveals the warped disk that has been widely measured and characterized by previous studies \citep[e.g.,][]{Gaia Collaboration2018, Poggio2018, Mackereth2019b}.
At R $>$ 11.0\,kpc, the $V_{Z}$ shows a rarefaction-compression pattern, with a rarefaction for the middle and south planes, and a compression for the middle and north planes.
In our results, this rarefaction-compression pattern is more obvious for the younger population and tends to be weak with age increases (see the bottom panels of Fig.\,6 and Appendix Fig.\,A6).

We note that the warp signal can be found for almost all the populations, although it is obviously weak for the thick disk, OAGE, H$\alpha$MP and H$\alpha$MR populations (see the bottom panels of Fig.\,5 and Appendix Fig.\,A5).
Our results clearly reveal that the warp is likely a long-lived feature, which is consistent with the results of Li et al. ({\color{blue}{2020}}).
For the older populations, the warp feature is still quite noisy, and a further large dataset of older stars is required to confirm this pattern, which will provide important clues to the origin of warp generation.

\subsection{Mapping the disk velocity dispersions for various populations}

\subsubsection{Radial velocity dispersion}

The radial velocity dispersion $\sigma_{R}$ of various populations, as a function of $R$ and $Z$, is shown in the upper panels of Figs.\,7--8 and Appendix Figs.\,A7--A8.
For all the RC sample stars, $\sigma_{R}$ decreases as $R$ increases at $R$ $\leq$ 13.0\,kpc and, then tends to be flat.
The gradient of the on-plane stars ($|Z|<0.5$\,kpc) is larger than that of the off-plane stars ($|Z| > 0.5$\,kpc).

\begin{figure*}[t]
\centering
\subfigure{
\includegraphics[width=16.5cm]{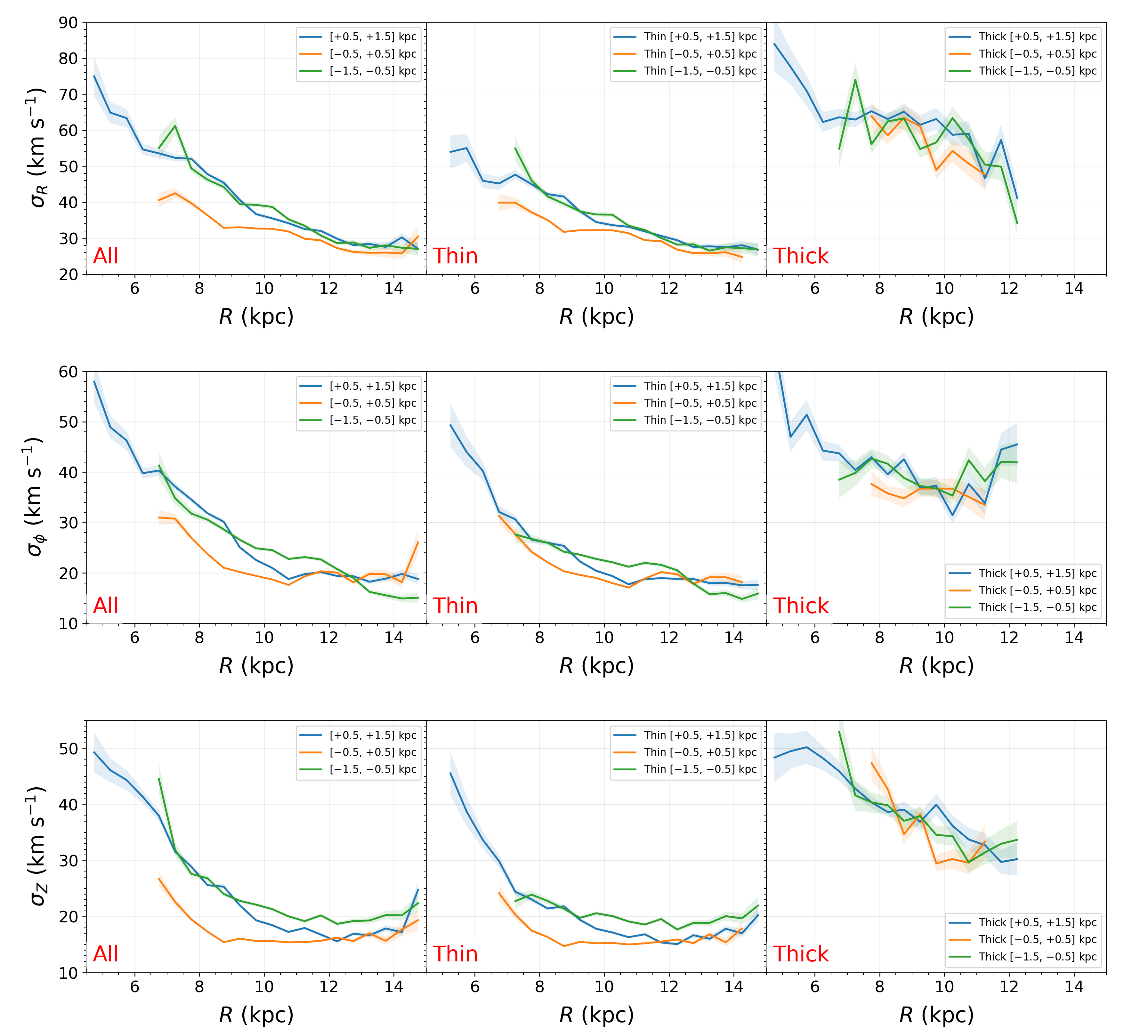}
}

\caption{Radial velocity dispersion (upper), azimuthal velocity dispersion (middle) and vertical velocity dispersion (bottom) of the whole RC sample stars (first column) and thin/thick disk populations (second and third columns) as a function of R for different disk heights (lines marked by different colors).
The number of stars in each bin is required to be greater than 50 and the width of the bin is set to 0.5\,kpc.}
\end{figure*}
%%\label{Fig.7}

The $\sigma_{R}$ shows clearly north-south differences at $R$ ranging from $\sim$ 9.0\,kpc to $\sim$ 11.0\,kpc, with the lower $Z$ plane has larger $\sigma_{R}$ than the higher $Z$ plane, and such asymmetry behavior can be found in the thin disk, YAGE, IAGE, L$\alpha$MP, L$\alpha$MR, I$\alpha$MP and I$\alpha$MR populations (see the upper panels of Fig\,8 and Appendix Fig.\,A8,).

In addition, we are also able to confirm the well-known population depended on dispersion relations in our results, as an example, $\sigma_{R}$ increases as [$\alpha$/Fe] increase for the same [Fe/H], while decreases with increasing [Fe/H] for the same [$\alpha$/Fe] (see the upper panels of Fig.\,7 and Appendix Fig.\,A7).
This general behavior is widely measured and characterized in the previous studies \citep[e.g.,][]{Quillen2001, Aumer2009, Casagrande2011, Minchev2014, Yu2018, Hayden2020, Han2020}.

The $\sigma_R$--$R$ gradient traced by the thick disk and H$\alpha$MR seems to be flatter at $R$ = 8$\sim$11\,kpc.
This behavior is also in agreement with predictions by chemo-dynamical models of the Galactic disk, where radial migration of the kinematic hotter stars is expected to smooth the gradient with time \citep[e.g.,][]{Minchev2018} and, the flaring nature of the disk may be another possible reason.
We note the decreasing trend in $\sigma_{R}$--$R$ at $R$ larger than $\sim$11.0\,kpc may be caused by the contamination of the thin disk stars where few stars are found in these distant bins.

The $\sigma_{R}$ at the solar position $\sigma_{R_0}$ and the scale length $R_{\sigma}$ of exponential decrease, are the key fundamental parameters to construct the dynamical models of the Galactic disk.
In Appendix Fig.\,B1, we determine the values of the $\sigma_{R_0}$ and $R_{\sigma}$, for both thin and thick disk populations at different disk heights.
To do so, we fit the profile of $\sigma_{R}$--$R$ with an exponential function:

\begin{equation}
\sigma_{R}(R) = \sigma_{R_0} exp \left( - \frac{R - R_{0}}{R_{\sigma}} \right)
\end{equation}
We use the python package emcee \citep{Foreman-Mackey2013} to produce an MCMC sampling of the posterior distribution of the two parameters, with defined the minimum of the least squares as the negative log-likelihood and, setting 100 `walkers', running 5,000 steps with moving the first 2,000 steps for each walker.
The corner plots of the posterior distribution of the MCMC samples of the two parameters (and their 1-$\sigma$ confidence intervals) are displayed in Appendix Fig.\,B2.
The fitting results are summarized in Table\,2.

\begin{figure*}[t]
\centering
\subfigure{
\includegraphics[width=16.5cm]{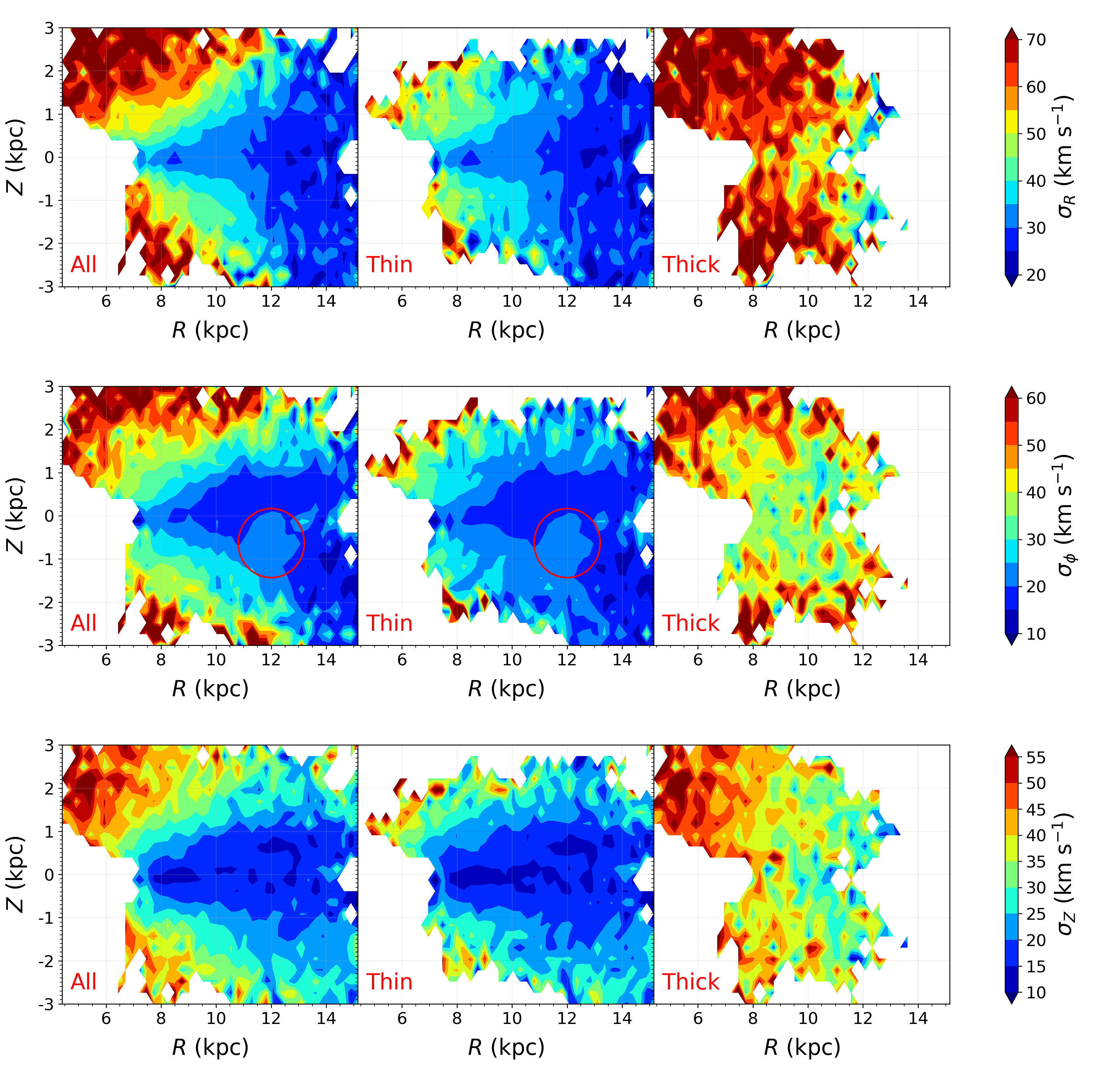}
}

\caption{Radial velocity dispersion (upper), azimuthal velocity dispersion (middle) and vertical velocity dispersion (bottom) of the whole RC sample stars (first column) and thin/thick disk populations (second and third columns) in $R$-$Z$ plane.
The color bars on the right of each panel represent the values of the velocities of different components.
There are at least 5 stars in each bin, and the $R$ and $Z$ bins both have a width of 0.2\,kpc.
The red circles in the middle panels mark a possible hotter substructure, which is discussed in section 4.1.}
\end{figure*}
%%\label{Fig.8}

For the thick disk population, the results for different heights, show sudden decreases at $R$ $>$ 12.5\,kpc, which are likely caused by the contamination of the flared [$\alpha$/Fe] rich thin disk stars \citep{Sun2020}, and therefore probably lead to the deviation of the fitting (see the upper panels of Appendix Fig.\,B1).

For thin disk population, the values of the two parameters ($\sigma_{R_0}$, $R_{\sigma}$) are (34.18\,$_{-0.16}^{+0.17}$ km\,s$^{-1}$, 20.31\,$_{-0.98}^{+1.07}$ kpc) at the disk plane ($|Z|$ = [0.0, 0.5]\,kpc), which is close to that found by Huang et al. ({\color{blue}{2016}}).
In addition, a strong signal might be linked to the Perseus Arm at around 10.25\,kpc for $|Z|$ = [0.0, 0.5]\,kpc since the $\sigma_{R}$ has an obvious increase at this region (see Appendix Fig.\,B1).
The Perseus Arm traced by our analysis located at R $\sim$ 10.25\,kpc is also in rough agreement with previous studies \citep[e.g.,][]{Xu2018, Chen2019, Sun2023}.

As expected, $\sigma_{R_0}$ of both thin and thick disk populations, increases as $|Z|$ increases, which is likely the result of the stellar migrations of the disk stars.
The values of $R_{\sigma}$ for these two populations show some differences, with the thin disk seeming to decrease as $|Z|$ increases, while the thick disk increases with increasing $|Z|$.
The increasing trend of the thin disk is likely again linked to the flaring nature of the outer disk.
While the decreasing trend of the thick disk may be caused by the contamination of the thick disk by the flared thin disk stars are stronger for the higher $|Z|$ than the lower $|Z|$.

\subsubsection{Azimuthal velocity dispersion}

The middle panels of Figs.\,7--8 and Appendix Figs.\,A9--A10 show the distributions of the azimuthal velocity dispersion $\sigma_{\phi}$ for different populations.

The results for the whole sample stars indicate that the $\sigma_{\phi}$ -- $R$ are quite complicated for different heights.
For $Z$ = [$-$1.5, $-$0.5]\,kpc, the $\sigma_{\phi}$ shows a global decreasing trend with increasing $R$.
For $Z$ = [$-$0.5, 0.5]\,kpc, the $\sigma_{\phi}$ decreases with $R$ at $R$ less than $\sim$ 10.0\,kpc, and shows no obvious change at $R$ ranging from $\sim$ 10.0 to $\sim$14.0\,kpc, and then displays a weak increasing trend beyond $R$ $\sim$ 14.0\,kpc.
For $Z$ = [0.5, 1.5]\,kpc, the $\sigma_{\phi}$ decreases with $R$ increases at $R$ less than $\sim$ 11.0\,kpc, and then shows flat pattern outwards.

\begin{table*}[t]
\caption{The $R_{\sigma}$ and $\sigma_{R_0}$ for the thin and thick disk populations at different $Z$ planes}

\centering

\setlength{\tabcolsep}{9.5mm}{
\begin{tabular}{lllllllll}
\hline 
\hline
\specialrule{0em}{5pt}{0pt}
$|Z|$ (kpc)        &   $\sigma_{R_0}$ (km\,s$^{-1}$)   &                    &     $R_{\sigma}$ (kpc)           &   \\ 
\specialrule{0em}{5pt}{0pt}                     
\hline
\specialrule{0em}{3pt}{0pt}

                    &                                    &    Thick disk      &                                   &   \\
\specialrule{0em}{3pt}{0pt}
\hline
\specialrule{0em}{3pt}{0pt}

$|Z|$ = [0.0,\,1.5] &      63.10 $_{-0.65}^{+0.64}$      &                    &      12.70 $_{-0.95}^{+1.12}$     &   \\[0.2cm]
$|Z|$ = [1.5,\,3.0] &      69.67 $_{-1.06}^{+1.07}$      &                    &      13.91 $_{-1.39}^{+1.72}$     &   \\ 
\specialrule{0em}{3pt}{0pt}                     
\hline
\specialrule{0em}{3pt}{0pt}                     

                    &                                    &    Thin disk       &                                   &   \\
\specialrule{0em}{3pt}{0pt}                     
\hline
\specialrule{0em}{3pt}{0pt}                     

$|Z|$ = [0.0,\,0.5] &      34.18 $_{-0.16}^{+0.17}$      &                    &      20.31 $_{-0.98}^{+1.07}$     &   \\[0.2cm]
$|Z|$ = [0.5,\,1.0] &      41.44 $_{-0.29}^{+0.29}$      &                    &      10.52 $_{-0.32}^{+0.33}$     &   \\[0.2cm]
$|Z|$ = [1.0,\,1.5] &      42.41 $_{-0.60}^{+0.61}$      &                    &      11.70 $_{-0.64}^{+0.73}$     &   \\[0.2cm]
$|Z|$ = [1.5,\,3.0] &      44.32 $_{-1.03}^{+1.04}$      &                    &      10.34 $_{-0.72}^{+0.84}$     &   \\
\specialrule{0em}{3pt}{0pt}                     
\hline

\end{tabular}}
\label{tab:datasets}
\end{table*}

\begin{figure*}[t]
\centering
\subfigure{
\includegraphics[width=8.8cm]{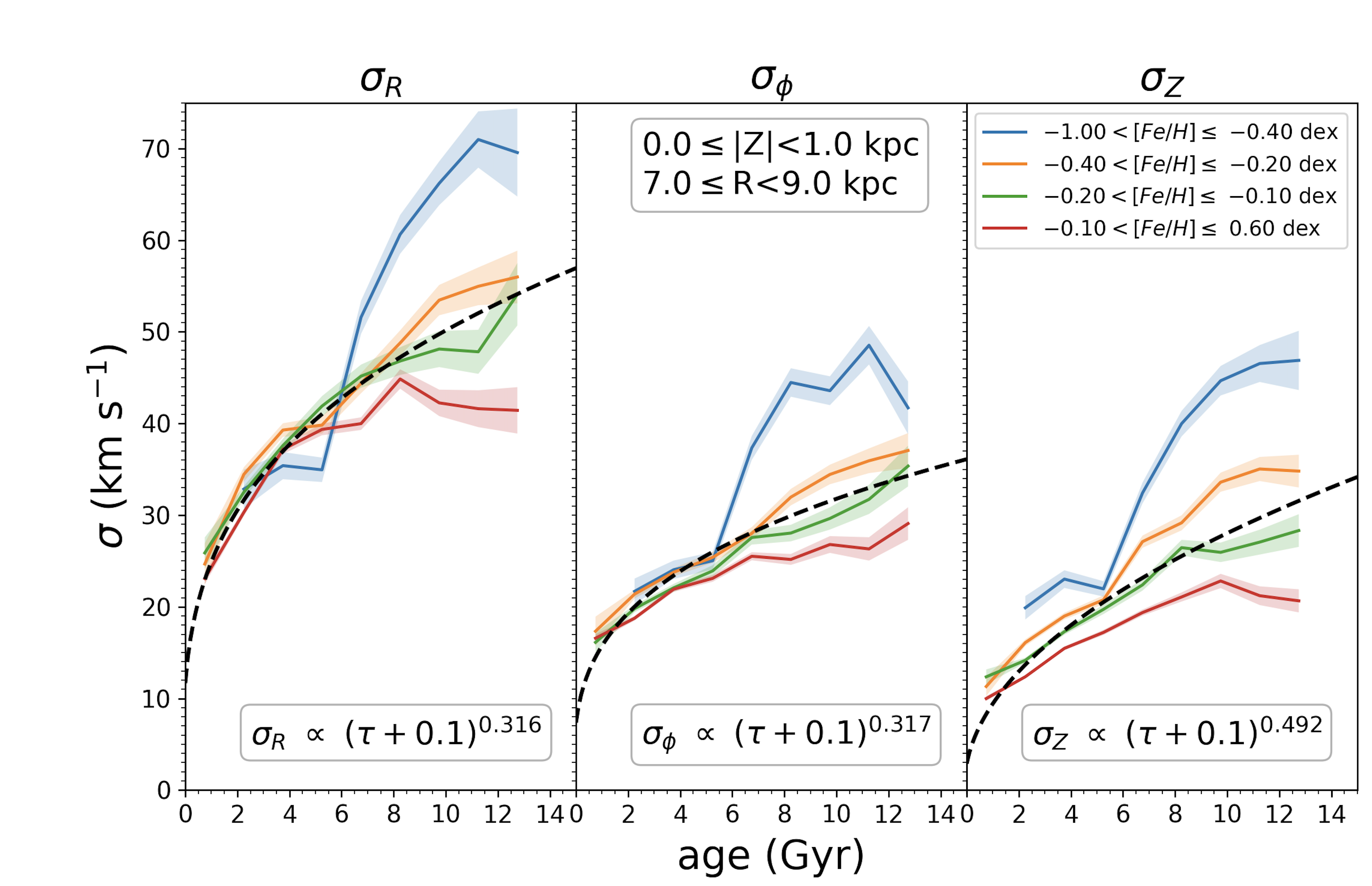}
}
\subfigure{
\includegraphics[width=8.8cm]{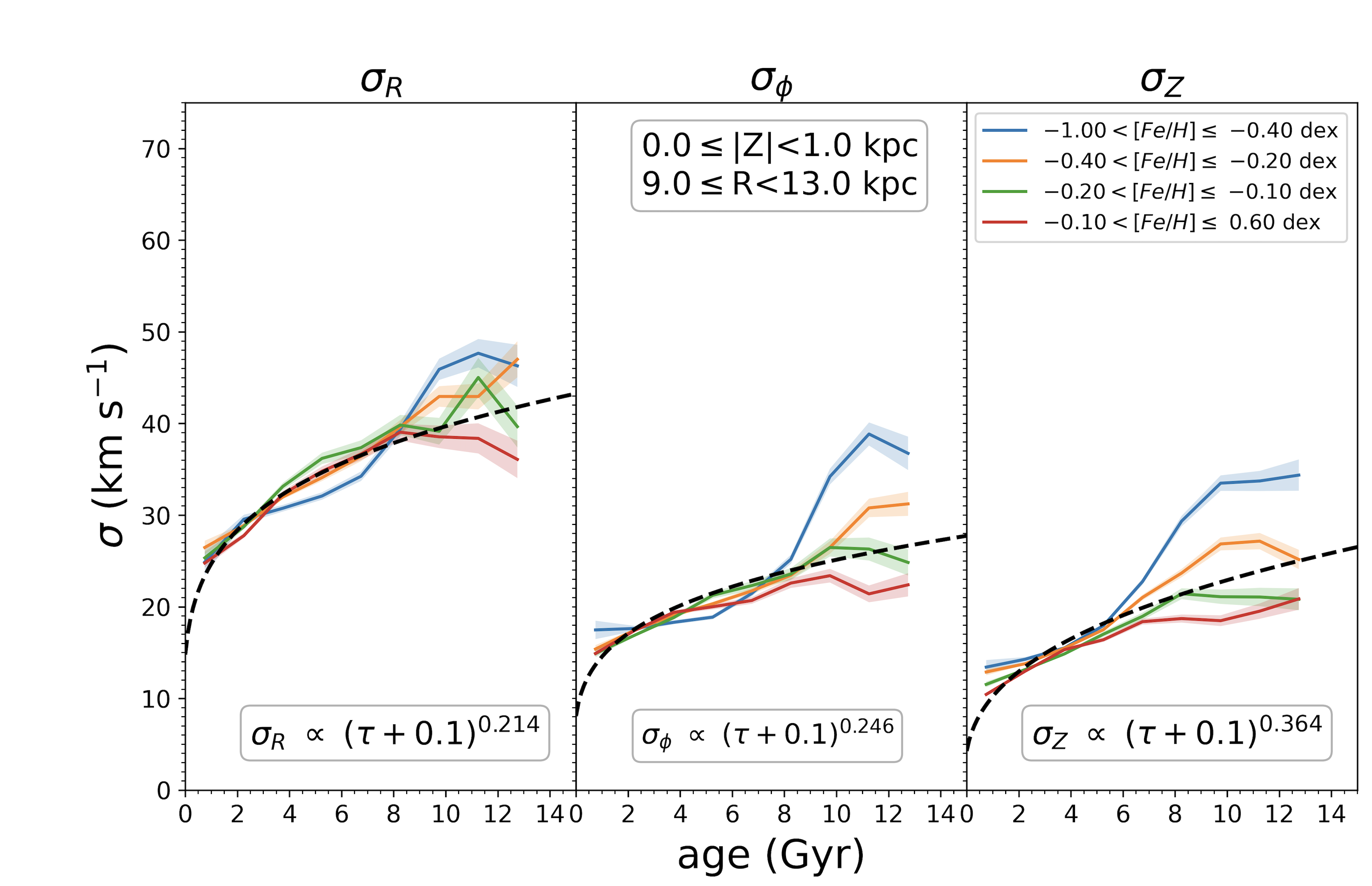}
}

\subfigure{
\includegraphics[width=8.8cm]{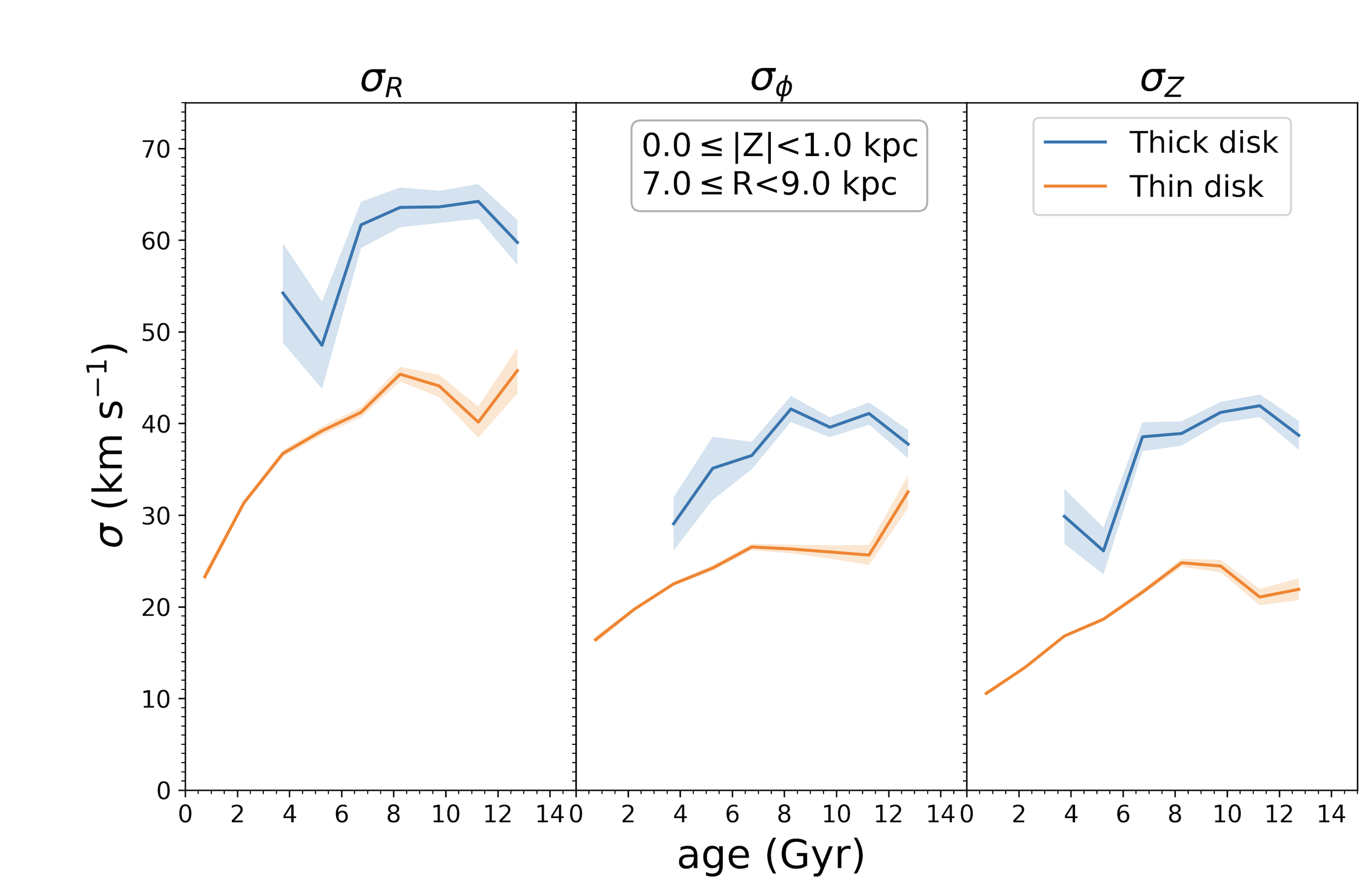}
}
\subfigure{
\includegraphics[width=8.8cm]{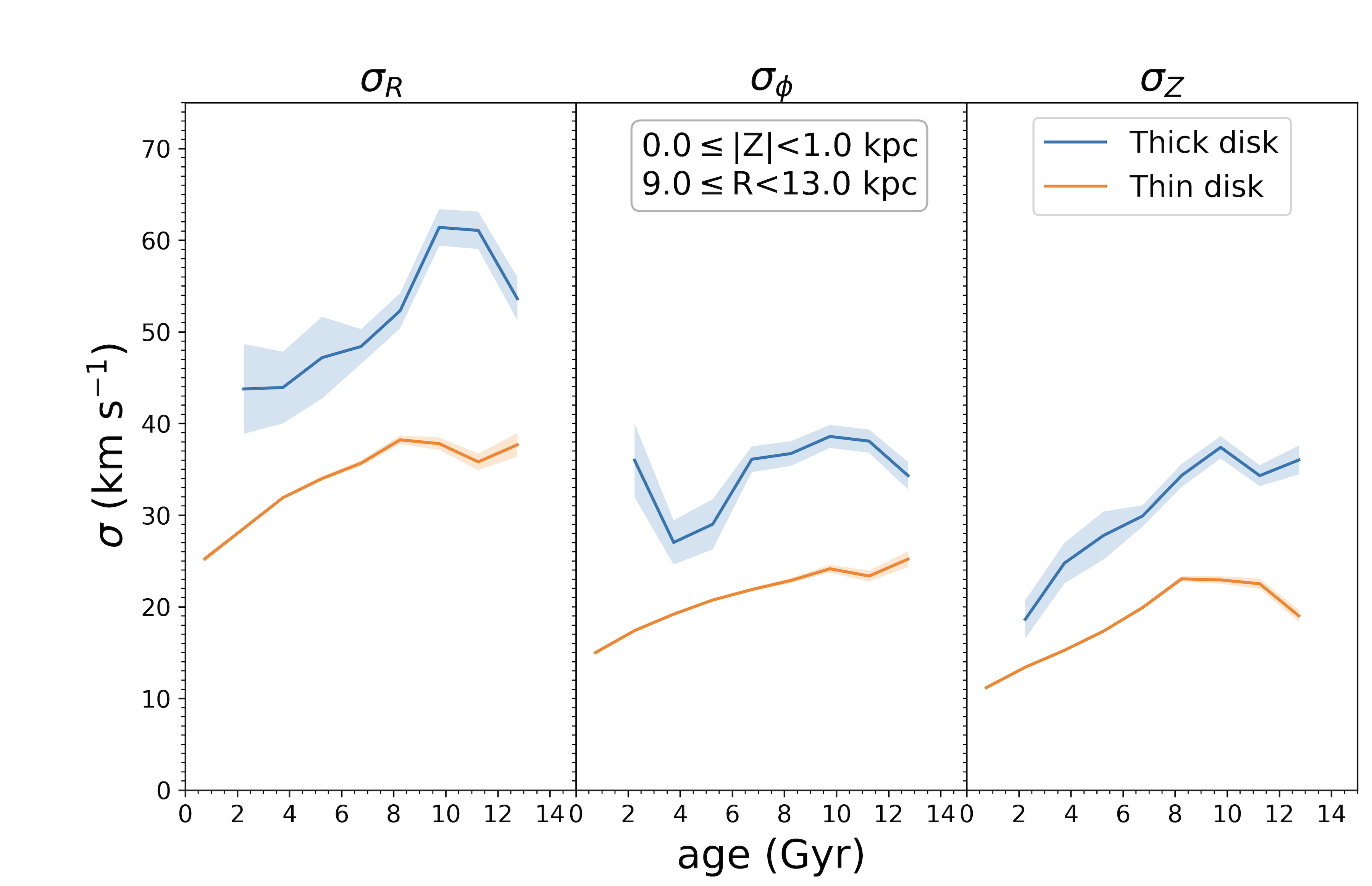}
}
\caption{The AVRs for the solar circle (top-left panel) and outer disk (top-right panel) regions, and the bottom row displays AVRs of the thin/thick disks.
There is a minimum of 20 stars per bin.}
\end{figure*}
%%\label{Fig.9}

The radial gradient of $\sigma_{\phi}$ for the whole sample stars is stronger than that for the thin disk stars.
And our results from populations of different ages indicate that this gradient becomes stronger with increasing age (see Appendix Fig.\,A9).

All the RC sample stars show an obvious kinematically hot region located at $R$ ranging from $\sim$ 11.0\,kpc to $\sim$ 13.0\,kpc at the lower disk plane (see the middle panels of Fig.\,8), which leads the Galactic north-south asymmetry of azimuthal heating \citep[e.g.,][]{Wang2019}.
We focus on this signal in various populations, finding that it is ubiquitous in the thin disk, YAGE, IAGE, L$\alpha$MP, L$\alpha$MR, I$\alpha$MP and I$\alpha$MR populations, whereas such behavior is not obvious in the thick disk, OAGE, H$\alpha$MP and H$\alpha$MR populations (see the middle panels of Fig.\,8 and Appendix Fig.\,A10).

\subsubsection{Vertical velocity dispersion}

The $\sigma_{Z}$ as a function of $R$ and $Z$, of the various populations, as shown in the bottom panels of Figs.\,7--8 and Appendix Figs.\,A11--A12.

\begin{figure*}[t]
\centering
\subfigure{
\includegraphics[width=8.8cm]{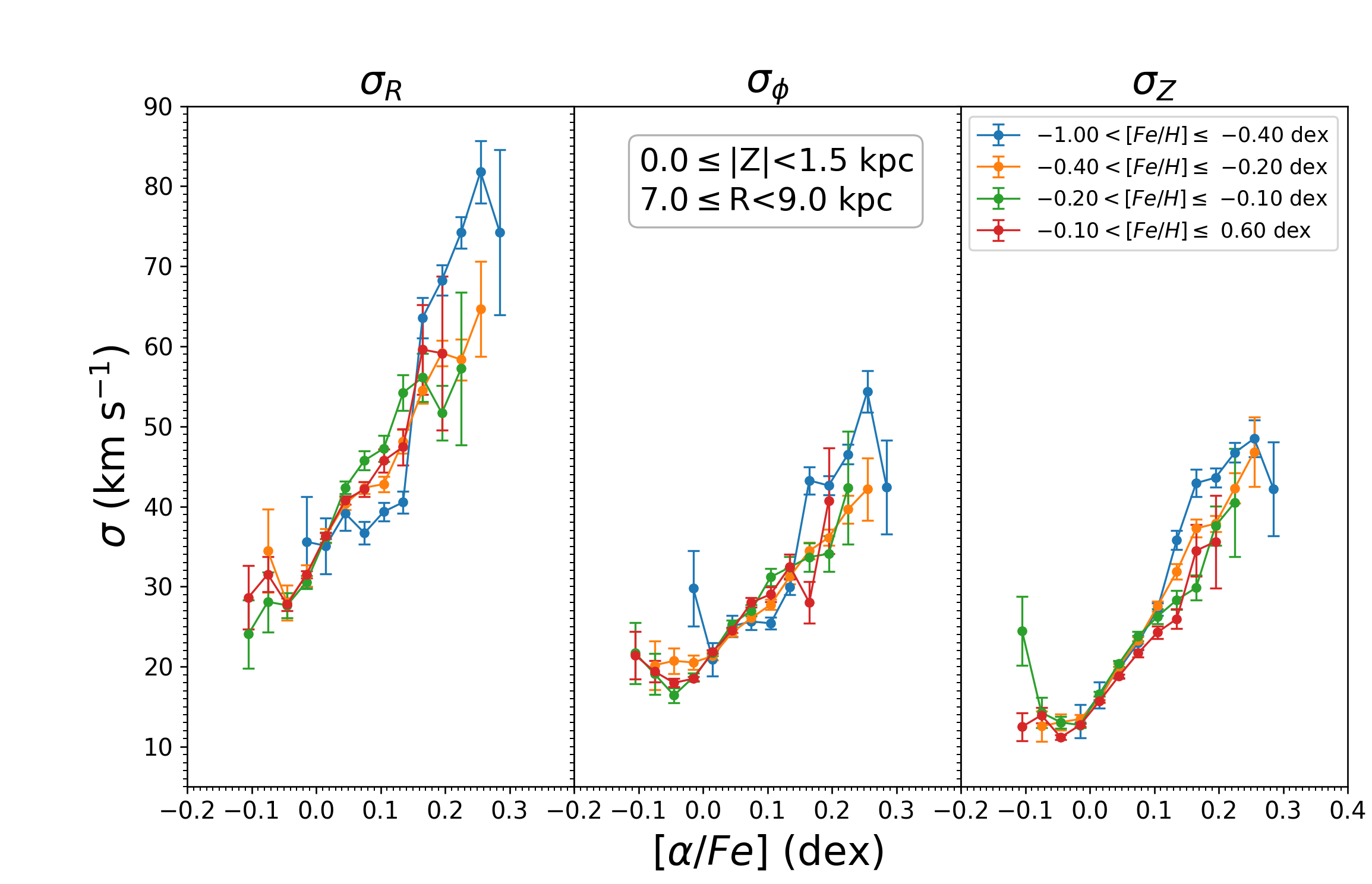}
}
\subfigure{
\includegraphics[width=8.8cm]{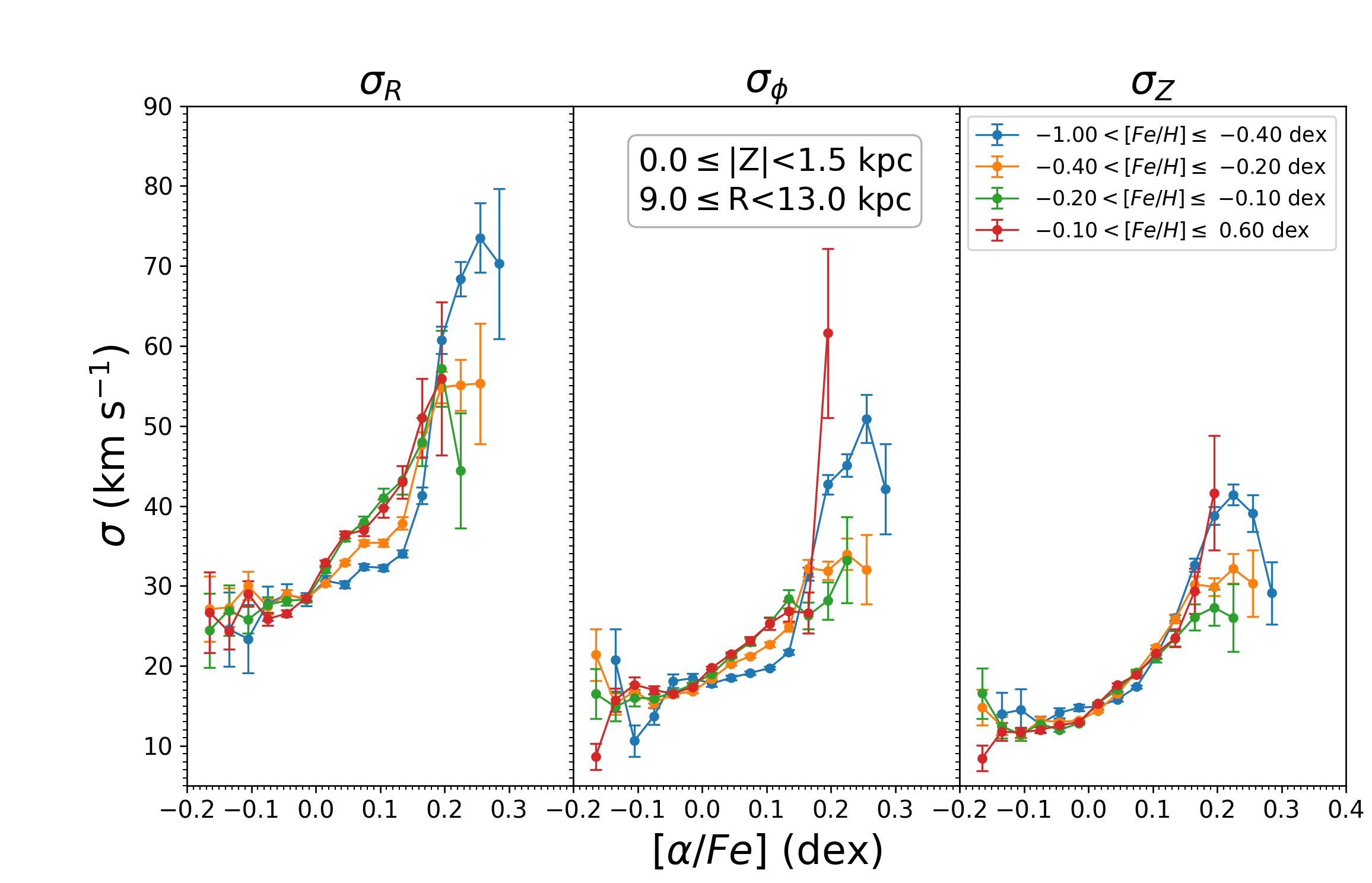}
}
\caption{The velocity dispersion as a function of [$\alpha$/Fe] and [Fe/H] for the solar circle (left panel) and outer disk (right panel) regions.
The number of stars in each bin is required to be greater than 10.}
\end{figure*}
%%\label{Fig.10}

\begin{table*}[t]

\caption{The slopes of the AVRs that are modeled as a power law $\sigma$ $\propto$ ($\tau + 0.1)^{\beta}$}

\setlength{\tabcolsep}{7.6mm}{
\centering
\begin{tabular}{cccccc}
\hline
\hline
\specialrule{0em}{5pt}{0pt}
Region                          &          [Fe/H] (dex)         &    $\beta_{R}$    & $\beta_{\phi}$    &     $\beta_{Z}$        \\

\specialrule{0em}{5pt}{0pt}
\hline
\specialrule{0em}{5pt}{0pt}
                                &           All stars               & 0.316 $\pm$ 0.008 & 0.317 $\pm$ 0.008 & 0.492 $\pm$ 0.009  \\[0.2cm]
7.0 $\leq$ $R$ $<$ 9.0\,kpc     & $-$1.00 $<$ [Fe/H] $\leq$ $-$0.40 & 0.656 $\pm$ 0.001 & 0.609 $\pm$ 0.001 & 0.710 $\pm$ 0.001  \\[0.2cm]
           \&                   & $-$0.40 $<$ [Fe/H] $\leq$ $-$0.20 & 0.287 $\pm$ 0.017 & 0.321 $\pm$ 0.018 & 0.497 $\pm$ 0.018  \\[0.2cm]
0.0 $\leq$ $|Z|$ $<$ 1.0\,kpc   & $-$0.20 $<$ [Fe/H] $\leq$ $-$0.10 & 0.275 $\pm$ 0.017 & 0.287 $\pm$ 0.018 & 0.409 $\pm$ 0.018  \\[0.2cm]
                                & $-$0.10 $<$ [Fe/H] $\leq$  \,\,0.60   & 0.258 $\pm$ 0.010 & 0.220 $\pm$ 0.011 & 0.354 $\pm$ 0.011  \\

\specialrule{0em}{5pt}{0pt}
\hline
\specialrule{0em}{5pt}{0pt}
                                &           All stars               & 0.214 $\pm$ 0.005 & 0.246 $\pm$ 0.005 & 0.364 $\pm$ 0.005  \\[0.2cm]
9.0 $\leq$ $R$ $<$ 13.0\,kpc    & $-$1.00 $<$ [Fe/H] $\leq$ $-$0.40 & 0.247 $\pm$ 0.014 & 0.337 $\pm$ 0.016 & 0.566 $\pm$ 0.015  \\[0.2cm]
           \&                   & $-$0.40 $<$ [Fe/H] $\leq$ $-$0.20 & 0.221 $\pm$ 0.009 & 0.230 $\pm$ 0.009 & 0.357 $\pm$ 0.010  \\[0.2cm]
0.0 $\leq$ $|Z|$ $<$ 1.0\,kpc   & $-$0.20 $<$ [Fe/H] $\leq$ $-$0.10 & 0.222 $\pm$ 0.011 & 0.238 $\pm$ 0.011 & 0.286 $\pm$ 0.011  \\[0.2cm]
                                & $-$0.10 $<$ [Fe/H] $\leq$  \,\,0.60   & 0.208 $\pm$ 0.009 & 0.169 $\pm$ 0.008 & 0.266 $\pm$ 0.009  \\
\specialrule{0em}{5pt}{0pt}
\hline
\specialrule{0em}{3pt}{0pt}
\end{tabular}
}
\label{tab:datasets}
\end{table*}

The results from the whole sample stars indicate that for $|Z|$ $\leq$ 0.5\,kpc, the $\sigma_{Z}$ shows a strong negative gradient at $R$ $\leq$ 9.0\,kpc , and tends to be flat at $R$ ranging from $\sim$9.0\,kpc to $\sim$12.0\,kpc, and then shows weak increasing trend outwards.
For $|Z|$ $\geq$ 0.5\,kpc, the $\sigma_{Z}$ displays a global decreasing trend at $R$ less than $\sim$12.0\,kpc, and then shows weak increasing outwards.
These results are largely consistent with previous observations \citep[e.g.,][]{Gaia Collaboration2018, Sanders2018}.
The increasing trend of $\sigma_{Z}$--$R$ indicates the flaring nature of the outer Galactic disk \citep[e.g.,][]{Bovy2016, Mackereth2017, Sarkar2018, Mackereth2019b}.
Based on our results, the signal of disk flaring is significant for the thin disk, YAGE, IAGE, L$\alpha$MP, L$\alpha$MR, I$\alpha$MP and I$\alpha$MR populations and much weaker for old populations.

\subsection{The age--velocity dispersion relations (AVRs) and [$\alpha$/Fe]--velocity dispersion relations}

Stellar age-velocity dispersion relations (AVRs) in the solar neighborhood have been widely confirmed to be valuable for constraining the structures and evolution of the Milky Way \citep[e.g.,][]{Stromberg1946, Wielen1977, Nordstrom2004, Yu2018, Sharma2021}.
Previous studies indicate that the nature of the increasing trend of the AVRs may be linked to the heating and perturbation histories of the Galactic disk \citep[e.g.,][]{Stromberg1946, Wielen1977, Nordstrom2004, Quillen2001, Aumer2009, Holmberg2009, Yu2018, Sharma2021}.
Quillen \& Garnett ({\color{blue}{2001}}) reported a jump in AVRs for stars older than $\sim$9.0\,Gyr may be linked to a thick disk component that is shaped by a minor merger.
However, other studies tend to suggest that stars older than $\sim$ 8.0\,Gyr generally display a simple power law in AVRs \citep[e.g.,][]{Aumer2009, Holmberg2009}.

In our results, the AVRs of various populations for different Galactic disk regions are shown in Fig.\,9.
The dashed line in the panel is the best fitting of AVRs for the whole sample stars in the region that is modeled with a simple power law as follows:

\begin{equation}
\sigma_{v} = \sigma_{v,0} (\tau + 0.1)^{\beta_{v}}
\end{equation}

Here we defined $\sigma_{v}$ changes with ($\tau + 0.1)^{\beta_{v}}$ since the contribution of stellar birth dispersion on AVRs is generally considered to be no less than 0.1\,Gyr \citep[e.g.,][]{Sharma2021}, and the $\sigma_{v,0}$ is the constant of the exponential growth of $\sigma_{v}$ with $\tau$.

\begin{figure*}[t]
\centering
\subfigure{
\includegraphics[width=16.5cm]{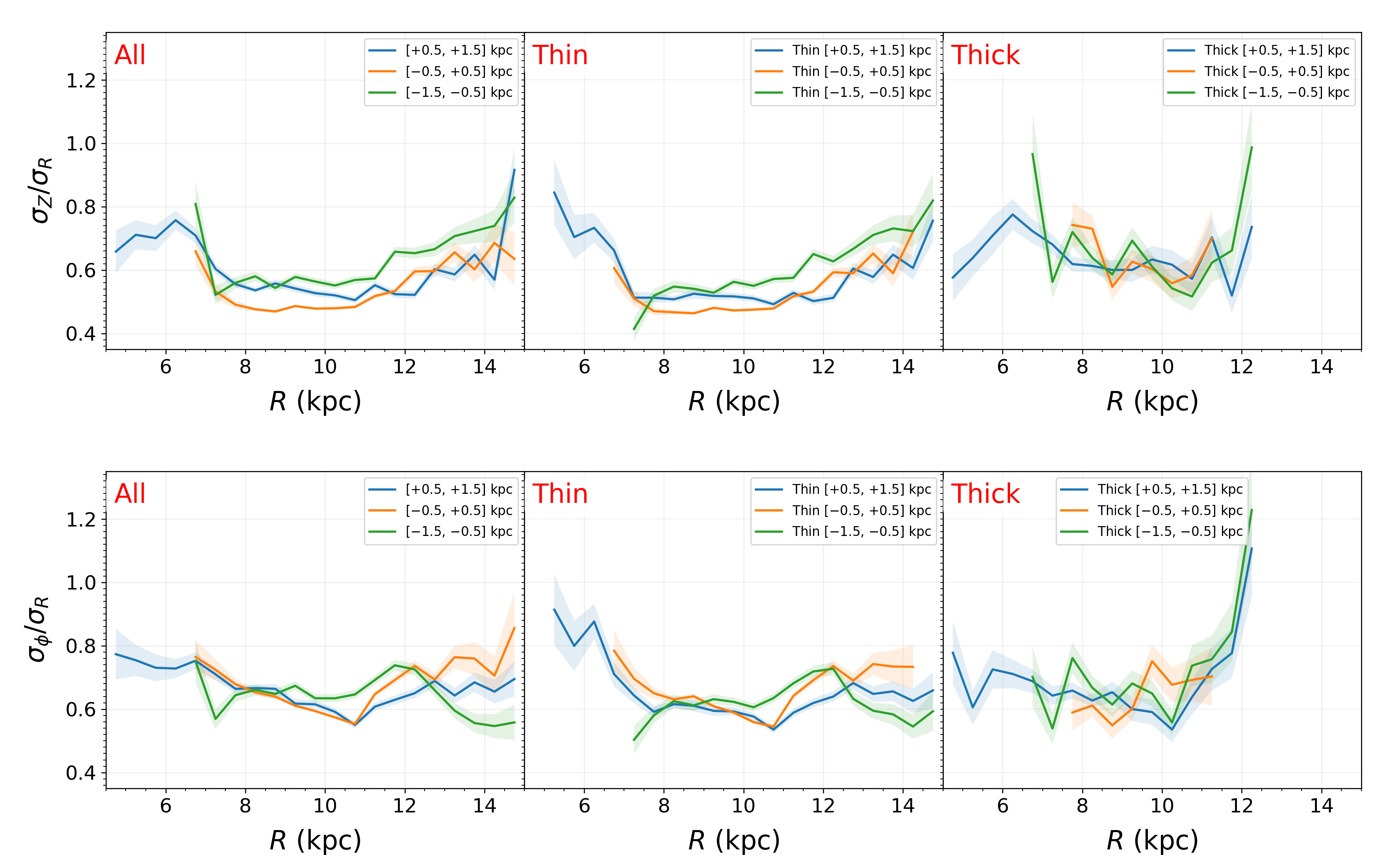}
}

\caption{Ratios of $\sigma_{Z}$/$\sigma_{R}$ (upper) and the $\sigma_{\phi}$/$\sigma_{R}$ (bottom) of the whole RC sample stars (first column) and thin/thick disk populations (second and third columns) as a function of $R$ for different disk heights (lines marked by different colors).
The number of stars in each bin is required to be greater than 50 and the width of the bin is set to 0.5\,kpc.}
\end{figure*}
%%\label{Fig.11}

The results in the upper panels of Fig.\,9 and Table 3, indicate that all mono-[Fe/H] populations show a global increasing trend in $\sigma$ as age increases, and the slopes of different velocity dispersions are displayed $\beta_{R}$ $<$ $\beta_{\phi}$ $<$ $\beta_{Z}$ for all stars without further binning on [Fe/H], with values $\beta_{R}$ = 0.316, $\beta_{\phi}$ = 0.317 and, $\beta_{Z}$ = 0.492 at the solar circle with $|Z|$ $<$ 1.0\,kpc and 7.0 $\leq$ $R$ $<$ 9.0\,kpc, and values $\beta_{R}$ = 0.214, $\beta_{\phi}$ = 0.246 and, $\beta_{Z}$ = 0.364 at the outer disk with 9.0 $\leq$ $R$ $<$ 13.0\,kpc and $|Z|$ $<$ 1.0\,kpc.

Generally, the AVR relations are not well separated for younger populations with age smaller than 7.0 Gyr, and the velocity dispersions significantly increase as decreasing [Fe/H] under the same age bins for old populations with age older than 7.0 Gyr.
In particular, a jump increase in AVRs for stars older than $\sim$ 7.0$-$9.0\,Gyr.
This jump increase could be caused by a broad variety of factors:
(i) as a thick disk component caused by a violent origin as suggested in previous studies \citep[e.g.,][]{Quinn1993, Abadi2003, Kazantzidis2008, Villalobos2008, Brook2004, Brook2005, Brook2007, Bournaud2009};
(ii) may be caused by the underestimation of the age of the older stars, especially for the metal-poor stars.
At present, we can not rule out any of these two factors since current observations and theories can not provide a high-precision measurement of stellar age \citep[e.g.,][]{Yu2018, Mackereth2019b, Huang2020, Spina2021}.
However, the results of AVRs for different disk volumes indicate that an obvious gap for thin and thick disks (see the bottom panels of Fig.\,9), which supports the jump of the AVRs at age around 7.0$-$9.0\,Gyr may be caused by the violent heating nature of thick disk.

We also fit the AVRs with a simple power law for various mono-[Fe/H] populations in Appendix Fig.\,C1, and the slopes ($\beta_{R}$, $\beta_{\phi}$, $\beta_{Z}$) are also summarised in Table 3.
For metal-poor populations, the AVRs show complicated oscillations and can not be fitted well by a simple power law function.
In addition, a global trend of $\beta_{R}$ $<$ $\beta_{\phi}$ $<$ $\beta_{Z}$ can be also clear seen for different mono-[Fe/H] populations (see Table 3).

We further investigate the $\sigma$--[$\alpha$/Fe] relations, as shown in Fig.\,10.
The results indicate that the three velocity dispersions increase as [$\alpha$/Fe] increases, with $\sigma_{R}$ $>$ $\sigma_{\phi}$ $>$ $\sigma_{Z}$ at the same [$\alpha$/Fe].
And the jump increase is obviously found at [$\alpha$/Fe] around 0.1$-$0.2\,dex, which further confirms the reliability of the jump increase in AVRs (see the upper panels of Fig.\,9).

\subsection{Ratios of velocity dispersions of various populations}

The ratios of different velocity dispersions can provide some important information to understand the heating processes \citep[e.g.,][]{Kuijken1991, Dehnen1998, Aumer2009, Sharma2014}.
We measure the ratios of $\sigma$, as a function of $R$ and $Z$, for various populations, as shown in Figs.\,11--12 and Appendix Figs.\,D1--D4.

\begin{figure*}[t]
\centering
\subfigure{
\includegraphics[width=16.5cm]{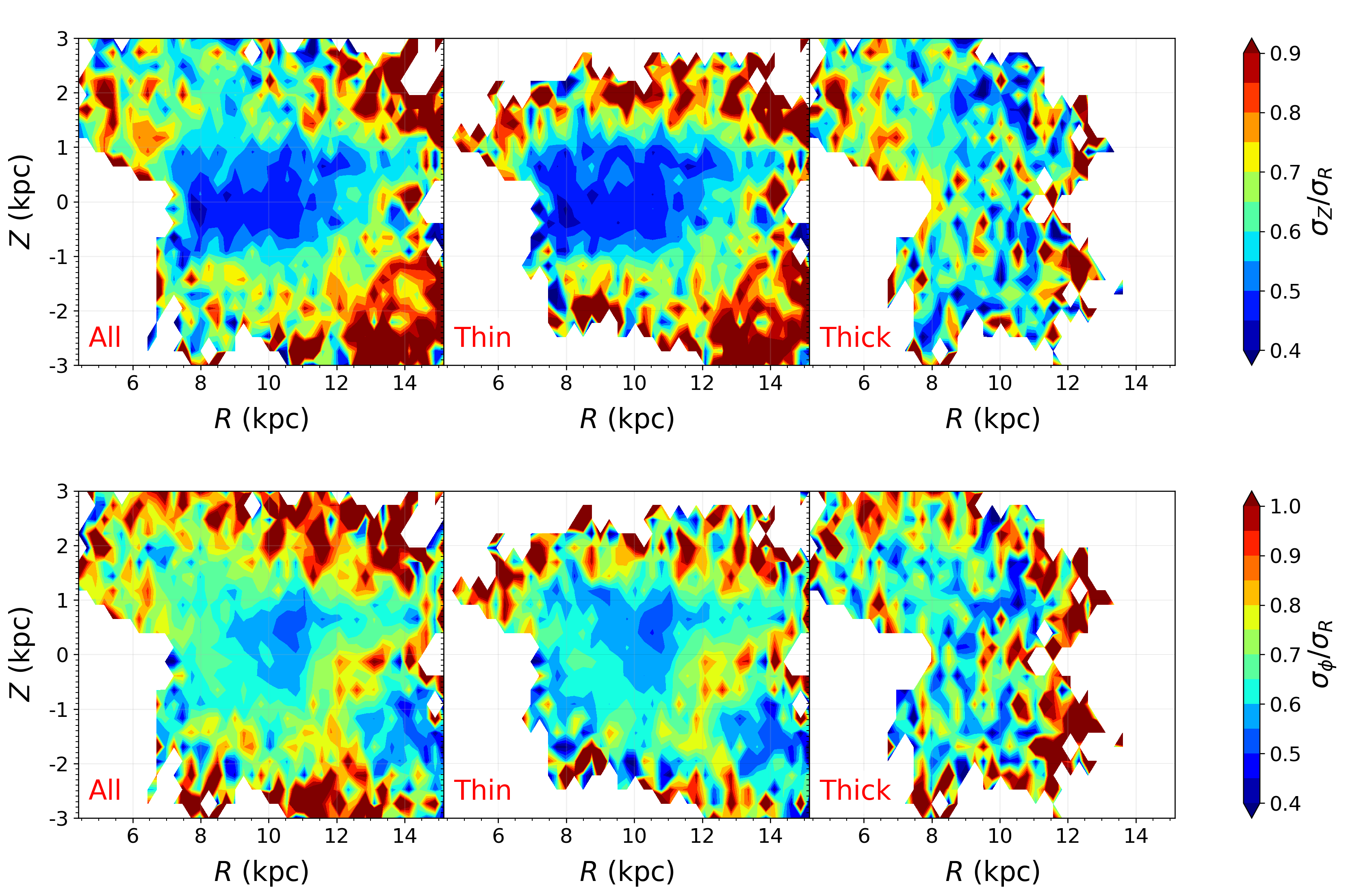}
}
\caption{Ratios of $\sigma_{Z}$/$\sigma_{R}$ (upper) and the $\sigma_{\phi}$/$\sigma_{R}$ (bottom) of the whole RC sample stars (first column) and thin/thick disk populations (second and third columns) in $R$-$Z$ plane.
The color bars on the right of each panel represent the values of the ratios of different components.
There are at least 5 stars in each bin, and both $R$ and $Z$ bins have a width of 0.2\,kpc.}
\end{figure*}
%%\label{Fig.12}

The $\sigma_{Z}/\sigma_{R}$ as function of $R$ and $Z$ for various populations are shown in the upper panels of Figs.\,11--12 and Appendix Figs.\,D1--D2.
The result of the whole sample stars indicates that for $Z$ = [$-$1.5, $-$0.5]\,kpc, the $\sigma_{Z}/\sigma_{R}$ displays a constant value of 0.5\,$\sim$\,0.6 at 8.0 $<$ $R$ $<$ 11.5\,kpc, and then increases as $R$ increases outwards, raise to larger than $\sim$ 0.6 at $R$ $\geq$ 11.5\,kpc.
For $Z$ = [$-$0.5, 0.5]\,kpc, the $\sigma_{Z}/\sigma_{R}$ decreases as $R$ increases at $R$ $\leq$ 8.0\,kpc, and shows a constant value with $\sim$ 0.5 at $R$ ranging from $\sim$ 8.0\,kpc to $\sim$ 11.0\,kpc, and then increases to over $\sim$ 0.6 at $R$ $>$ 12.5\,kpc. 
For $Z$ = [0.5, 1.5]\,kpc, the $\sigma_{Z}/\sigma_{R}$ decreases as $R$ increases at $R$ $\leq$ 7.0\,kpc, and then shows a constant value around 0.5$\sim$0.6 at 7.0 $<$ $R$ $\leq$ 14.0\,kpc, and then rise again to over 0.6 at $R$ $>$ 14.0\,kpc.
At $R$ $\geq$ 9.0\,kpc, the $\sigma_{Z}/\sigma_{R}$ is larger for $Z$ = [$-$1.5, $-$0.5]\,kpc than that for $Z$ = [0.5, 1.5]\,kpc, implying south disk suffered more vertical heating than the north disk (see the upper panels of Fig.\,11--12).
This behavior is extremely clear in the younger populations, e.g., the populations of the thin disk, YAGE, IAGE, L$\alpha$MP, L$\alpha$MR, I$\alpha$MP and I$\alpha$MR (see the upper panels of Fig.\,11--12 and Appendix Fig.\,D1--D2).

For the thin disk population, the $\sigma_{Z}/\sigma_{R}$ decreases as $R$ increases at $R$ $\leq$ 7.0\,kpc, and then displays a global smooth increase outwards, with the value rising from $\sim$ 0.4 at $R$ $\sim$ 7.0\,kpc to $\sim$ 0.7 at $R$ $\sim$ 14.0\,kpc.
This is roughly consistent with the results of Sharma et al. ({\color{blue}{2014}}), whose reported $\sigma_{Z}/\sigma_{R}$ $\sim$ is $\sim$ 0.44 at age $\sim$ 3.0\,Gyr, and rise to $\sim$ 0.65 at age $\sim$ 10.0\,Gyr.

For the thick disk population, the $\sigma_{Z}/\sigma_{R}$ is almost constant with value around 0.5 $-$ 0.6 at 7.0 $<$ $R$ $<$ 11.5\,kpc, and then increases as $R$ increases at $R$ $\geq$ 11.5\,kpc, in which the value of $Z$ = [$-$1.5, $-$0.5]\,kpc seems to rise to nearly 1.0 at $R$ $\sim$ 12.5\,kpc.
At solar radius, $\sigma_{Z}/\sigma_{R}$ ranges from $\sim$ 0.6 to $\sim$ 0.8, which is roughly consistent with previous studies \citep[e.g.,][]{Sharma2014, Anguiano2018, Hayden2020}, which suggested $\sigma_{Z}/\sigma_{R}$ is 0.8 for Shu-model, is 0.68 for Gaussian model \citep[e.g.,][]{Sharma2014}, and is $\sim$ 0.7 for a series of observations \citep[e.g.,][]{Anguiano2018, Hayden2020}.
In addition, our different age populations results indicate that $\sigma_{Z}/\sigma_{R}$ shows an increasing trend with age, which is in rough agreement with the results of recent observations \citep[e.g.,][]{Sharma2014} and simulations \citep[e.g.,][]{Aumer2009}.

The $\sigma_{\phi}/\sigma_{R}$ as function of $R$ and $Z$ for various populations are shown in the bottom panels of Figs.\,11--12 and Appendix Figs.\,D3--D4.
The result of the whole sample stars indicates that for $Z$ = [$-$0.5, 0.5]\,kpc and $Z$ = [0.5, 1.5]\,kpc, the $\sigma_{\phi}/\sigma_{R}$ decreases as $R$ increases at $R$ $\leq$ 11.0\,kpc, with the value from $\sim$ 0.78 at $R$ $\sim$ 7.0\,kpc to $\sim$ 0.57 at $R$ $\sim$ 11.0\,kpc, and then increases outwards.
For $Z$ = [$-$1.5, $-$0.5]\,kpc, the $\sigma_{\phi}/\sigma_{R}$ shows a constant value of $\sim$ 0.65 at $R$ ranging from $\sim$ 8.0\,kpc to $\sim$ 11.0\,kpc, and then increases as $R$ increases at $R$ ranging form $\sim$ 11.0\,kpc to $\sim$ 12.0\,kpc, then decreases outwards.

For thin disk, stars with $Z$ = [$-$0.5, 0.5]\,kpc and $Z$ = [0.5, 1.5]\,kpc, the $\sigma_{\phi}/\sigma_{R}$ shows decreasing trend at $R$ $\leq$ 11.0\,kpc, and then increases as $R$ increases outwards.
Stars with $Z$ = [$-$1.5, $-$0.5]\,kpc, the $\sigma_{\phi}/\sigma_{R}$ increases as $R$ increases at $R$ $\leq$ 12.5\,kpc, and then decreases outwards.
For thick disk stars, stars with $|Z|$ $\geq$ 0.5\,kpc, the $\sigma_{\phi}/\sigma_{R}$ shows a global weak decreasing trend at $R$ $\leq$ 10.25\,kpc, with values from $\sim$ 0.7 at $R$ $\sim$ 8\,kpc to $\sim$ 0.6 at $R$ $\sim$ 10.25\,kpc, and then increases outwards.
While stars with $Z$ = [$-$0.5, 0.5]\,kpc, the $\sigma_{\phi}/\sigma_{R}$ shows a global weak increasing trend as $R$ increases, with the value rising from $\sim$ 0.6 to $\sim$ 0.7.

It is worth noting that, our results of thin and thick disk populations in the solar neighborhood are well consistent with previous studies \citep[e.g.,][]{Anguiano2018, Hayden2020}, as an example, Anguiano et al. ({\color{blue}{2018}}) reported the $\sigma_{\phi}/\sigma_{R}$ of the thin and thick disk populations in the solar neighborhood are respectively $\sim$ 0.67 and $\sim$ 0.7.
Their thick disk result is roughly consistent with our result, but their thin disk results are slightly larger than our thin disk result of $\sigma_{\phi}/\sigma_{R}$ $\sim$ 0.6, while our result is in good agreement with the result that predicted by the rotation curve \citep[i.e., $\sigma_{\phi}/\sigma_{R}$ = 0.5 $\sim$ 0.6, e.g.,][]{Kuijken1991, Evans1993}, and is also consistent with the result reported value by Dehnen \& Binney ({\color{blue}{1998}}) very well.
We mentioned that the innermost bin with $R$ less than around 6--7\,kpc can be observed only for Z = [0.5,\,1.5]\,kpc.
This means that the innermost disk stars we observed tend to be distributed in high Galactic latitude regions, where these stars are actually kinematically hot.
In addition, the severe sampling shortage in this area significantly reduces the reliability of the results, and hence, these results need to be further confirmed by more complete observations in the future.

\section{Discussion}

\subsection{A perturbation signal in the outer southern disk}
As marked in the middle panels of Fig.\,8, a significant substructure (marked with red circles) in the outer ($R$ between $\sim$ 11.0 and $\sim$ 13.0\,kpc) southern disk ($Z$ between $-$1.0 and 0.0\,kpc) is found for $\sigma_{\phi}$ in $R-Z$ diagram, in particular for the thin disk population.
This substructure is kinematically hotter than the background by $\sim$ 5.0--10.0\,km\,s$^{-1}$.
More interestingly, it is spatially associated with the north-south asymmetry structure in $V_{\phi}$ map (marked with red circles in Fig.\,6), implying a possible similar origin of the two substructures.
If it is true, this hotter substructure would rotate faster than its surrounding stars by $\sim$ 5.0--10.0 km\,s$^{-1}$.
By the shape of this substructure in Fig.\,8, we consider this substructure may be linked to a minor merger/perturbation from south to north.
Since this signal is ubiquitous in the thin disk, YAGE, IAGE, L$\alpha$MP,
L$\alpha$MR, I$\alpha$MP and I$\alpha$MR populations, we can further identify this minor merger/perturbation event may have occurred within 4.0\,Gyr.
One possible explanation could be the result of a minor merger/perturbation from satellite galaxies, like the Sagittarius, that are thought to leave footprints both in the velocity and spatial distributions of the disk \citep[e.g.,][]{Gomez2012a, Gomez2012b, Laporte2018, Carrillo2019}.

\subsection{The flat or increasing trend in $\sigma_{Z}$ -- $R$}

For the whole sample stars with $Z$ = [$-$0.5, 0.5]\,kpc, the $\sigma_{Z}$ displays a flat trend at $R$ ranging from $\sim$\,9.0\,kpc to $\sim$\,12.0\,kpc and, then shows slight increase as increasing $R$ at $R$ $\geq$ 12.0\,kpc, such behavior is also clearly seen in the thin disk, YAGE, L$\alpha$MP and L$\alpha$MR populations.
This is an interesting result since the vertical velocity dispersion of the Galactic disk is generally suggested to decrease as $R$ increases, that is, ${\sigma_{Z} (R)}^{2}$ $\propto$ $\rho (R)$ $h_{Z}$ (see the equation 10 of Mackereth et al. {\color{blue}{2019b}}), where the $\rho (R)$ and $h_{Z}$ respectively present the surface density and the scale height of the Galactic disk.
Based on the observations of external galaxies with spiral arms and bulges, the $h_{Z}$ is generally confirmed to be a constant value \citep[e.g.,][]{Yoachim2006}, which leads to the $\sigma_{Z}$ decreases with increasing $R$.
However, our results clearly show the flattening or increasing trend in $\sigma_{Z}$ -- $R$ for the Galactic middle plane, especially for our thin disk, YAGE and L$\alpha$MR populations, the results show a clearly increasing trend in this relation.
Similar behavior can be also found in the APOGEE thin disk population \citep[e.g.,][]{Mackereth2019b}.

The $\sigma_{Z}$ increases with increasing $R$ is known as the disk flare considerably \citep[e.g.,][]{Bovy2016}, and the flare signal is also widely found in previous studies \citep[e.g.,][]{Minchev2015, Mackereth2017, Sanders2018, Mackereth2019b, Sun2020}.
Minchev et al. ({\color{blue}{2015}}) suggested that the relation of $h_{Z}$ -- $R$ may be distinguished for different populations, which may shape the different trend in $\sigma_{Z}$ -- $R$ for various populations.
In addition, the bending of the disk may also contribute to this trend of $\sigma_{Z}$ -- $R$ since previous studies \citep[e.g.,][]{Hunter1969, Khoperskov2017} also found the bending of the disk may result in the increasing trend of the $\sigma_{Z}$.

\subsection{The AVRs and the [$\alpha$/Fe]--velocity dispersion relations}

The global increasing trend in AVRs is the result of Galactic dynamical heating, including the scattering of giant molecular clouds (GMCs) \citep[e.g.,][]{Spitzer1951, Spitzer1953, Barbanis1967, Jenkins1992} and the spiral arms \citep[e.g.,][]{Bissantz2003, Combes2014}, the resonance of the Galactic bar and the perturbation of substructures \citep[][]{Antoja2018, Hunt2018}.
Both theories \citep[e.g.,][]{Lacey1984, Hanninen2002} and observations \citep[e.g.,][]{Wielen1977, Seabroke2007} suggest that the velocity dispersions show a power-law increasing trend with age, in which, $\sigma_{Z}$ $\propto$ $\tau^{\beta_{Z}}$.
Therefore, the jump increase in AVRs at age $\sim$ 7.0$-$9.0\,Gyr is a surprising result, and such behavior is also mentioned by previous observations \citep[e.g.,][]{Quillen2001, House2011, Lagarde2021}, while other studies tend to give a simple power-law profile in AVRs \citep[][]{Aumer2009, Holmberg2009, Sharma2021}.

Quillen \& Garnett ({\color{blue}{2001}}) and House et al. ({\color{blue}{2011}}) reported that the jump increase in AVRs may be related to a thick disk component caused by merger events.
However, the results of Aumer \& Binney ({\color{blue}{2009}}) and Holmberg \& Andersen ({\color{blue}{2009}}) trend to support the long-term heating.
Although we can not fully rule out the possibility of the jump found in AVRs is due to the underestimated age of the older stars, our results tend to suggest that this jump is more likely to be related to violent events, like the merger events as suggested by Quinn et al. ({\color{blue}{1993}}) and Kazantzidis et al. ({\color{blue}{2008}}).
The results of the chemically thin and thick disk populations indicate that there is an obvious gap in AVRs for thin and thick disk populations (Fig.\,9), and the youngest thick disk stars (around 4--6\,Gyr) generally have larger dispersions than that of the oldest thin disk stars as old as 12--14\,Gyr, indicating different heating histories experienced by thin and thick disks.
This jump, as well as the almost constant large values of velocity dispersions of all ages may imply that the thick disk stars are likely rapidly heated to such large velocity dispersions in a short time scale either by merger and accretion \citep[e.g.,][]{Quinn1993, Abadi2003, Grand2016, Kazantzidis2008, Belokurov2018, Deason2018, Helmi2018, Kruijssen2019, Mackereth2019a}, or born in the chaotic mergers of gas-rich systems and/or turbulent interstellar medium (ISM) \citep[e.g.,][]{Brook2004, Brook2007, Brook2012, Wisnioski2015}.

Using RAVE data, Minchev et al. ({\color{blue}{2014}}) reported that the $\sigma$ -- [$\alpha$/Fe] displays a downturn at highest-[$\alpha$/Fe] in the solar neighborhood, and suggested this behavior may be linked to a major merger event.
Such signal is also found in the simulations \citep[e.g.,][]{Minchev2013}, which suggest that a major merger at high red-shift ($z$ $\sim$ 2.0) caused stellar radial merger to the solar neighborhood, and so as result in the migrators often have colder kinematics than the same age stars that formed locally \citep[e.g.,][]{Minchev2012, Vera-Ciro2014}.
However, this behavior is less obvious in the latter new observations \citep[e.g.,][]{Guiglion2015, Hayden2018, Hayden2020, Lagarde2021} from improved data.
In our analysis, the results indicate that only the most metal-poor population display such decreasing trend in velocity dispersions -- [$\alpha$/Fe].
While this dramatic decrease does not exist in other metallicity populations, which is in rough agreement with the result of the high-resolution spectroscopy \citep{Hayden2020, Lagarde2021}.
Therefore, we suggest that the decreasing trend in $\sigma$ at highest-[$\alpha$/Fe] of Minchev et al. ({\color{blue}{2014}}) and our metal-poor population, may be caused by the larger uncertainties in chemistry and kinematics.

\subsection{Ratio of velocity dispersions}

The values of $\sigma_{Z}/\sigma_{R}$ and $\sigma_{\phi}/\sigma_{R}$ contain rich informations of the Galactic disk heating.
In the Galactic disk heating, two main heating agents are widely considered, that is, the GMCs \citep[e.g.,][]{Spitzer1953, Lacey1984} and spiral arms \citep[e.g.,][]{Barbanis1967}.
The general understanding is considered that the GMCs is contributed to the heating in all directions, which leads to the $\sigma_{R}$, $\sigma_{\phi}$ and $\sigma_{Z}$ increases at the same time \citep{Lacey1984, Hanninen2002}, whereas the spiral arms act to heat disk in-plane \citep[][]{Jenkins1990}, which dominates the rising of $\sigma_{R}$ and $\sigma_{\phi}$.

For thick disk population, the $\sigma_{Z}/\sigma_{R}$ is around 0.5$-$0.8 at $R$ around 7.0$-$11.5\,kpc.
This result is roughly consistent with that of Mackereth et al. ({\color{blue}{2019b}}), and is also in good agreement with the ratio expected from the GMC heating theoretically \citep[e.g.,][]{Lacey1984, Hanninen2002}, which reported 0.4--0.8 for $\sigma_{Z}/\sigma_{R}$.
At $R$ $\sim$ 7.0$-$11.5\,kpc, the $\sigma_{R}$ and $\sigma_{\phi}$ of thick disk populations have different values at different $R$ bins, but the values of the $\sigma_{Z}/\sigma_{R}$ and $\sigma_{\phi}/\sigma_{R}$ are very similar, and such behavior is also clearly seen in OAGE and H$\alpha$MP populations.
We consider this behavior likely because these stars are possibly formed in the chaotic mergers of gas-rich systems and turbulent ISM, which is in rough agreement with the discussion of Mackereth et al. ({\color{blue}{2019b}}).
At $R$ $\geq$ 11.5\,kpc, the $\sigma_{Z}/\sigma_{R}$ of thick disk population increases with $R$, and approaches a value of $\sim$ 1.0 at $R$ $\sim$ 12.5\,kpc, which may be caused by many possible reasons.
The first possible reason is that the bending of the disk, which is considered likely to rise the $\sigma_{Z}$ \citep[e.g.,][]{Hunter1969, Martig2014}, even results in the value of $\sigma_{Z}/\sigma_{R}$ to approach $\sim$ 1.0 \citep{Khoperskov2017}.
The second possible reason is the weakening of the effect of spiral arms heating the outer disk, such behavior is also considered for the APOGEE low [$\alpha$/Fe] population by Mackereth et al. ({\color{blue}{2019b}}).
The third possible reason is the contamination of the flared thin disk stars.
Besides, Li et al. ({\color{blue}{2020}}) also suggested that the thick disk is likely to have a weak warped signal, which may be another possible reason leading to the increasing trend of $\sigma_{Z}/\sigma_{R}$--$R$ in the outer disk region.

For the thin disk population, the $\sigma_{Z}/\sigma_{R}$ increases with $R$ at $R$ larger than $\sim$ 8.0\,kpc, with $\sigma_{Z}/\sigma_{R}$ rising from $\sim$ 0.4 to $\sim$ 0.8, which is roughly consistent with the result of Mackereth et al. ({\color{blue}{2019b}}).
The GMCs and spiral arms, along with the flaring, warping and bending of the Galactic disk are likely contributed to the observed trend of $\sigma_{Z}/\sigma_{R}$ -- $R$ relations of the thin disk population.

The results of various populations indicate that at $R$ larger than $\sim$ 11.0\,kpc, the $\sigma_{Z}/\sigma_{R}$ displays an increasing trend with increasing $R$, and such behavior can be well detected for all populations except the H$\alpha$MR.
We consider that this behavior may be linked to the flared stars and/or the disk warp.

In addition, a dip of $\sigma_{\phi}/\sigma_{R}$--$R$ at R $\sim$ 10.5\,kpc may again point to the existence of the Perseus arm.

However, we still can not rule out the contribution of other heating agents (i.e., satellite mergers, e.g., Quinn et al. {\color{blue}{1993}}; Villalobos \& Helmi {\color{blue}{2008}}; accretion events, e.g., Abadi et al. {\color{blue}{2003}}; interacting with a central bar, e.g., Minchev \& Famaey {\color{blue}{2010}}) to this behavior.

\section{Conclusions}

In this paper, we use nearly 140,000 RC stars selected from the LAMOST and {\it Gaia} to calculate the kinematics of different Galactic disk populations mainly between 6.0 $\leq$ R $\leq$ 15.0\,kpc and $|Z|$ $\leq$ 3.0\,kpc in very high accuracy. We find that:
\\
\\
$\bullet$ Young/$\alpha$-poor and old/$\alpha$-enhanced populations show significant different behaviors in the mean velocity, velocity dispersion, and age-velocity dispersion relation.
The wave-like behaviors in radial velocity, the north-south discrepancy in azimuthal velocity, and the warp in vertical velocity of young/$\alpha$-poor populations are stronger than those of old/$\alpha$-enhanced populations.
The north-south asymmetry in velocity dispersions is significant for the young/$\alpha$-poor populations, whereas the old/$\alpha$-enhanced populations show much weaker behavior.
The AVRs of young/$\alpha$-poor populations show an increasing trend with age, whereas the old/$\alpha$-enhanced populations show a flat trend with age.
\\
\\
$\bullet$ Mean velocity results reveal the oscillation(s) on a kilo-parsec scale, north-south rotational asymmetry, and warp of the disk, and these behaviors may also show moderate to significant population and spatial dependences.
\\
\\
$\bullet$ The results of velocity dispersions display a signal of spiral arms in the $\sigma_{R}$--$R$ plane of the thin disk stars with $|Z|$\,=\,[0.0,\,0.5]\,kpc.
Additionally, there is a flare signal observed in the outer region (i.e., R $\geq$ 12 kpc) in the vertical velocity dispersion, as well as a north-south azimuthal velocity dispersion asymmetry in the disk.
These flare and north-south heating asymmetry behaviors are linked to young or alpha-poor populations, and the north-south heating asymmetry is likely a result of a minor merger/perturbation from south to north within 4\,Gyr.
\\
\\
$\bullet$ The results of AVRs and [$\alpha$/Fe] -- velocity dispersion relations show a global increasing trend, revealing the dynamical heating of disk stars by scattering of GMCs and spiral arms.
A jump detected for thick disk population is likely caused by merger and accretion, or thick disk stars born in the chaotic mergers of gas-rich systems and/or turbulent ISM.
\\
\\
$\bullet$ The results of $\sigma_{Z}$/$\sigma_{R}$ and $\sigma_{\phi}$/$\sigma_{R}$ indicate that different populations exhibit distinct behaviors.
A north-south asymmetry is observed in $\sigma_{Z}$/$\sigma_{R}$, and this behavior may be linked to young or alpha-poor populations.
Additionally, a dip of $\sigma_{\phi}$/$\sigma_{R}$ at R $\sim$ 10.5\,kpc may be linked to the signal of the spiral arm.
For thick disk, the $\sigma_{Z}$/$\sigma_{R}$ is $\sim$ 0.5--0.8 at R $\sim$ 7.0--11.5 kpc, likely pointing to the heating of GMCs.
Moreover, $\sigma_{Z}$/$\sigma_{R}$ increases with $R$ at $R$ $\geq$ 11.5 kpc, suggesting the presence of a bending disk, the weakening of the effect of spiral arms heating in the outer disk, and the disk warping.
For the thin disk population, $\sigma_{Z}$/$\sigma_{R}$ increases with $R$, rising from $\sim$ 0.4 to $\sim$ 0.8.
This implies the GMCs and spiral arms, and the flaring, warping, and bending of the Galactic disk may contribute to the observed relation of the $\sigma_{Z}$/$\sigma_{R}$--$R$ for thin disk stars.

\section*{Acknowledgements}
We thank the anonymous referee for very useful suggestions to improve the work.
This work is supported by the NSFC projects 12133002, and National Natural Science Foundation of China grants 12073070, and 12003027.

Guoshoujing Telescope (the Large Sky Area Multi-Object Fiber Spectroscopic Telescope LAMOST) is a National Major Scientific Project built by the Chinese Academy of Sciences. Funding for the project has been provided by the National Development and Reform Commission. LAMOST is operated and managed by the National Astronomical Observatories, Chinese Academy of Sciences. The LAMOST FELLOWSHIP is supported by Special Funding for Advanced Users, budgeted and administrated by Center for Astronomical Mega-Science, Chinese Academy of Sciences (CAMS)

\appendix

%\clearpage
%\newpage

\section{Velocity field of mono-age and mono-[$\alpha$/Fe]-[Fe/H] populations}

This appendix present the mean velocities: $V_{R}$ (Fig.\,{\color{blue}{A1}} and {\color{blue}{A2}}), $V_{\phi}$ (Fig.\,{\color{blue}{A3}} and {\color{blue}{A4}}) and $V_{Z}$ (Fig.\,{\color{blue}{A5}} and {\color{blue}{A6}}), and the velocity dispersions: $\sigma_{R}$ (Fig.\,{\color{blue}{A7}} and {\color{blue}{A8}}), $\sigma_{\phi}$ (Fig.\,{\color{blue}{A9}} and {\color{blue}{A10}}) and $\sigma_{Z}$ (Fig.\,{\color{blue}{A11}} and {\color{blue}{A12}}), of mono-age and mono-[$\alpha$/Fe]-[Fe/H] populations.

\begin{figure*}[t]
\centering

\subfigure{
\includegraphics[width=16.5cm]{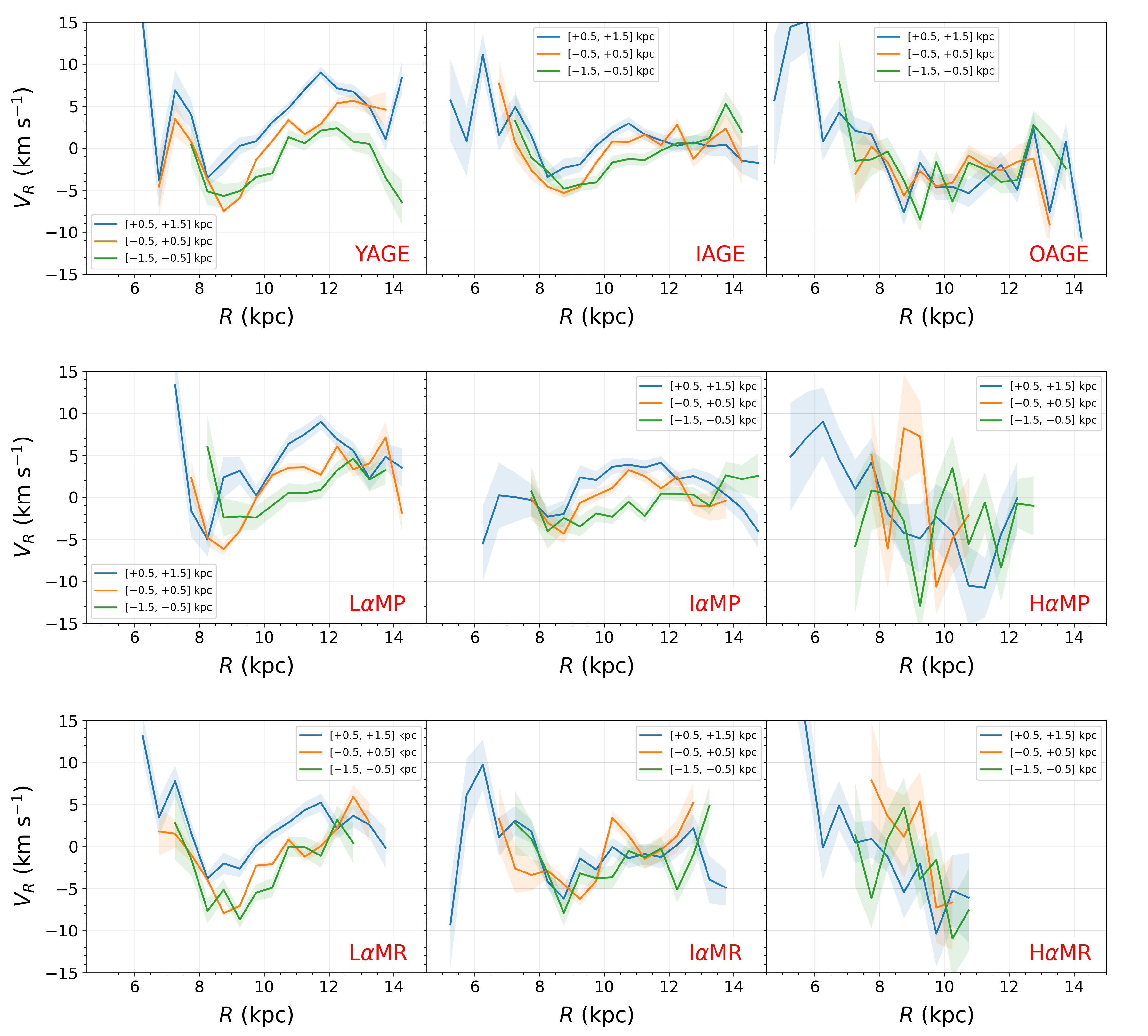}
}

\caption{Radial velocity of mono-age, mono-[$\alpha$/Fe]-[Fe/H] populations as a function of $R$ for different disk heights (lines marked by different colors).
The number of stars of each bin is required to be greater than 50 and the width of the bin is set to 0.5\,kpc.}
\end{figure*}
%%\label{Fig.A1}

\begin{figure*}[t]
\centering

\subfigure{
\includegraphics[width=16.5cm]{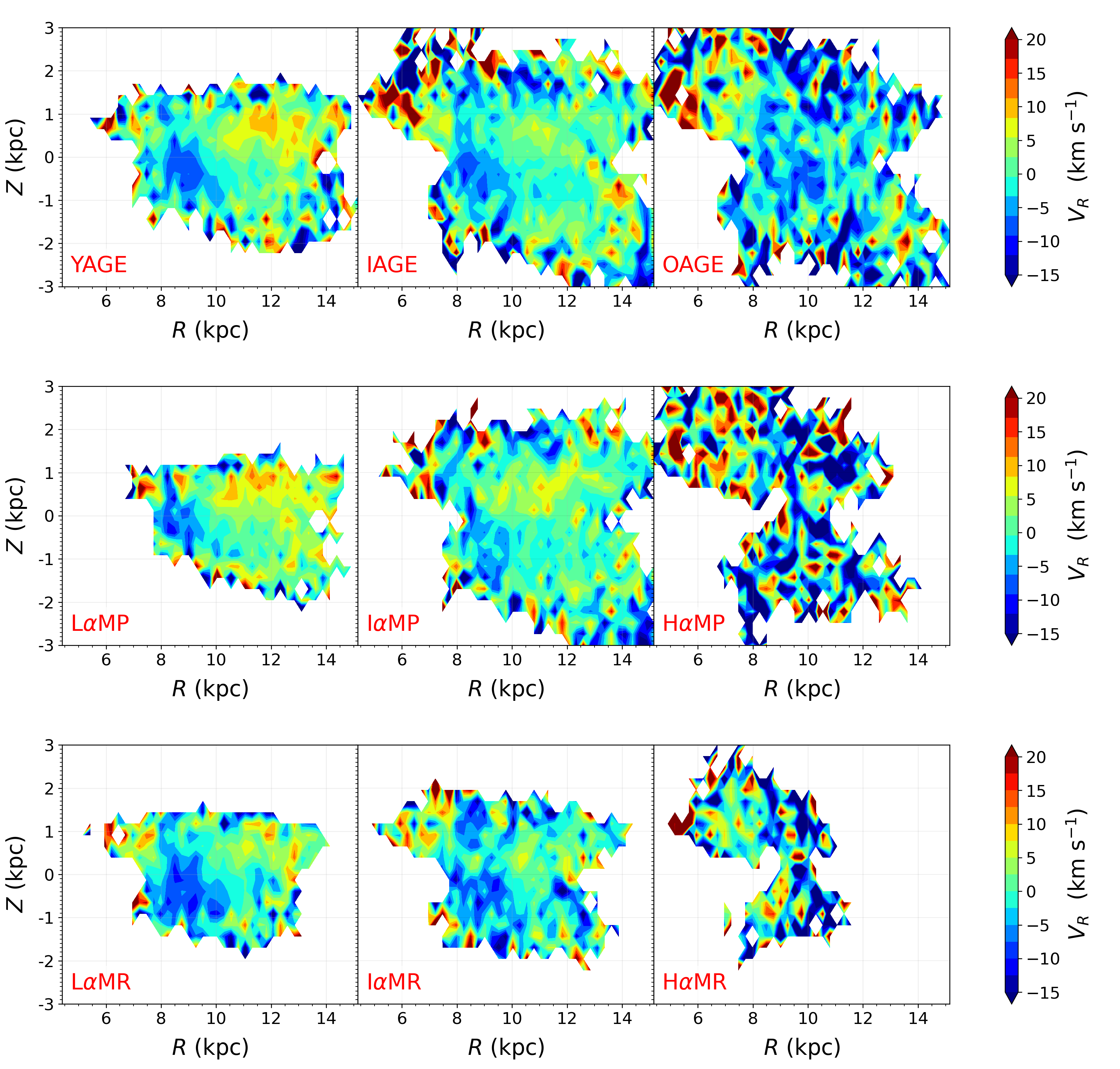}
}
\caption{Radial velocity of mono-age, mono-[$\alpha$/Fe]-[Fe/H] populations in $R$-$Z$ plane. The color bars on the right of each panel represent the values of the radial velocity, there are at least 5 stars in each bin, and both $R$ and $Z$ bins have a width of 0.2 kpc.}
\end{figure*}
%%\label{Fig.A2}

\begin{figure*}[t]
\centering

\subfigure{
\includegraphics[width=16.5cm]{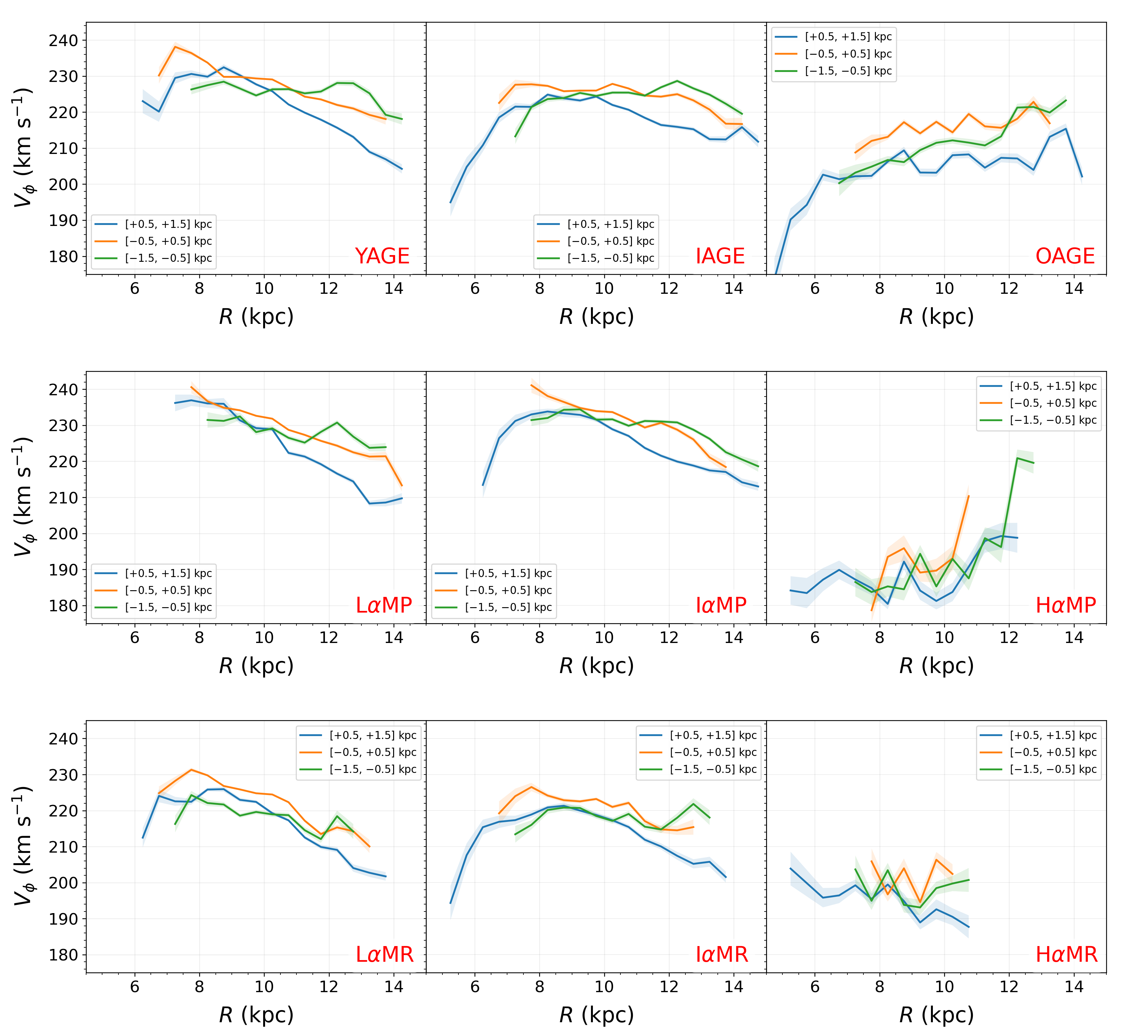}
}

\caption{Similar to Appendix Fig.\,A1. but for azimuthal velocity of mono-age and mono-[$\alpha$/Fe]-[Fe/H] populations.}
\end{figure*}
%%\label{Fig.A3}

\begin{figure*}[t]
\centering

\subfigure{
\includegraphics[width=16.5cm]{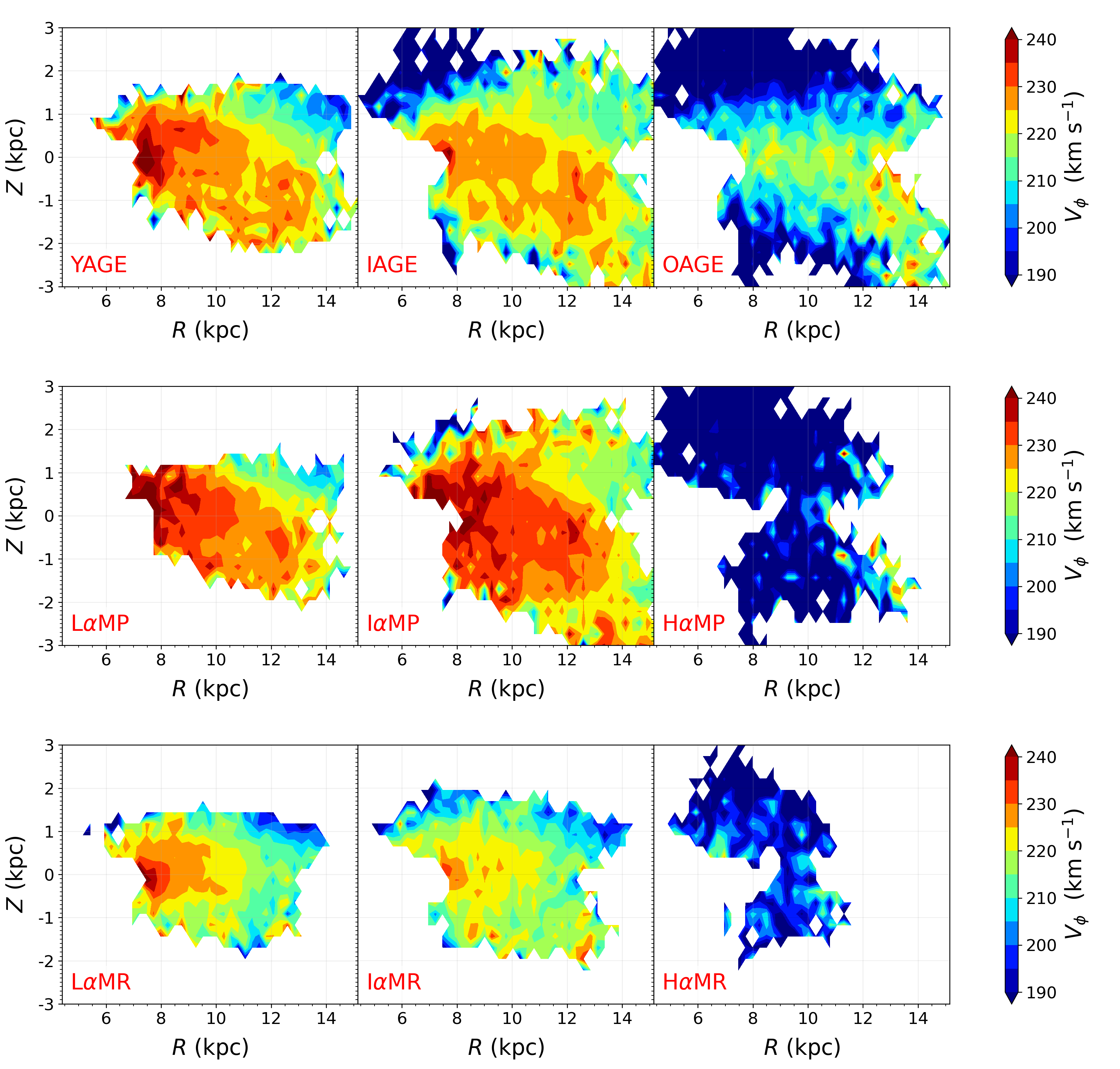}
}
\caption{Similar to Appendix Fig.\,A2 but for the azimuthal velocity of mono-age and mono-[$\alpha$/Fe]-[Fe/H] populations.}
\end{figure*}
%%\label{Fig.A4}

\begin{figure*}[t]
\centering

\subfigure{
\includegraphics[width=16.5cm]{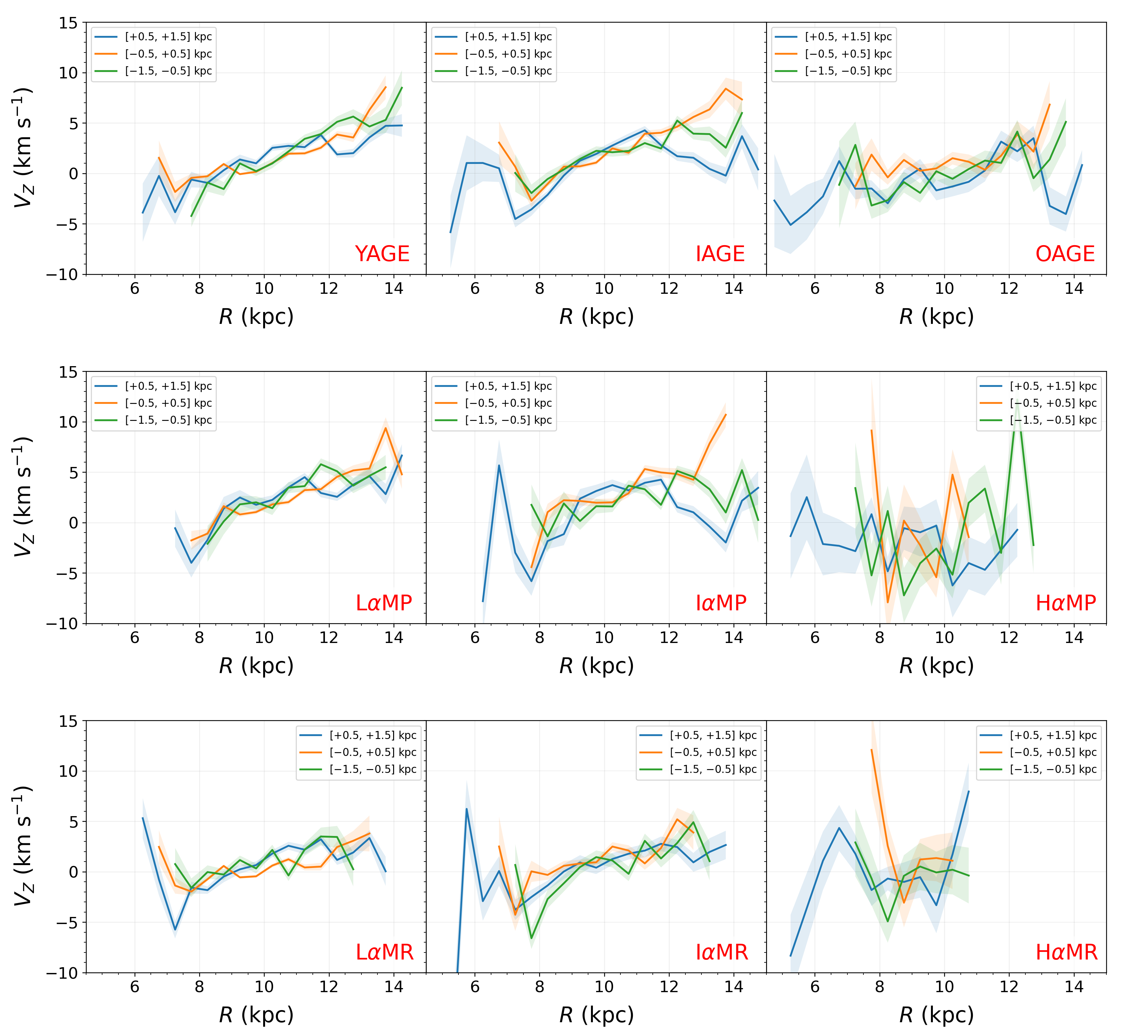}
}
\caption{Similar to Appendix Fig.\,A1 but for the vertical velocity of mono-age and mono-[$\alpha$/Fe]-[Fe/H] populations.}
\end{figure*}
%%\label{Fig.A5}

\begin{figure*}[t]
\centering
\subfigure{
\includegraphics[width=16.5cm]{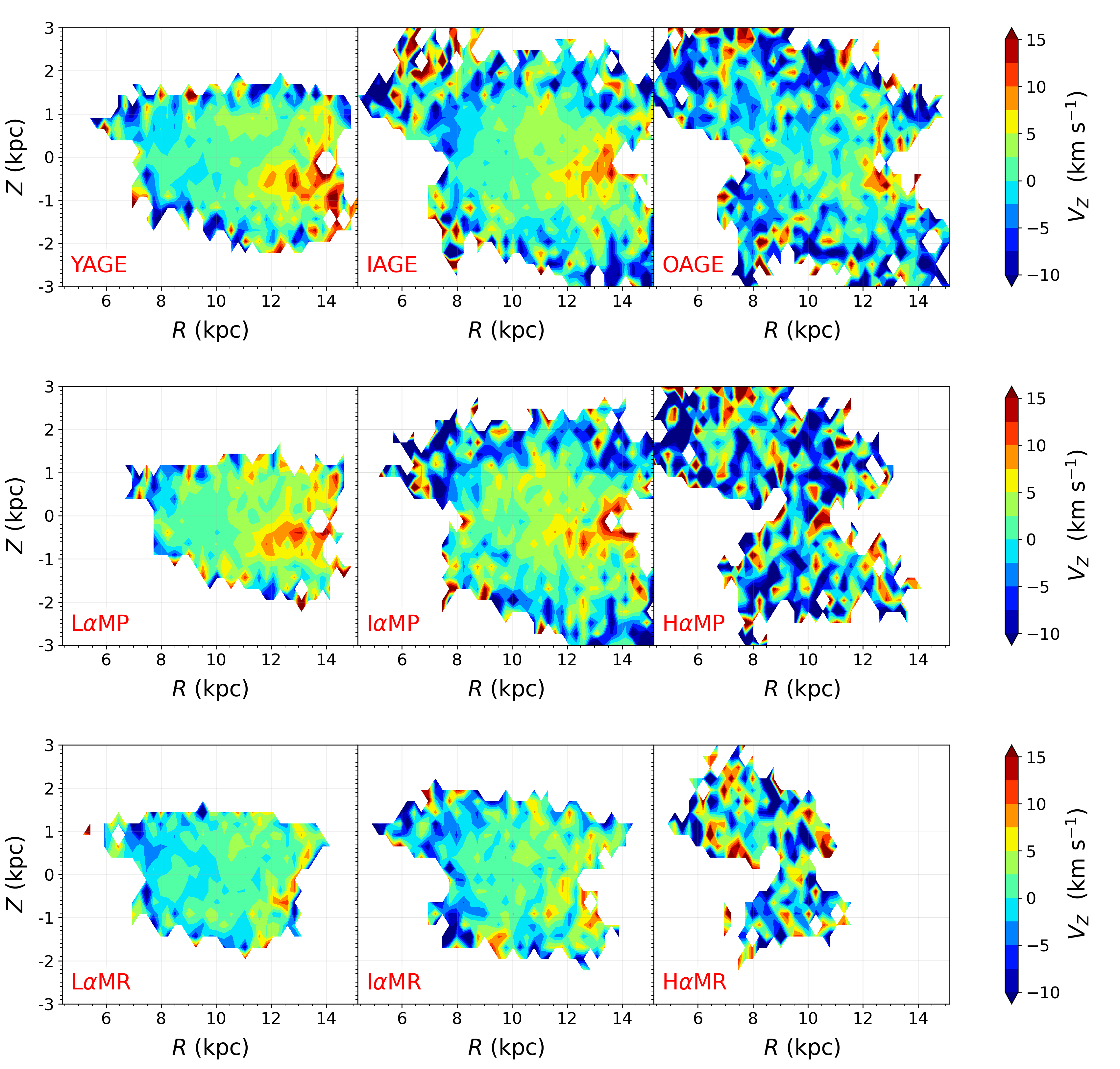}
}
\caption{Similar to Appendix Fig.\,A2 but for the vertical velocity of mono-age and mono-[$\alpha$/Fe]-[Fe/H] populations.}
\end{figure*}
%%\label{Fig.A6}

\begin{figure*}[t]
\centering

\subfigure{
\includegraphics[width=16.5cm]{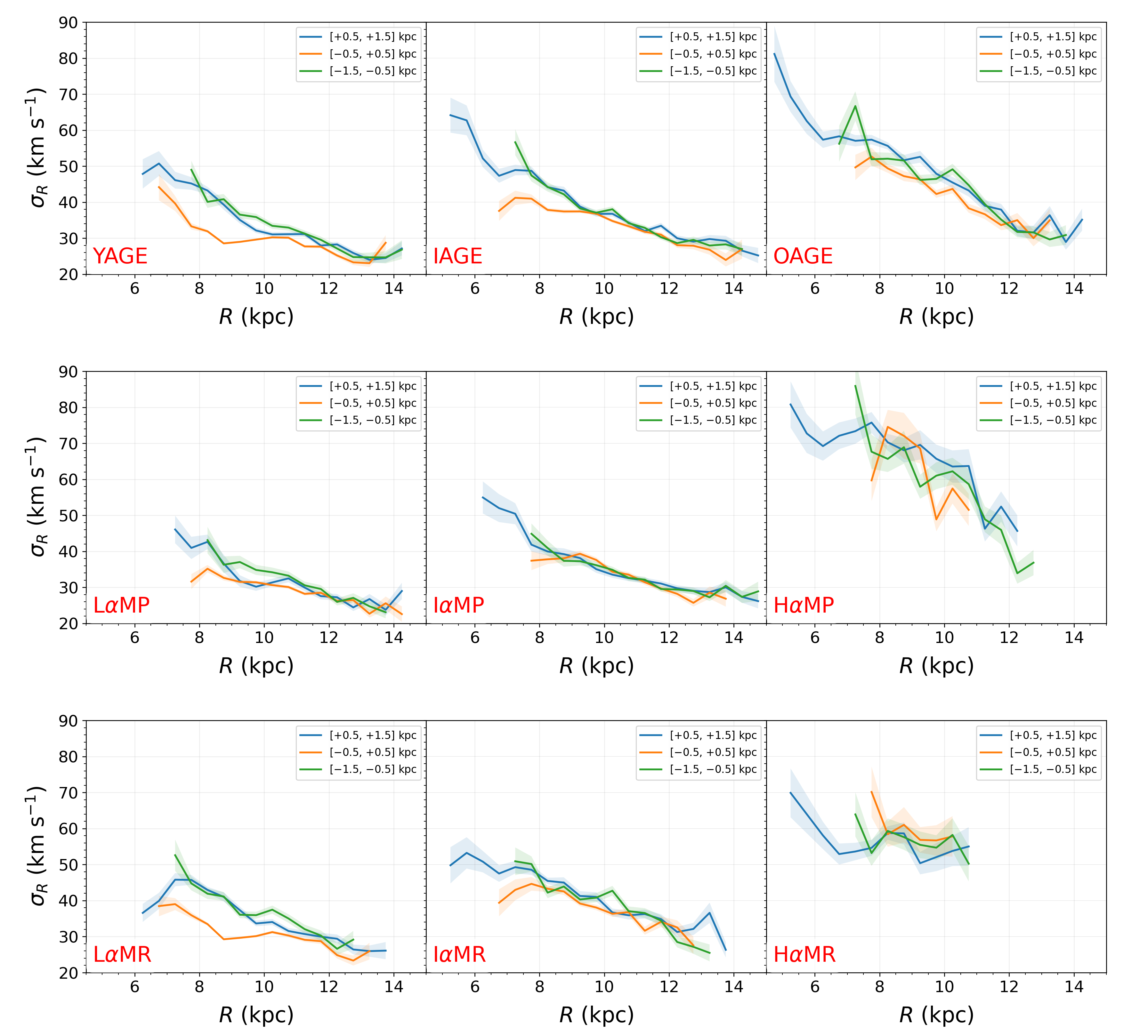}
}

\caption{Radial velocity dispersion of mono-age and mono-[$\alpha$/Fe]-[Fe/H] populations as a function of R for different disk heights (lines marked by different colors).
The number of stars of each bin is required to be greater than 50 and the width of the bin is set to 0.5\,kpc.}

\end{figure*}
%%\label{Fig.A7}

\begin{figure*}[t]
\centering
\subfigure{
\includegraphics[width=16.5cm]{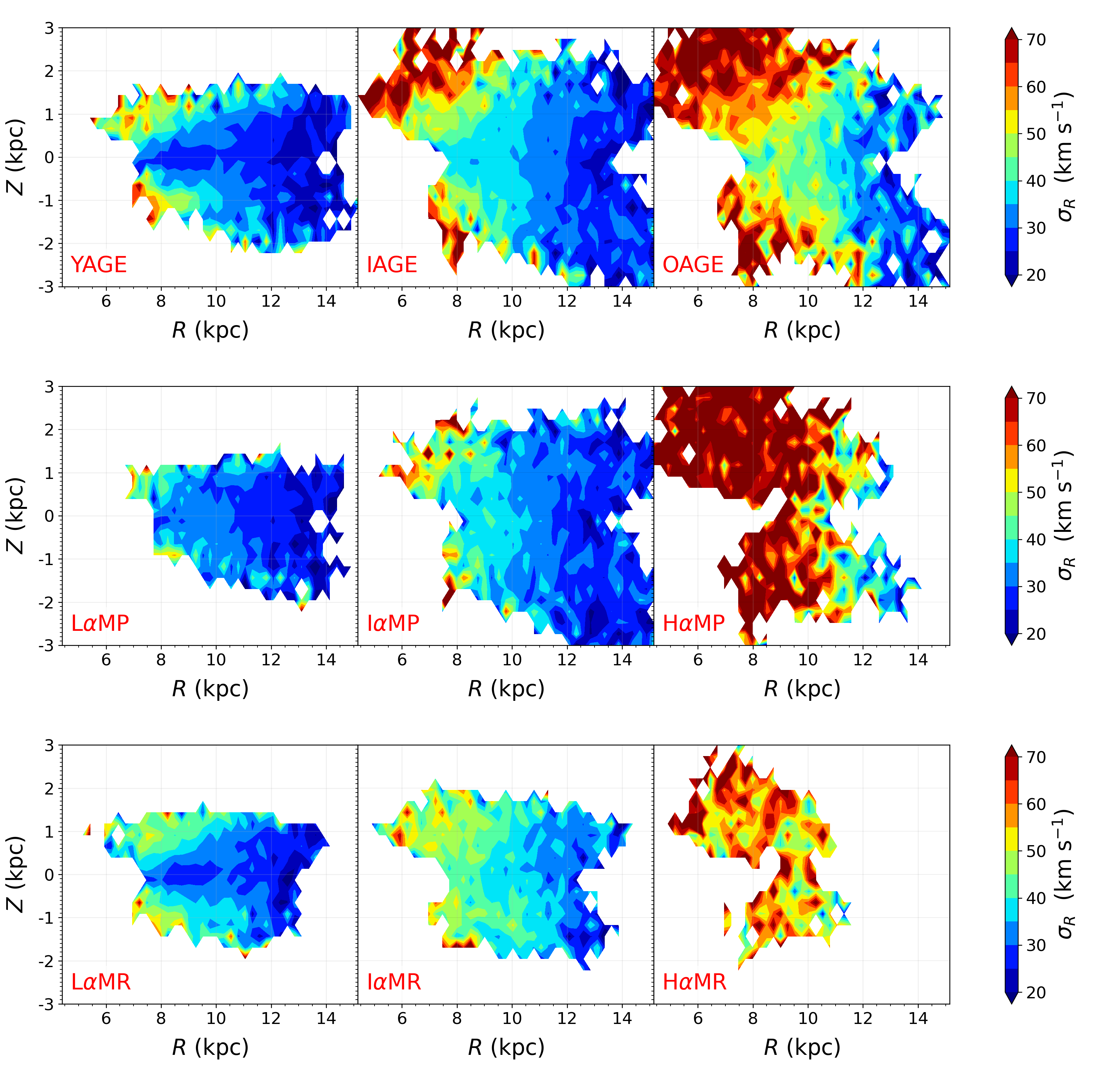}
}
\caption{Radial velocity dispersion of mono-age and mono-[$\alpha$/Fe]-[Fe/H] populations in $R$-$Z$ plane. The color bars on the right of each panel represent the values of the radial velocity dispersion, there are at least 5 stars in each bin, and the $R$ and $Z$ bins both have a width of 0.2 kpc.}
\end{figure*}
%%\label{Fig.A8}

\begin{figure*}[t]
\centering
\subfigure{
\includegraphics[width=16.5cm]{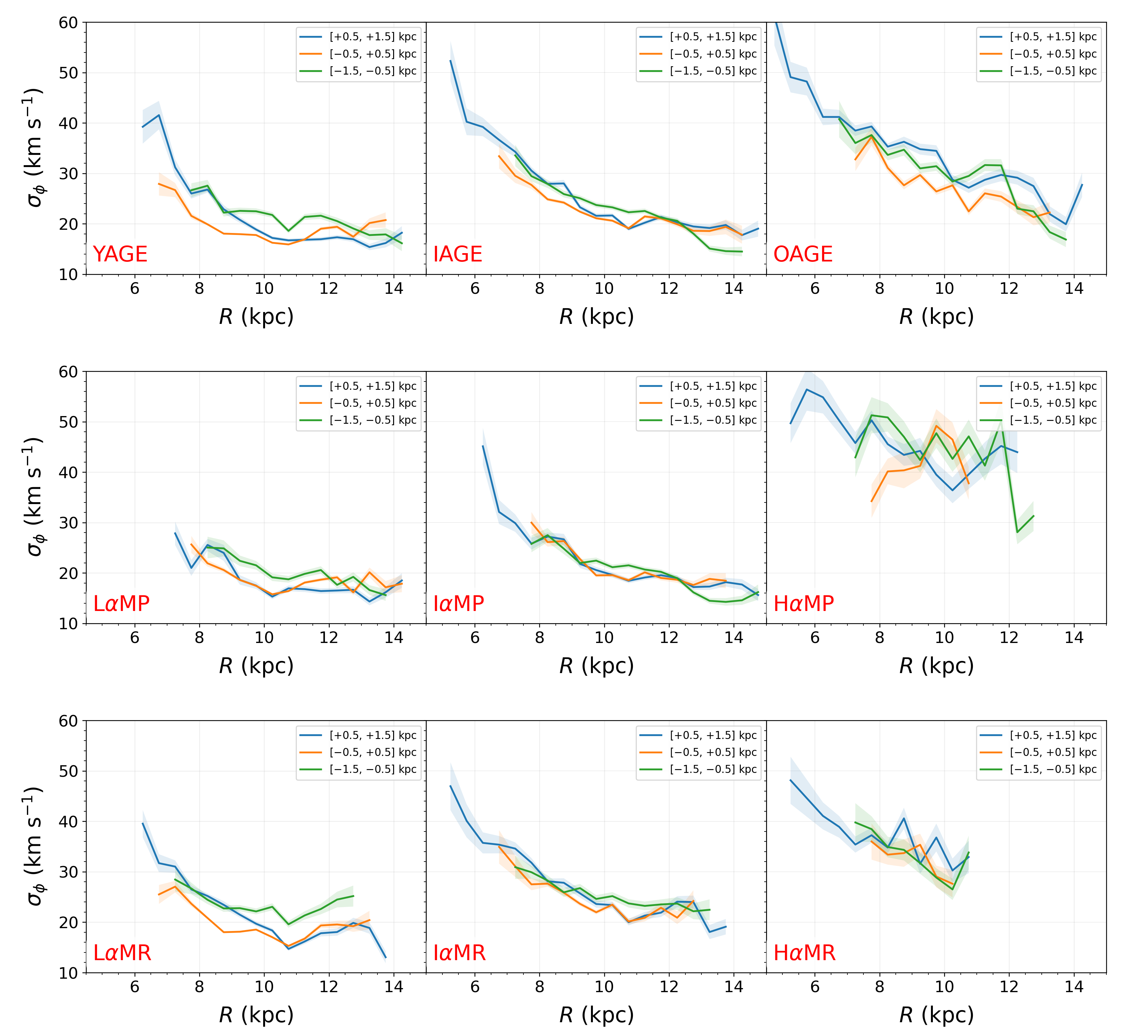}
}

\caption{Similar to the Appendix Fig.\,A7 but for the azimuthal velocity dispersion of mono-age and mono-[$\alpha$/Fe]-[Fe/H] populations.}
\end{figure*}
%%\label{Fig.A9}

\begin{figure*}[t]
\centering
\subfigure{
\includegraphics[width=16.5cm]{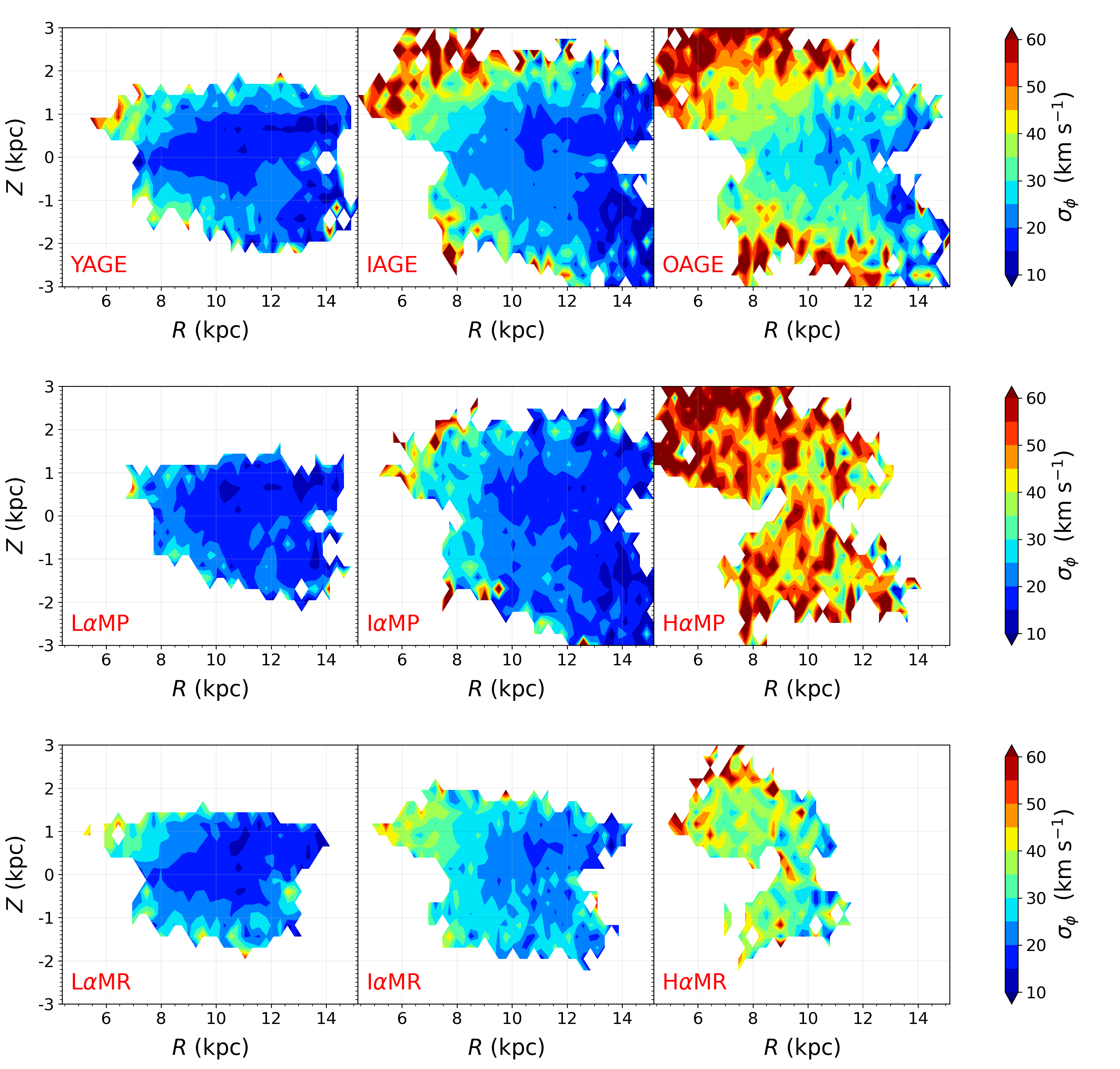}
}
\caption{Similar to the Appendix Fig.\,A8 but for the azimuthal velocity dispersion of mono-age and mono-[$\alpha$/Fe]-[Fe/H] populations.}
\end{figure*}
%%\label{Fig.A10}

\begin{figure*}[t]
\centering

\subfigure{
\includegraphics[width=16.5cm]{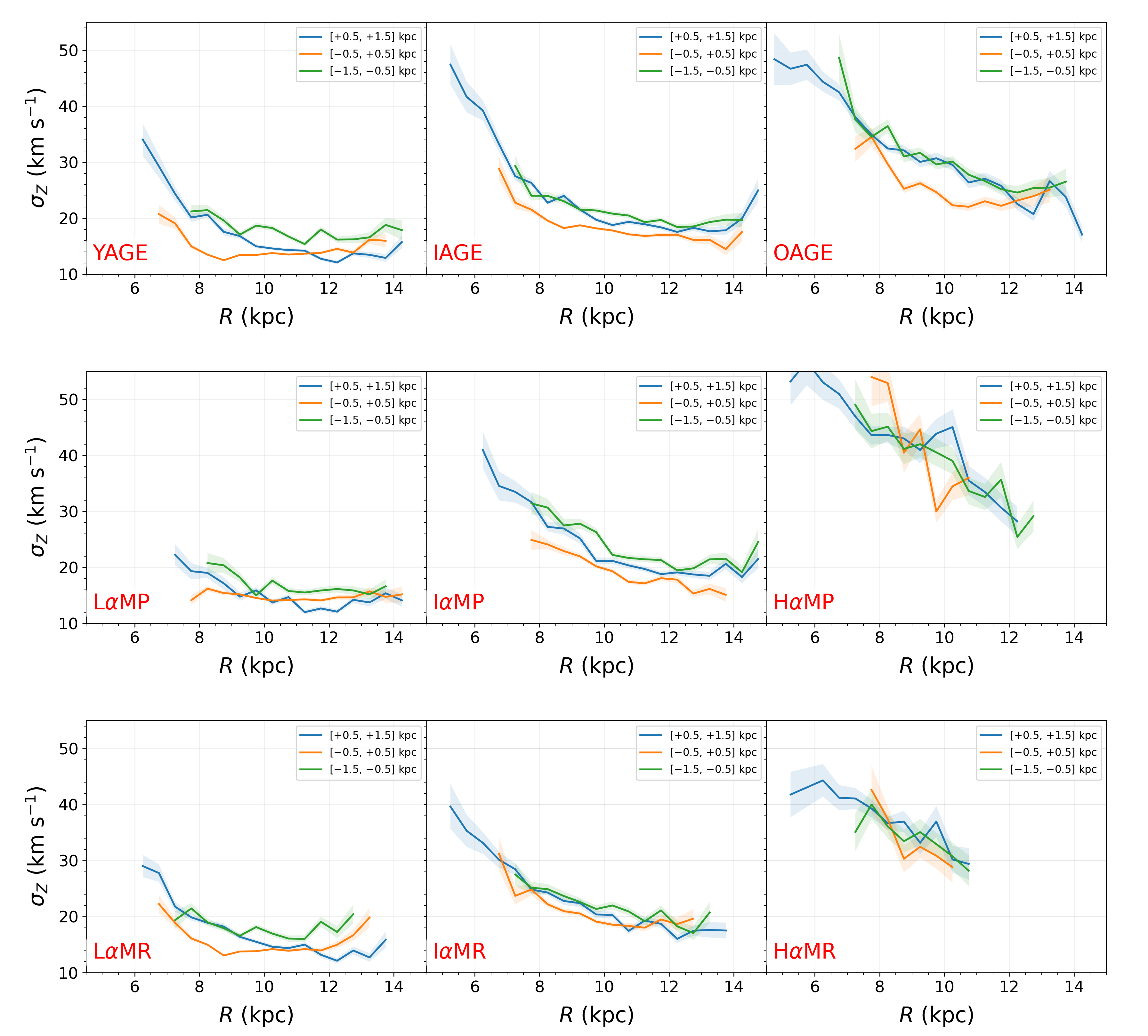}
}

\caption{Similar to the Appendix Fig.\,A7 but for the vertical velocity dispersion of mono-age and mono-[$\alpha$/Fe]-[Fe/H] populations.}
\end{figure*}
%%\label{Fig.A11}

\begin{figure*}[t]
\centering
\subfigure{
\includegraphics[width=16.5cm]{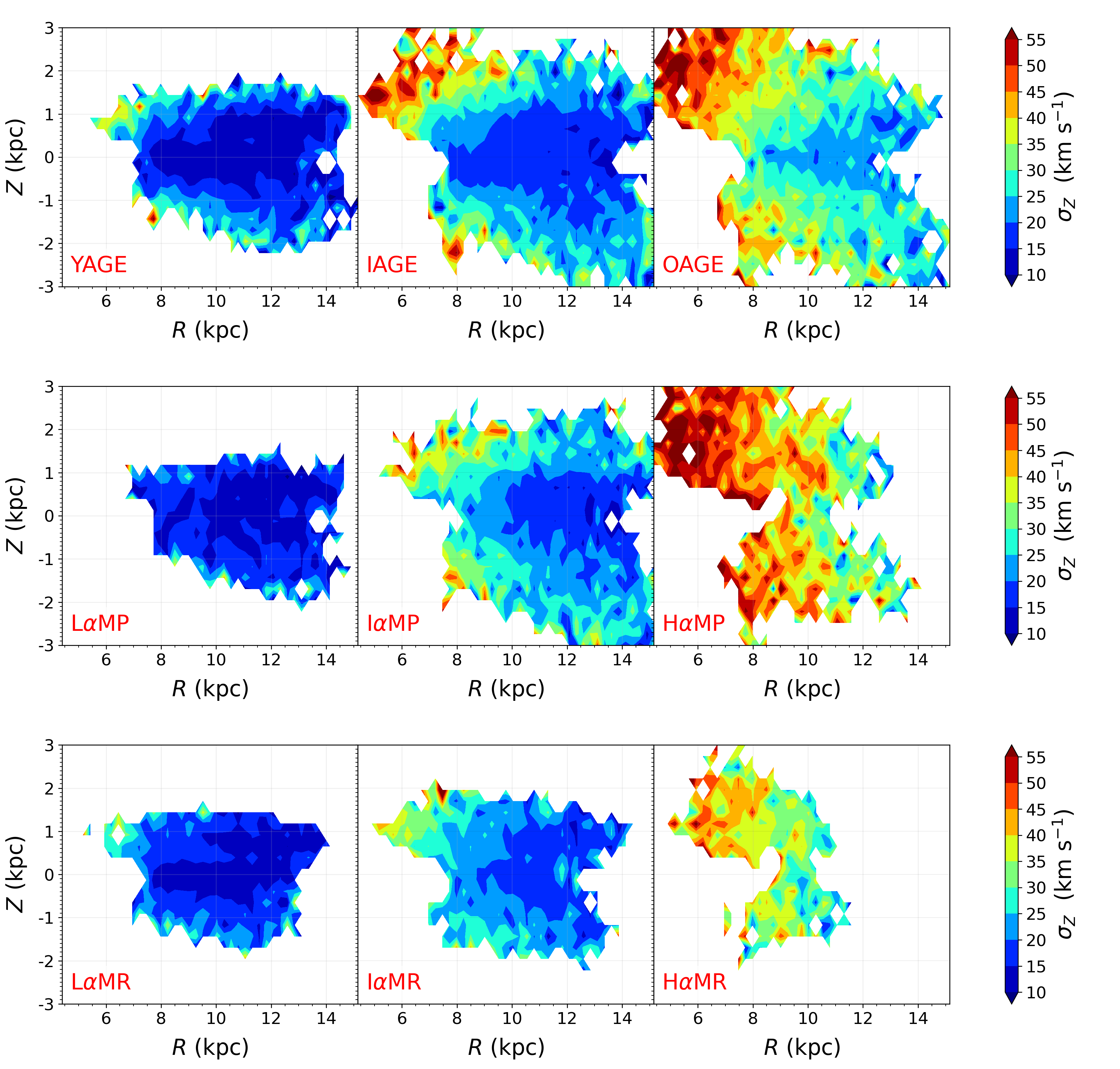}
}
\caption{Similar to the Appendix Fig.\,A8 but for the vertical velocity dispersion of mono-age and mono-[$\alpha$/Fe]-[Fe/H] populations.}
\end{figure*}
%%\label{Fig.A12}

\section{Determination of the $\sigma_{R_0}$ and $R_{\sigma}$ of the thin/thick disks from the profiles of the $\sigma_{R}$ -- $R$}

This appendix present the fitting results (Fig.\,{\color{blue}{B1}}) of the profiles of $\sigma_{R}$ -- $R$ of the thin/thick disks for different Galactic $Z$ planes, and the example posterior PDF of the $\sigma_{R_0}$ and $R_{\sigma}$ (Fig.\,{\color{blue}{B2}}).

\begin{figure*}[t]
\centering
\subfigure{
\includegraphics[width=8.8cm]{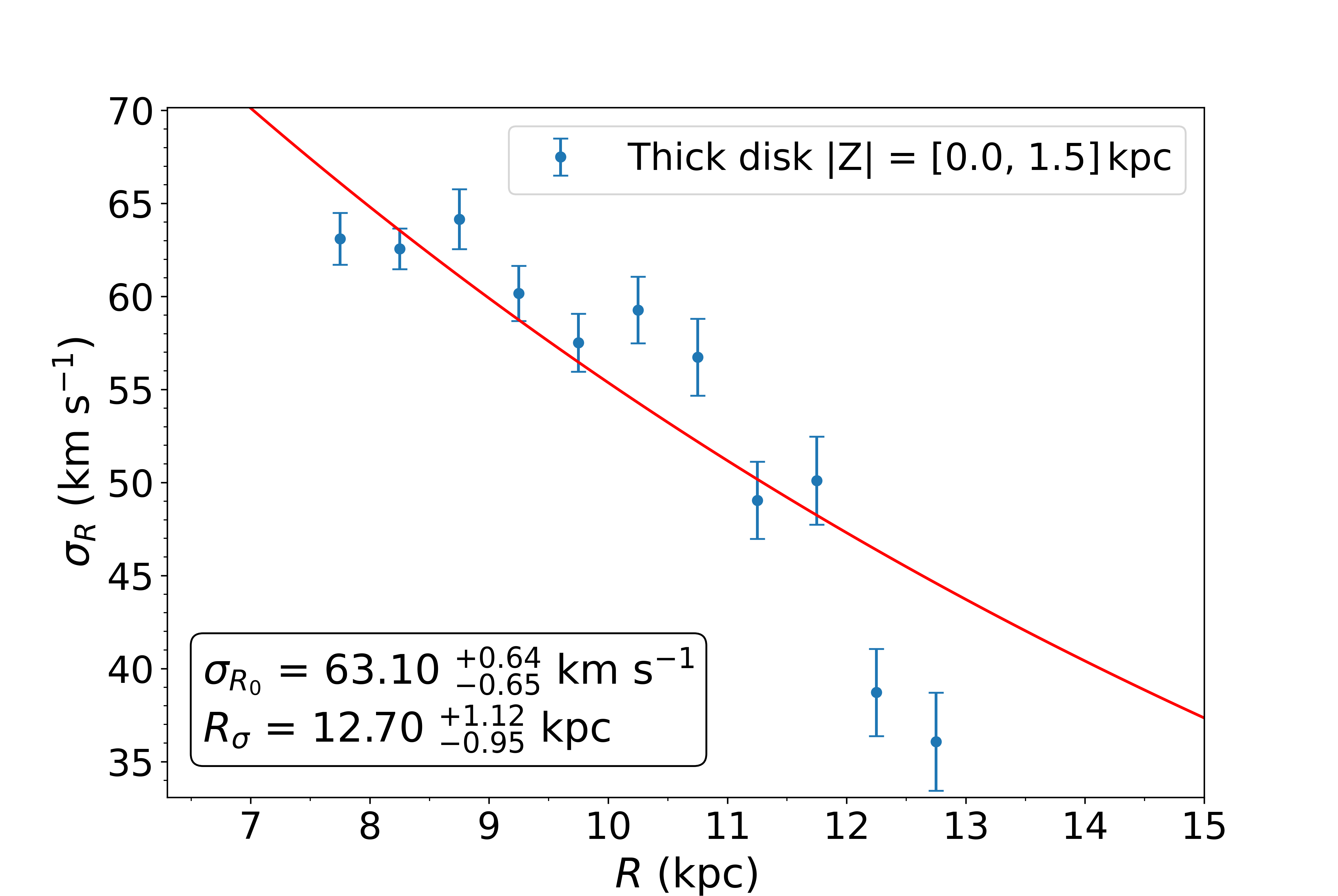}
}
\subfigure{
\includegraphics[width=8.8cm]{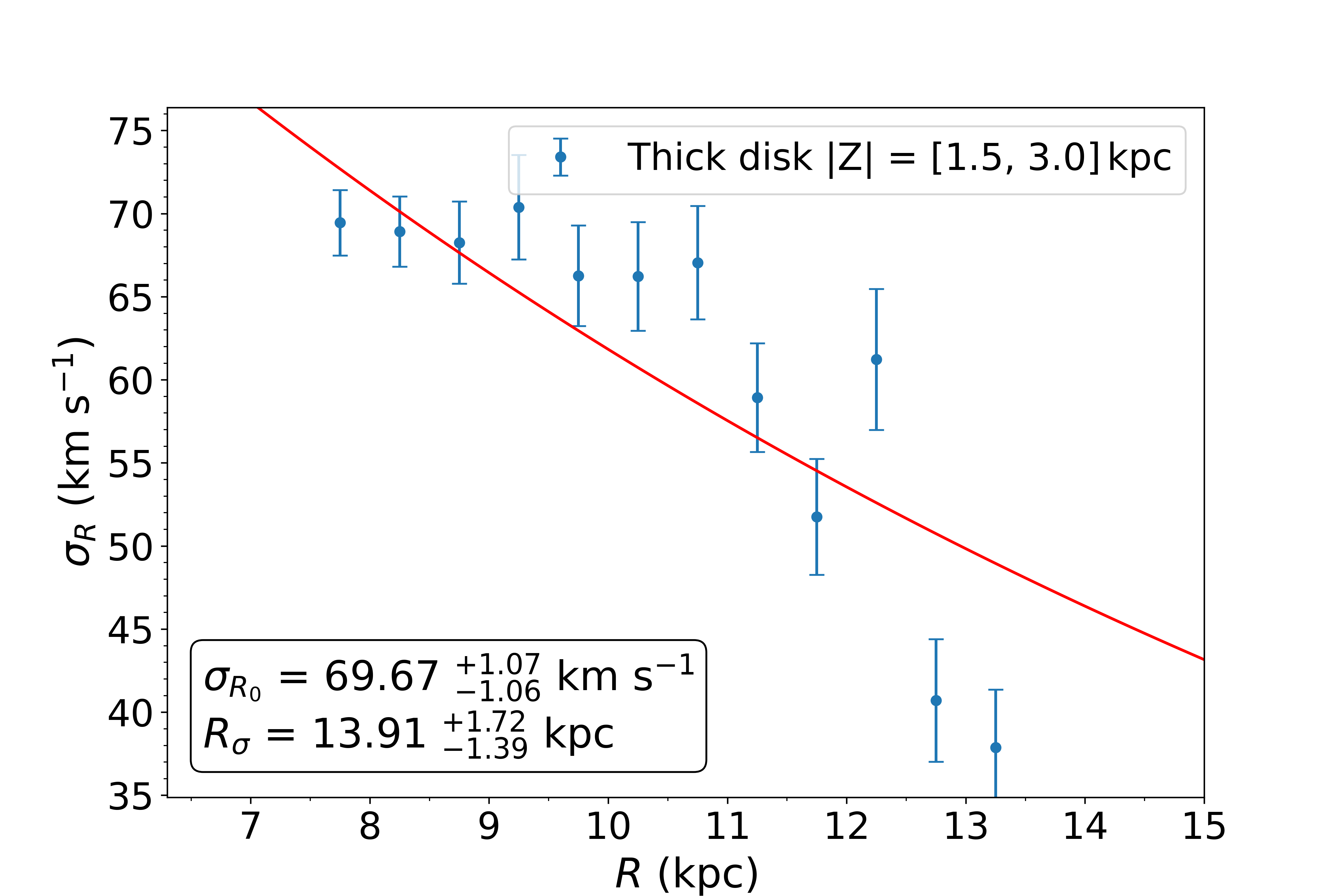}
}

\subfigure{
\includegraphics[width=8.8cm]{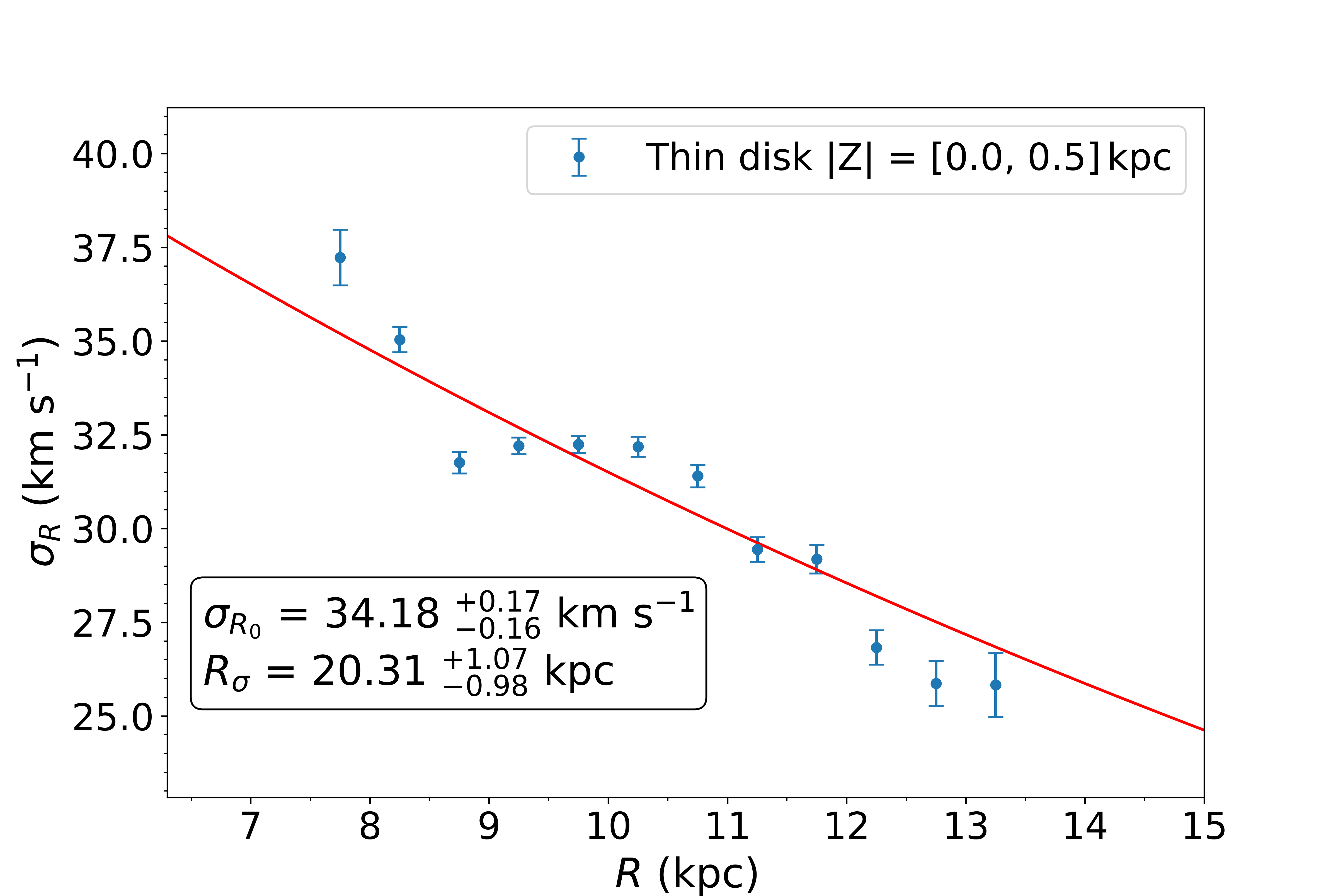}
}
\subfigure{
\includegraphics[width=8.8cm]{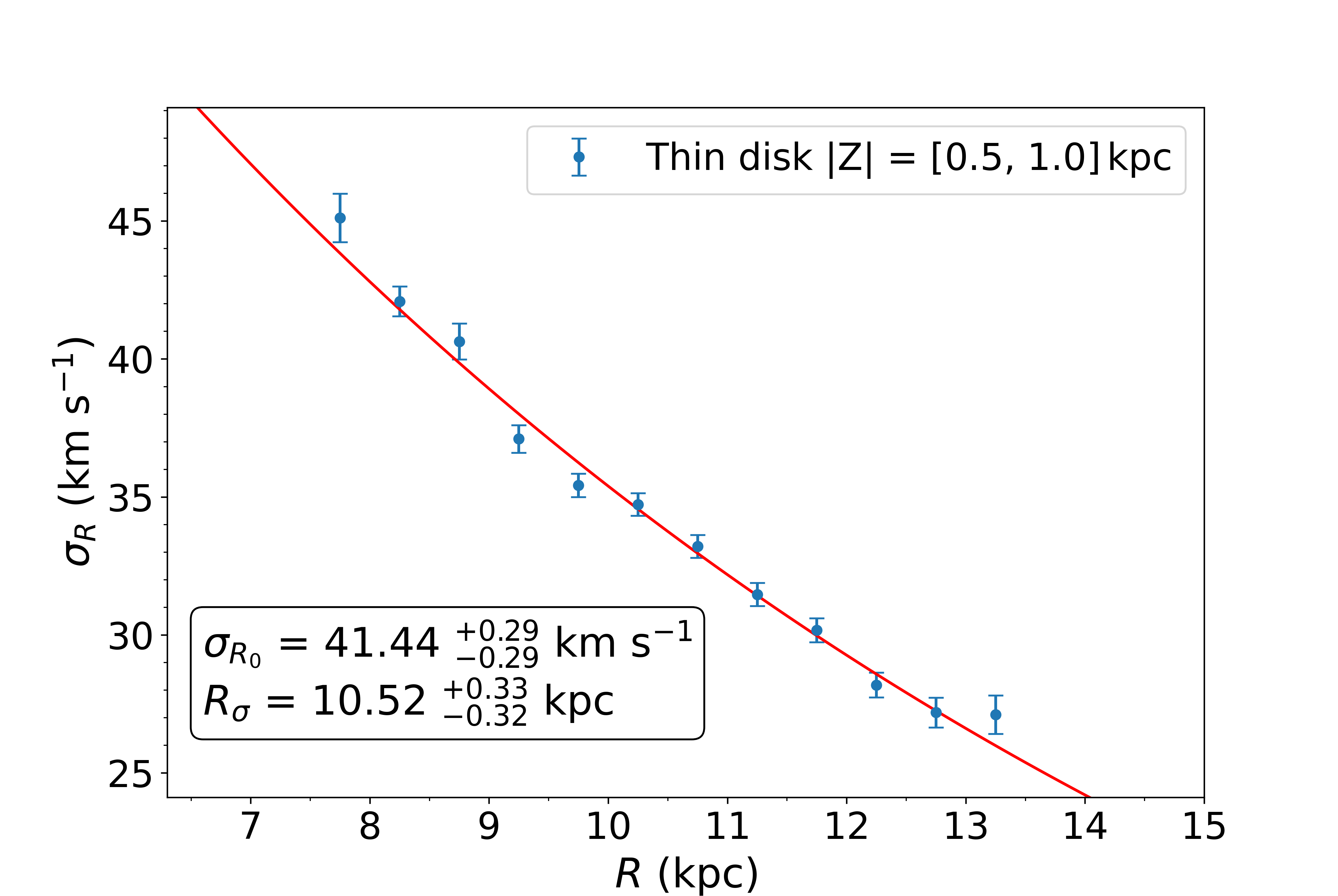}
}

\subfigure{
\includegraphics[width=8.8cm]{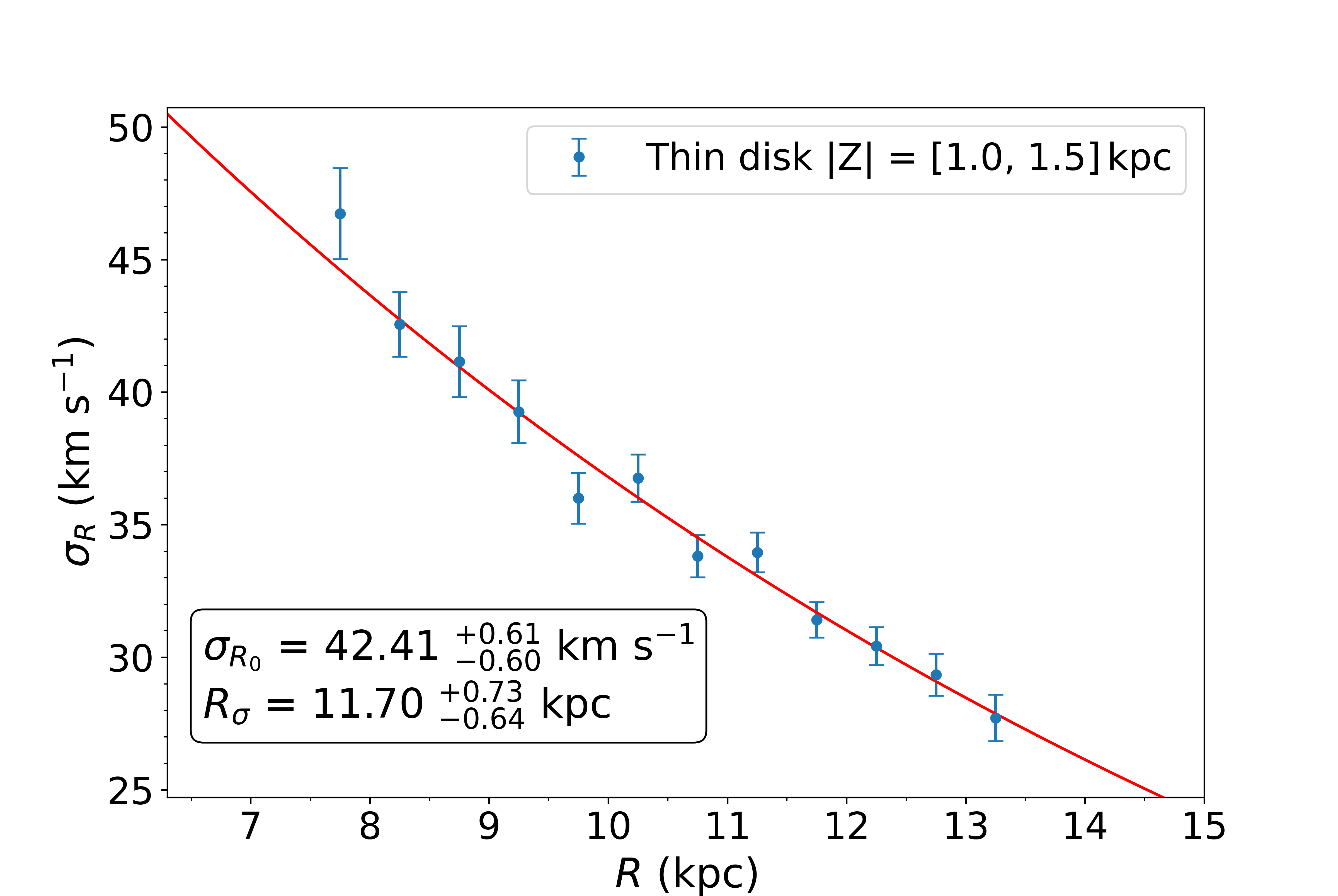}
}
\subfigure{
\includegraphics[width=8.8cm]{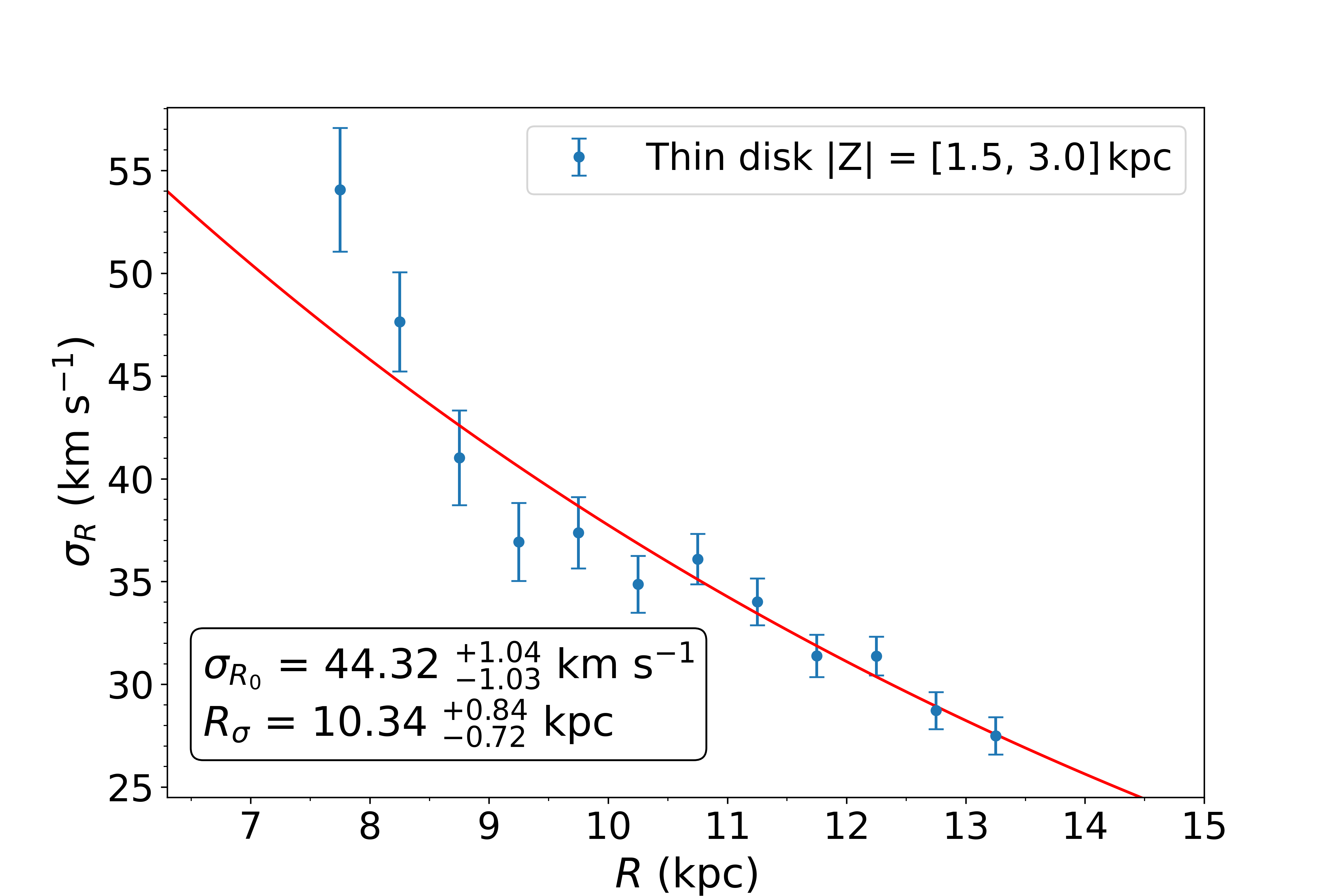}
}

\caption{Profile of radial velocity dispersion of the mono-disk populations for different Galactic $Z$ planes, blue dots and orange line represent the RC data and exponential best fit to the data points, respectively. There is a minimum of 50 stars per bin}
\end{figure*}
%%\label{Fig.B1}

\begin{figure*}[t]
\centering
\subfigure{
\includegraphics[width=7.2cm]{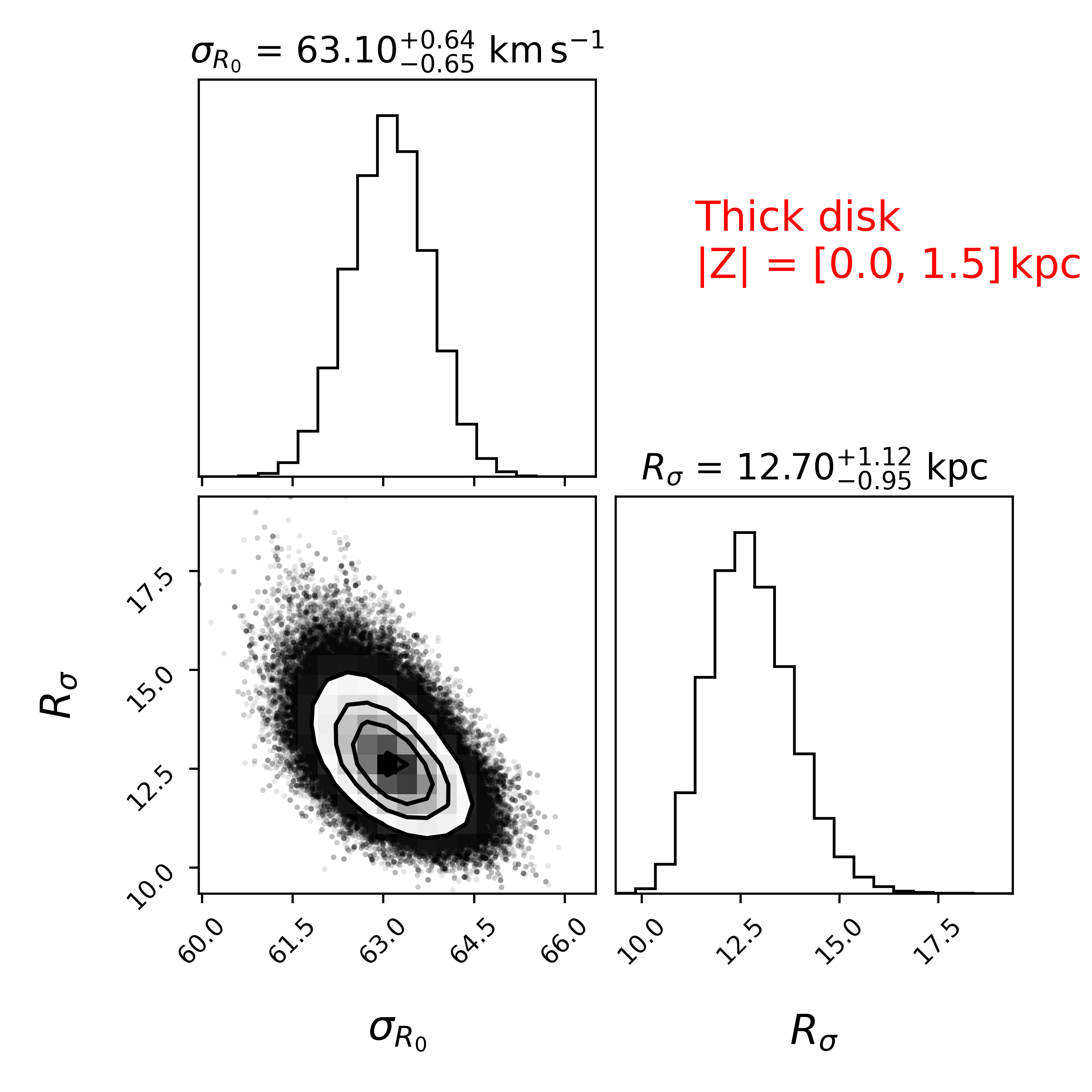}
}
\hspace{0.6cm}
\subfigure{
\includegraphics[width=7.2cm]{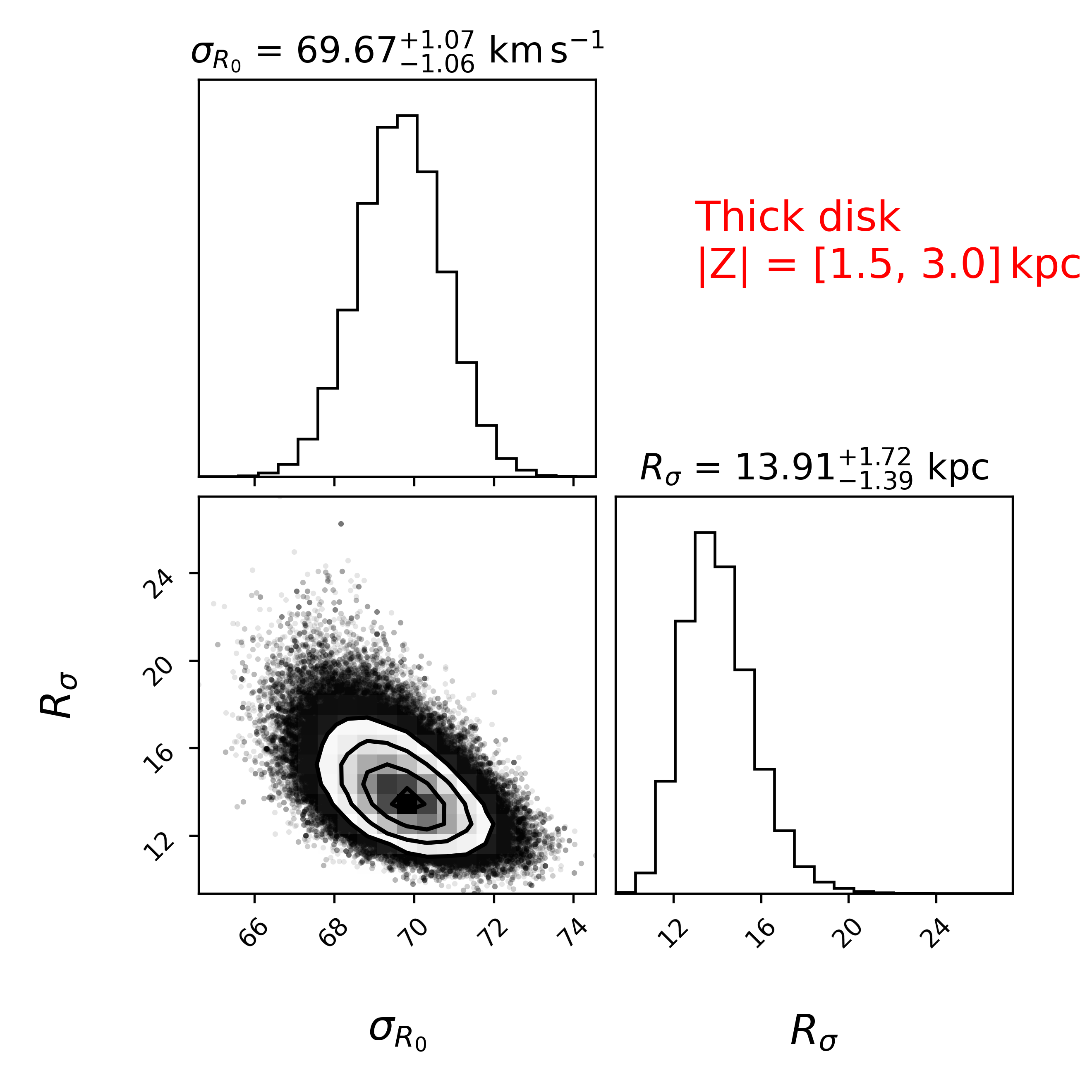}
}

\subfigure{
\includegraphics[width=7.2cm]{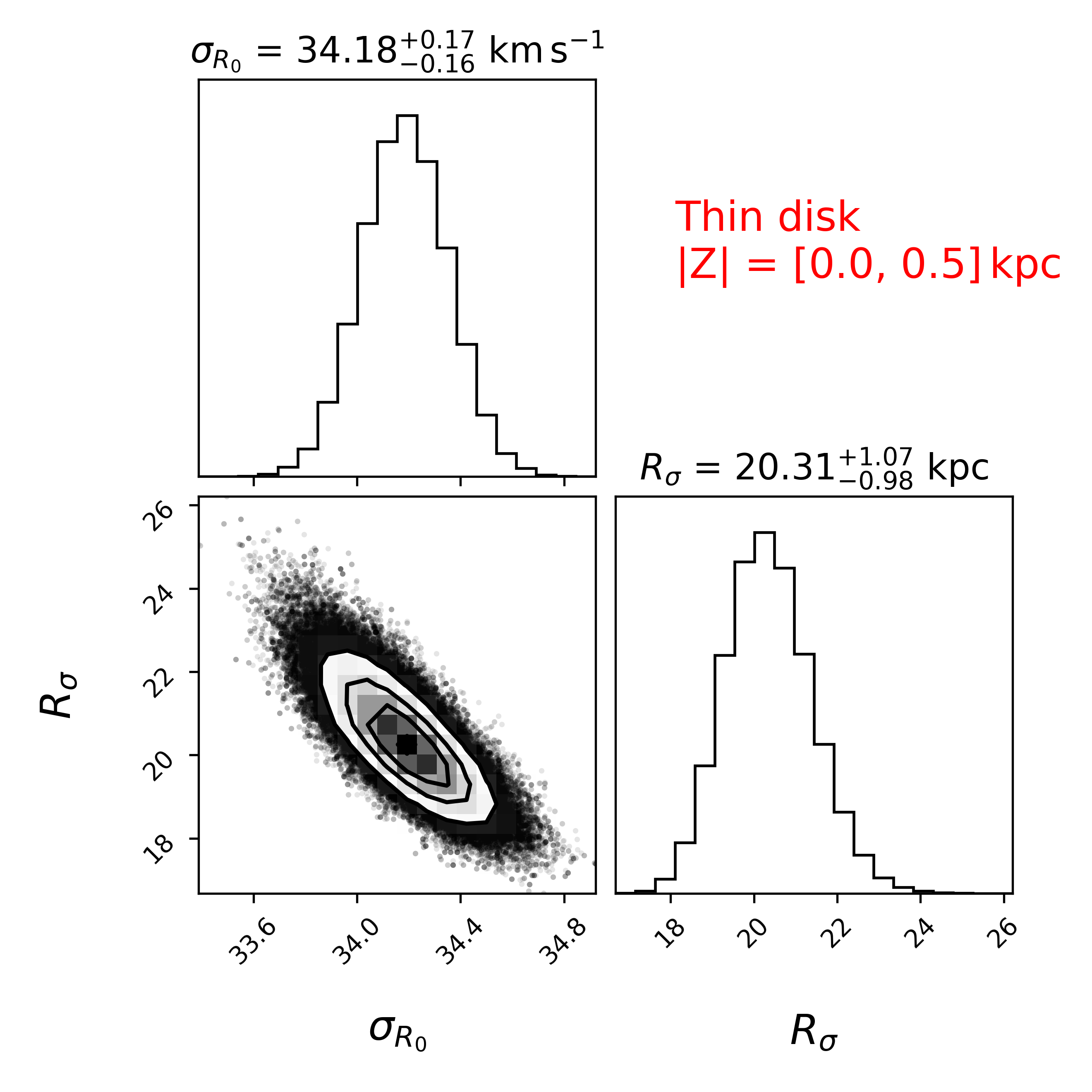}
}
\hspace{0.6cm}
\subfigure{
\includegraphics[width=7.2cm]{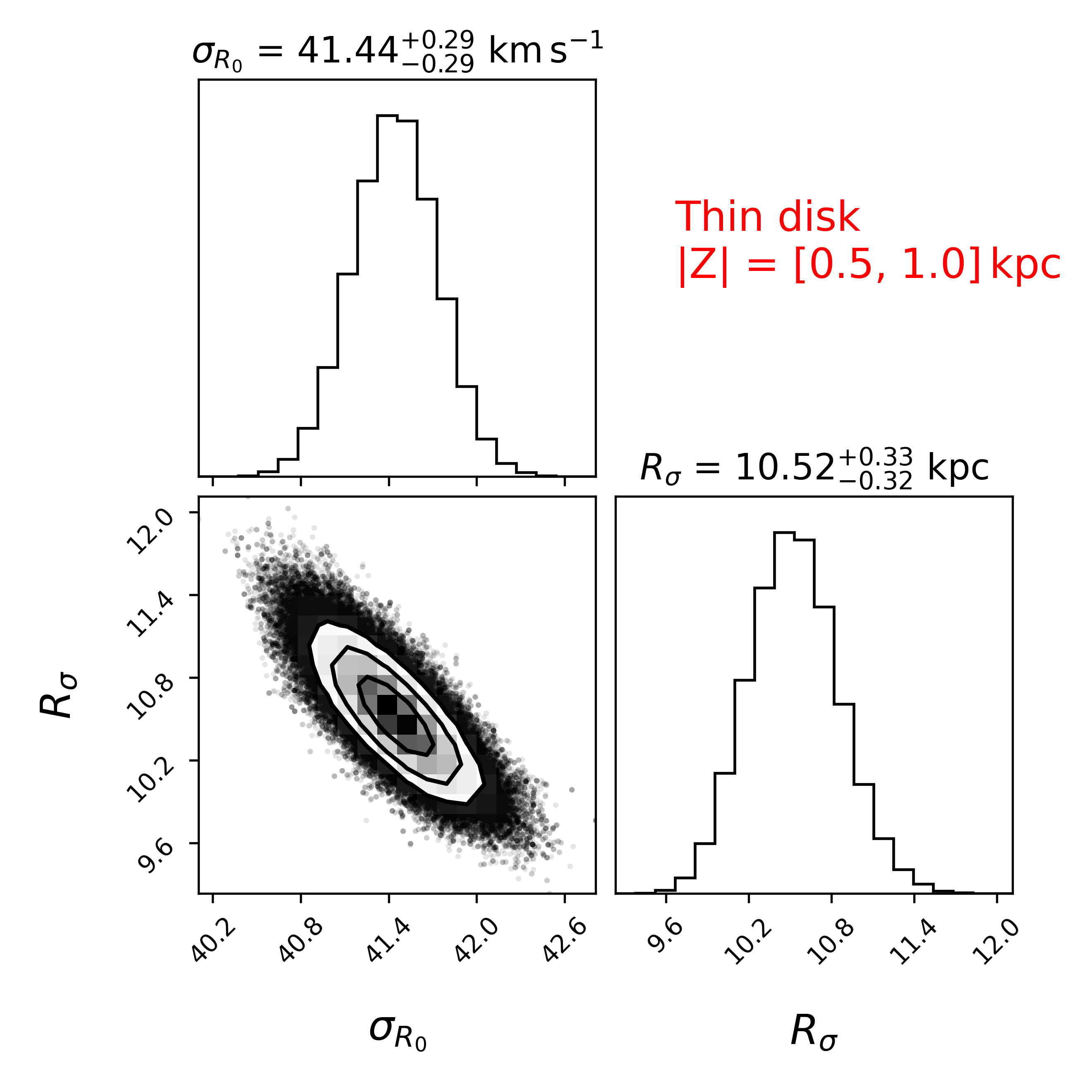}
}

\subfigure{
\includegraphics[width=7.2cm]{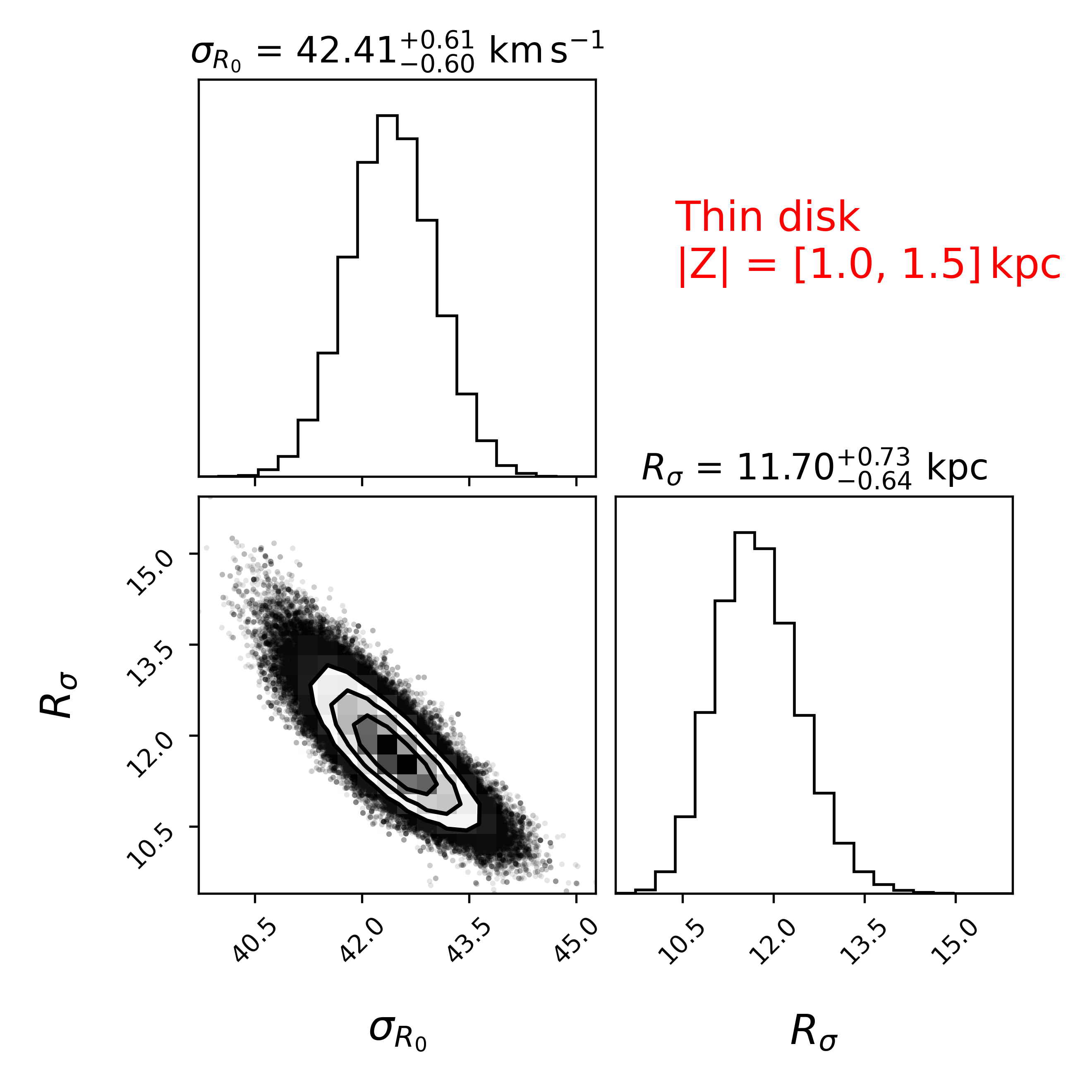}
}
\hspace{0.6cm}
\subfigure{
\includegraphics[width=7.2cm]{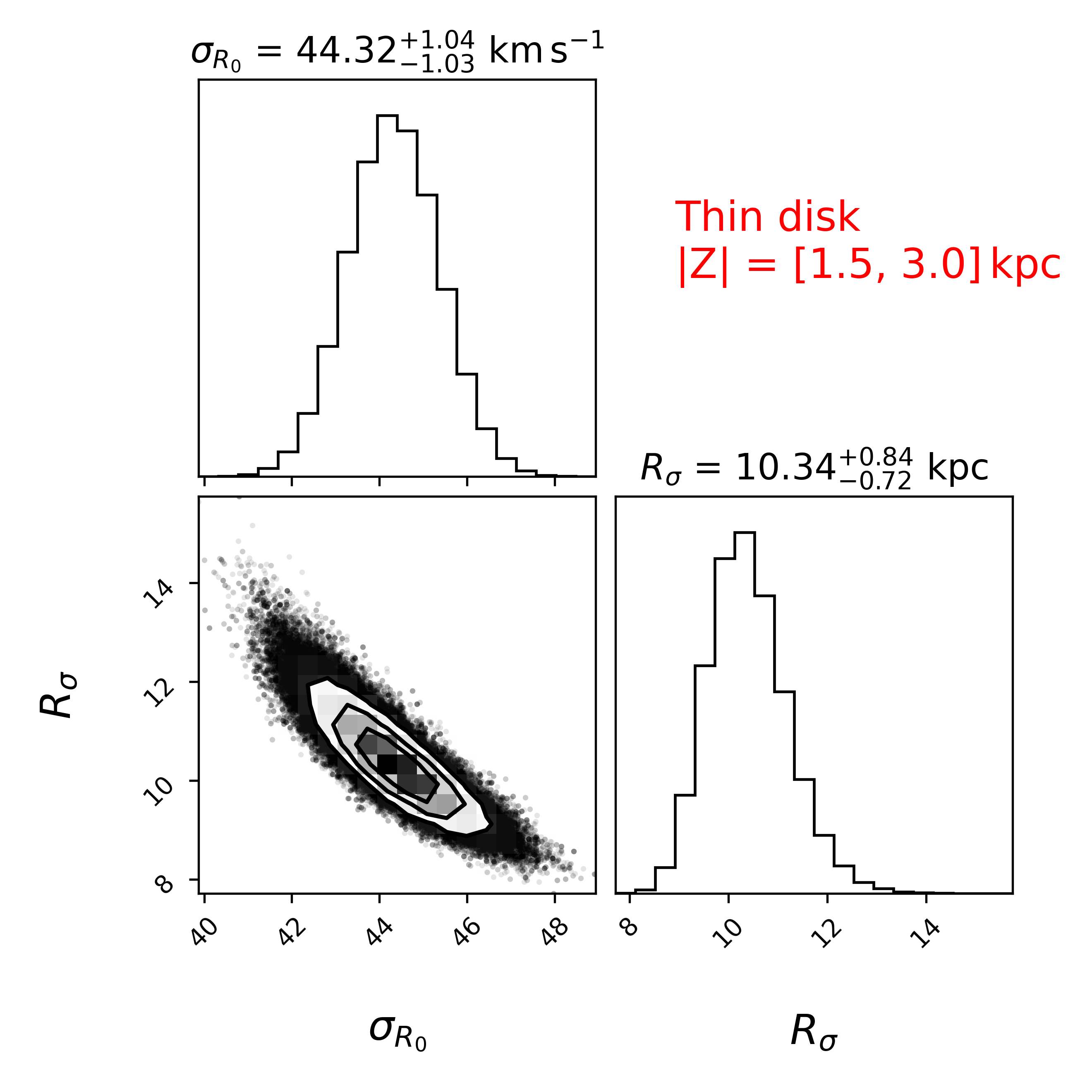}
}

\caption{The corner plots of posterior distribution of MCMC samples for $\sigma_{R_0}$ and $R_{\sigma}$.
The shadowed areas circled from inside to outside represent the confidence intervals of 1-$\sigma$, 2-$\sigma$ and 3-$\sigma$, respectively.}
\end{figure*}
%%\label{Fig.B2}

\section{AVRs of various [Fe/H] populations in the different Galactic disk regions}

This appendix displays AVRs (Fig.\,{\color{blue}{C1}}) of various [Fe/H] populations in the solar circle (left panel) and outer disk (right panel) regions.

\begin{figure*}[t]
\centering
\subfigure{
\includegraphics[width=8.8cm]{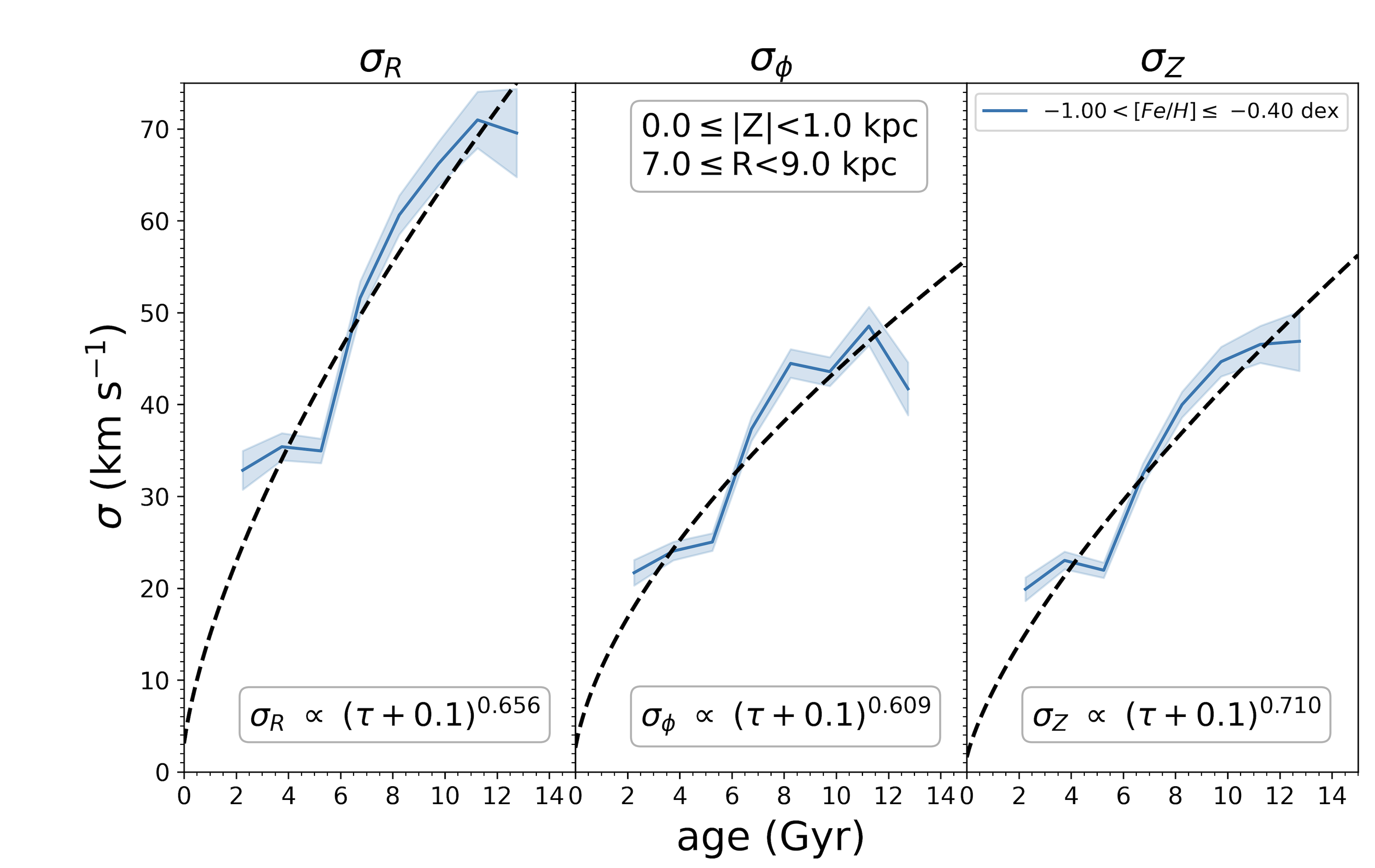}
}
\subfigure{
\includegraphics[width=8.8cm]{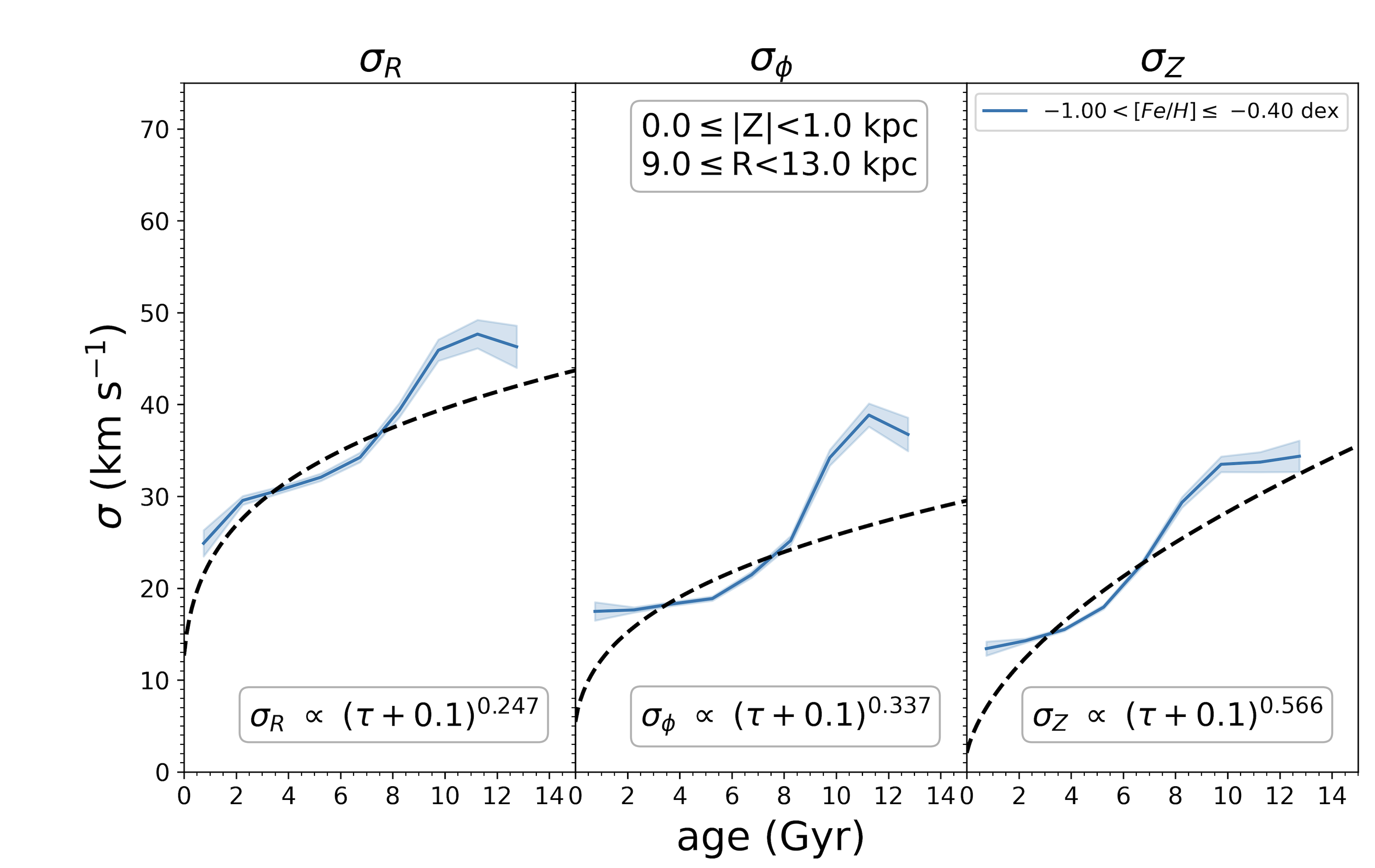}
}

\subfigure{
\includegraphics[width=8.8cm]{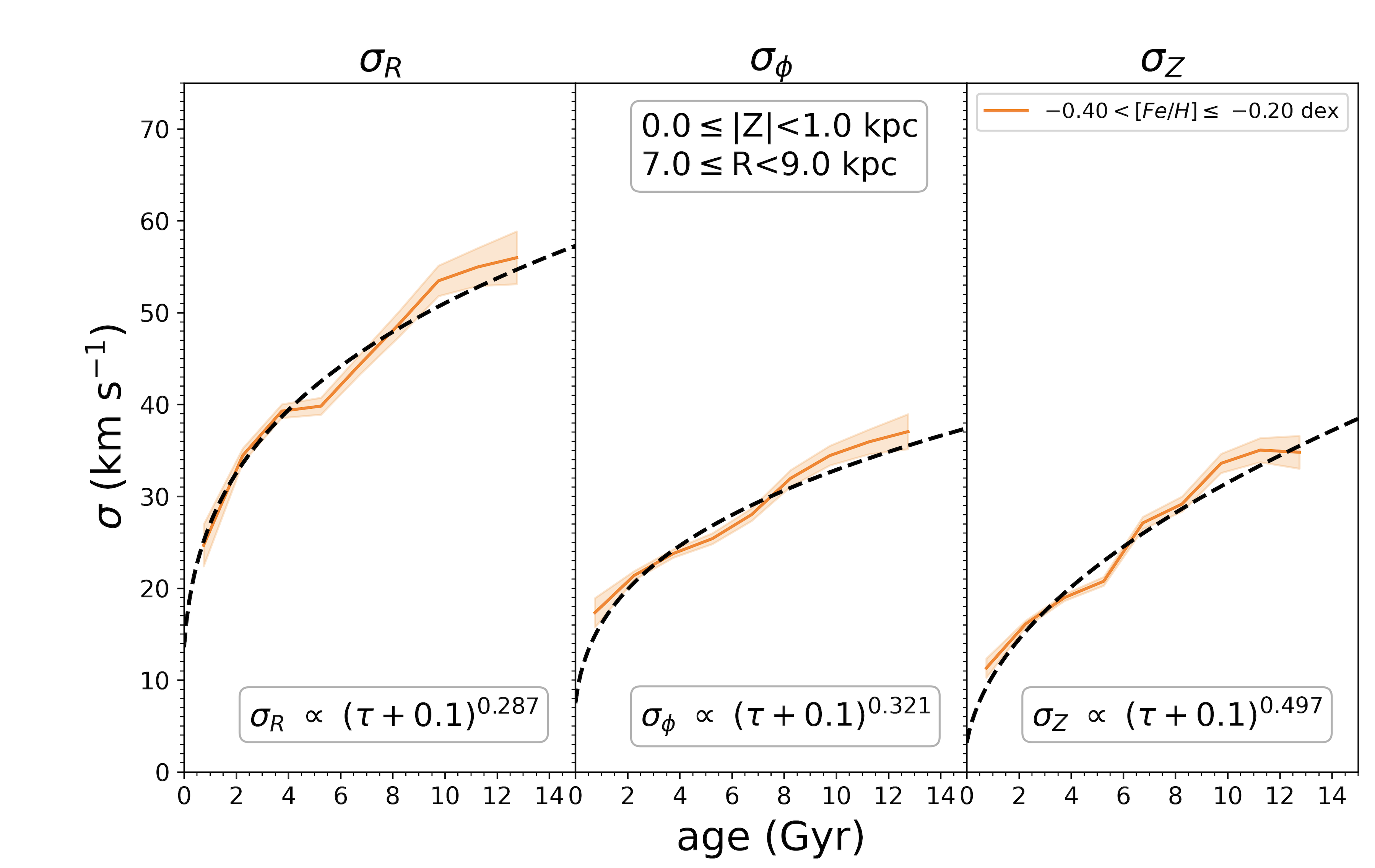}
}
\subfigure{
\includegraphics[width=8.8cm]{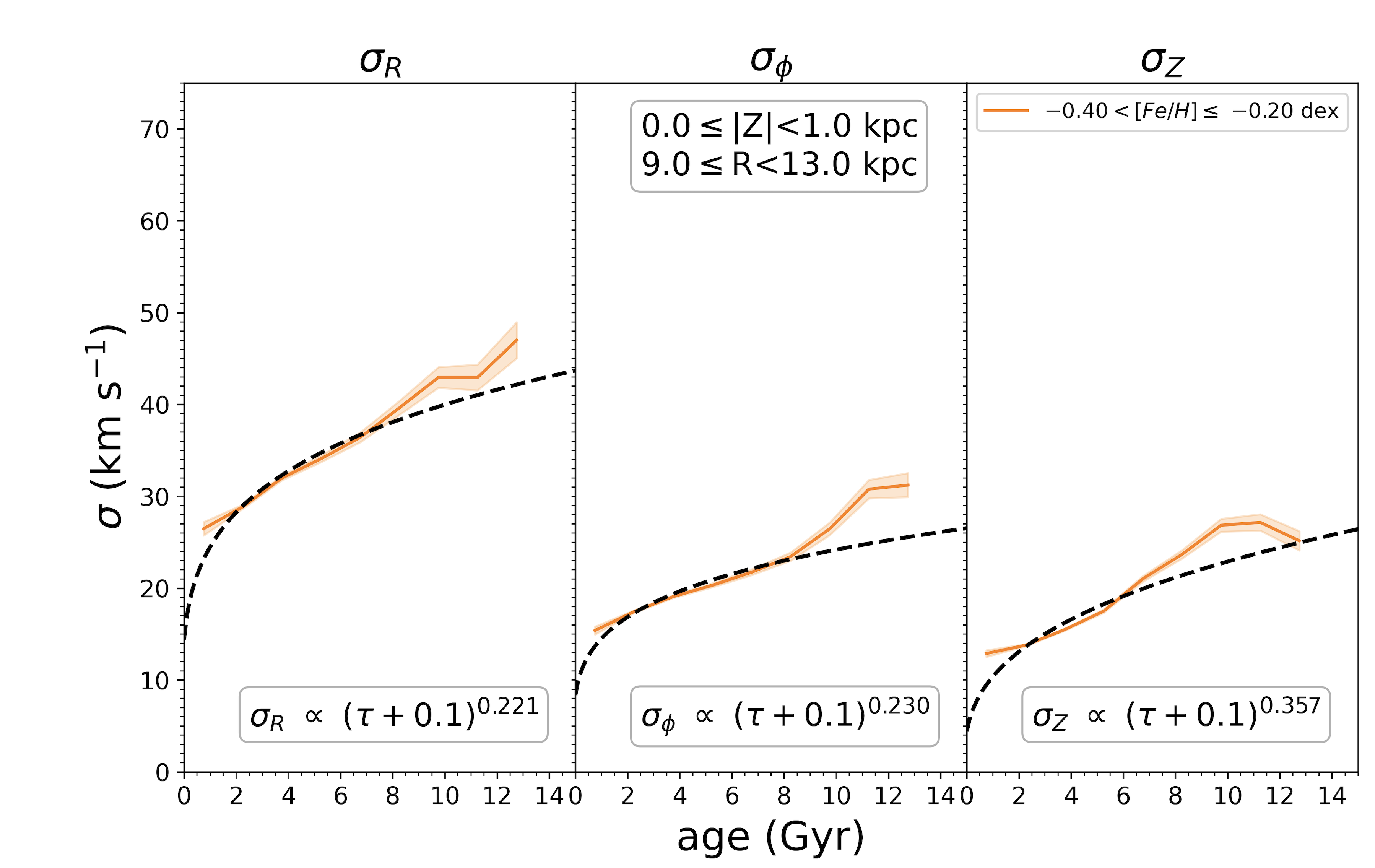}
}

\subfigure{
\includegraphics[width=8.8cm]{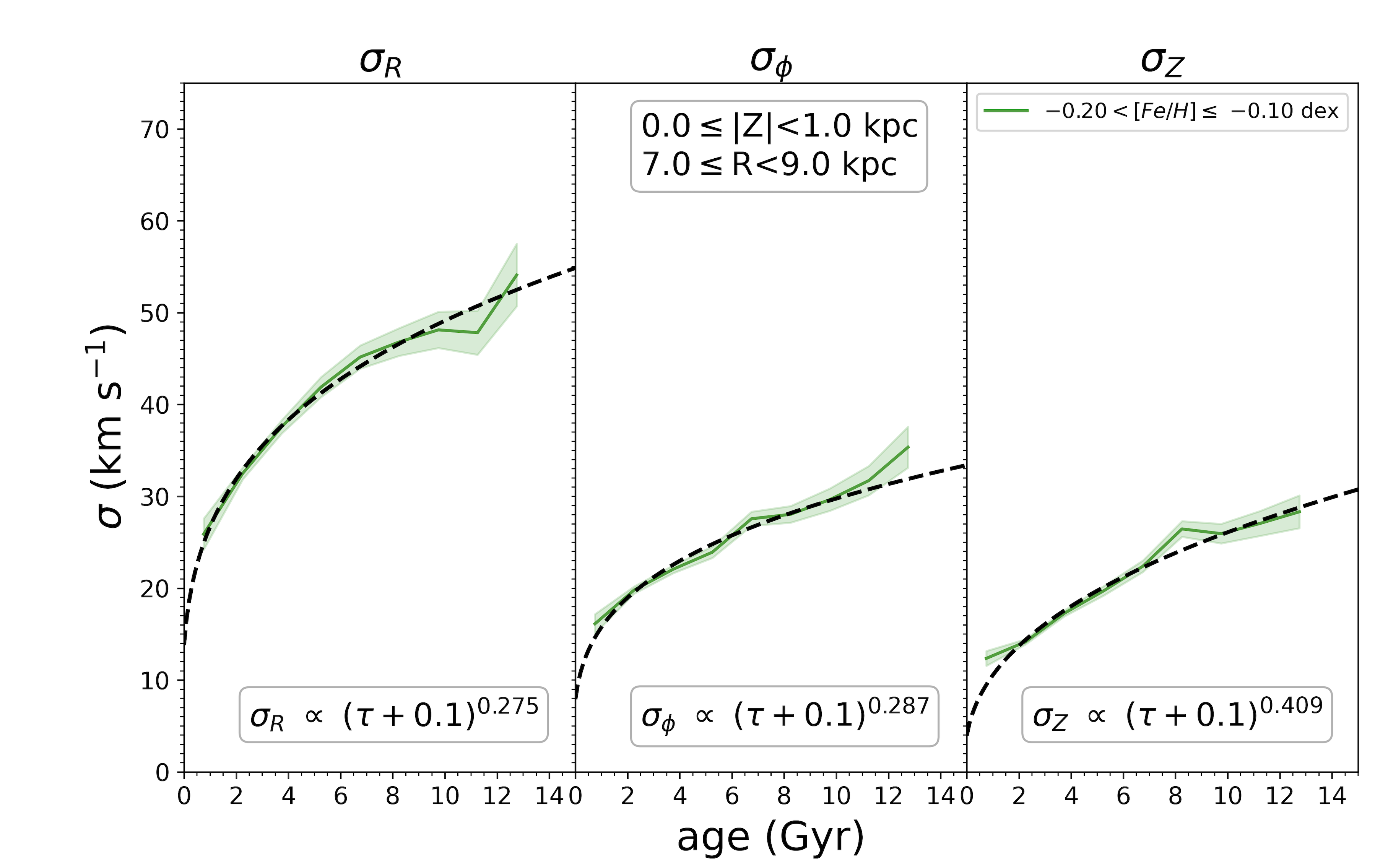}
}
\subfigure{
\includegraphics[width=8.8cm]{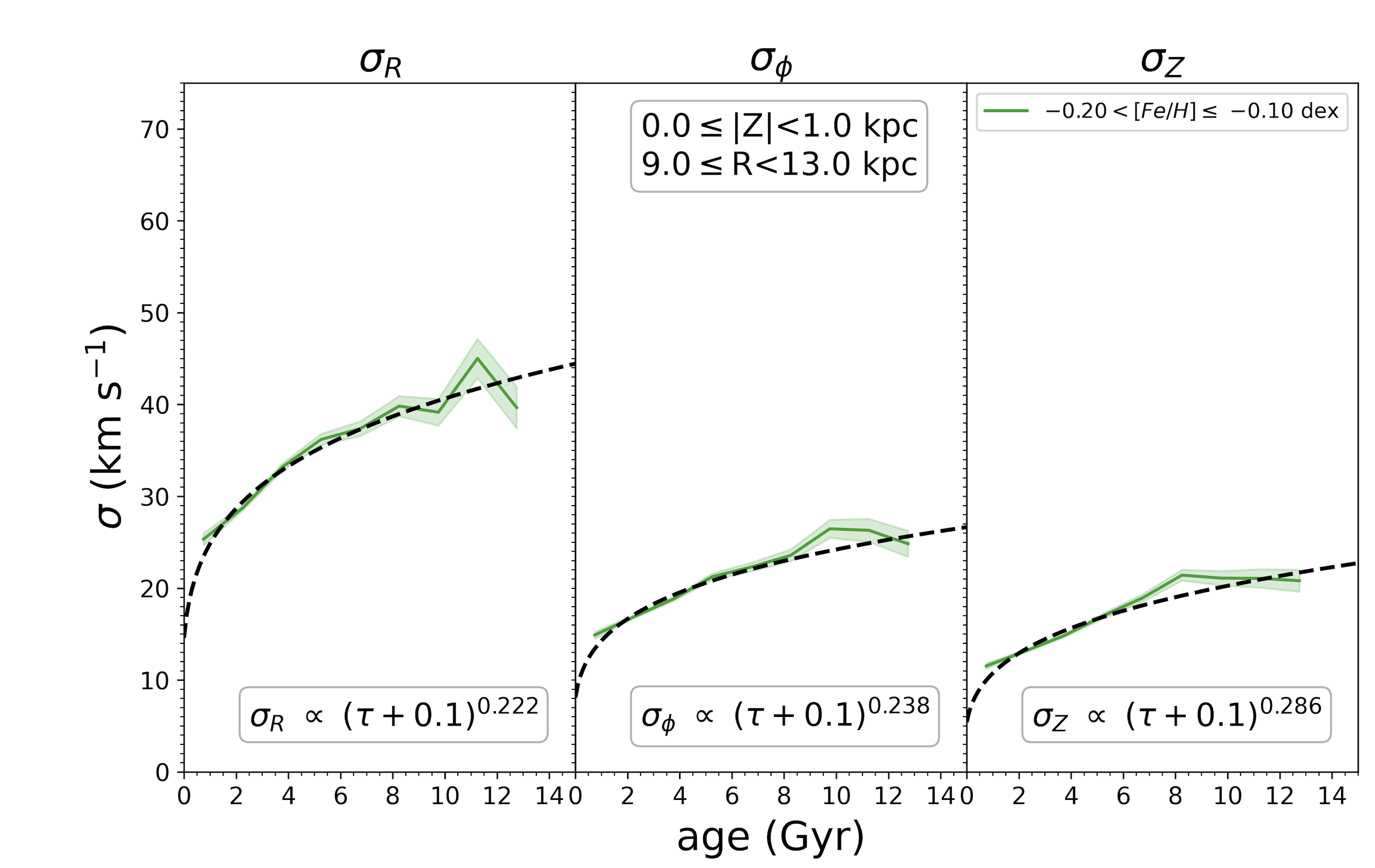}
}

\subfigure{
\includegraphics[width=8.8cm]{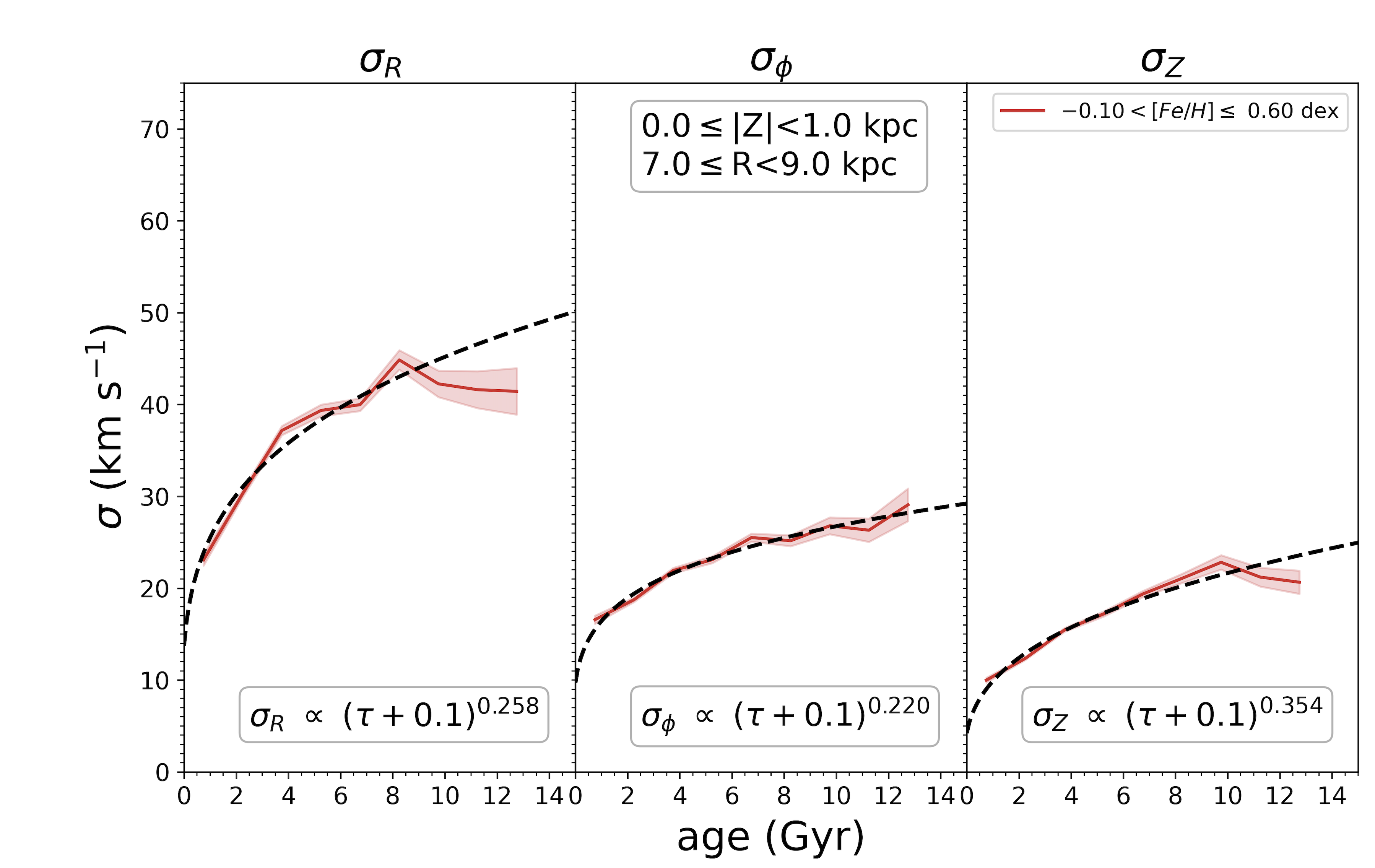}
}
\subfigure{
\includegraphics[width=8.8cm]{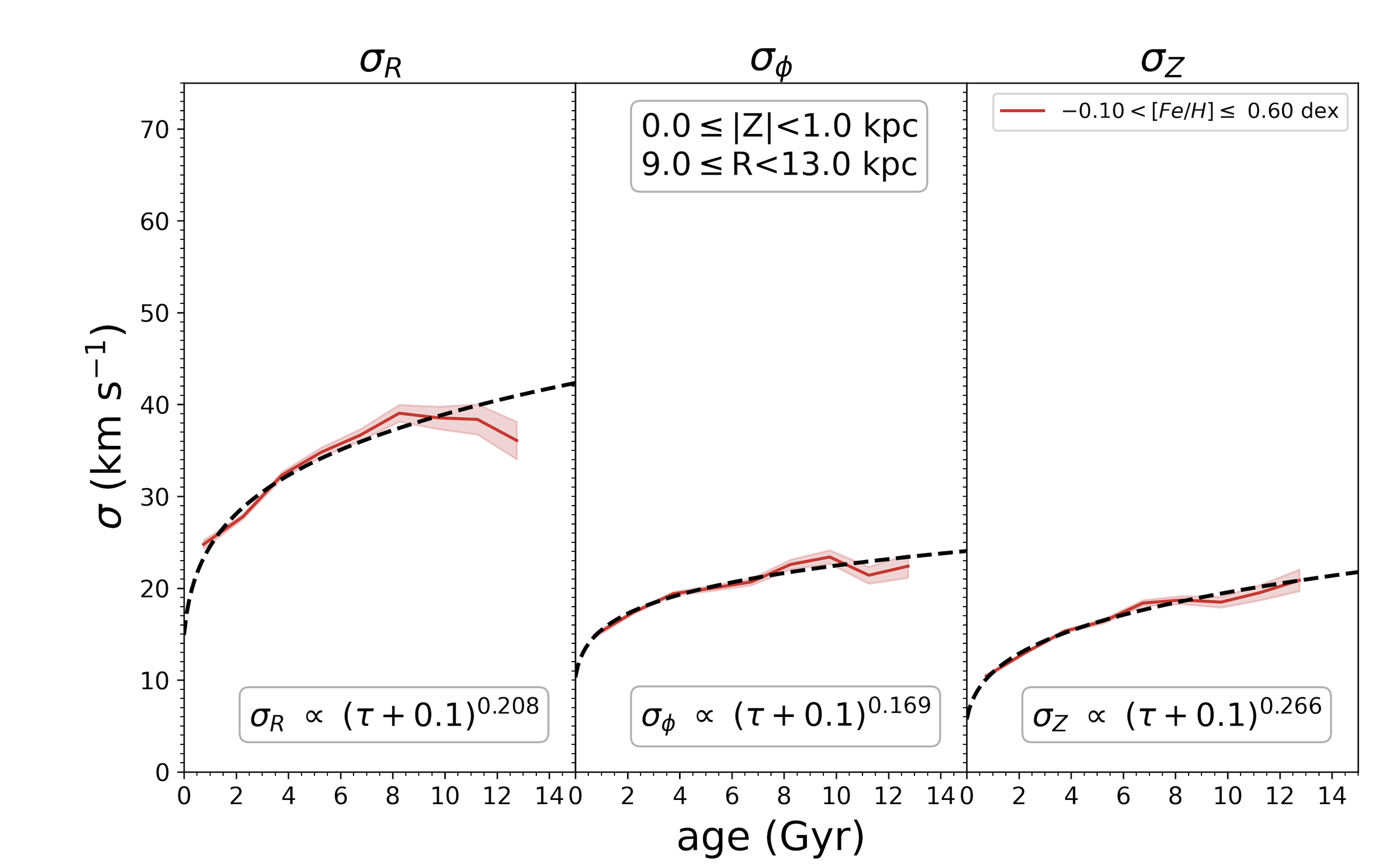}
}

\caption{The AVRs of various [Fe/H] populations for the solar circle (left panel) and outer disk (right panel) regions. There is a minimum of 20 stars per bin.}
\end{figure*}
%%\label{Fig.C1}

\section{Ratios of velocity dispersions of mono-age and mono-[$\alpha$/Fe]-[Fe/H] populations}
This appendix present the ratio of the vertical velocity dispersion compared to the radial velocity dispersion (Fig.\,{\color{blue}{D1}} and {\color{blue}{D2}}), and the ratio of the of azimuthal velocity dispersion compared to the radial velocity dispersion (Fig.\,{\color{blue}{D3}} and {\color{blue}{D4}}), of mono-age and mono-[$\alpha$/Fe]-[Fe/H] populations.

\begin{figure*}[t]
\centering
\subfigure{
\includegraphics[width=16.5cm]{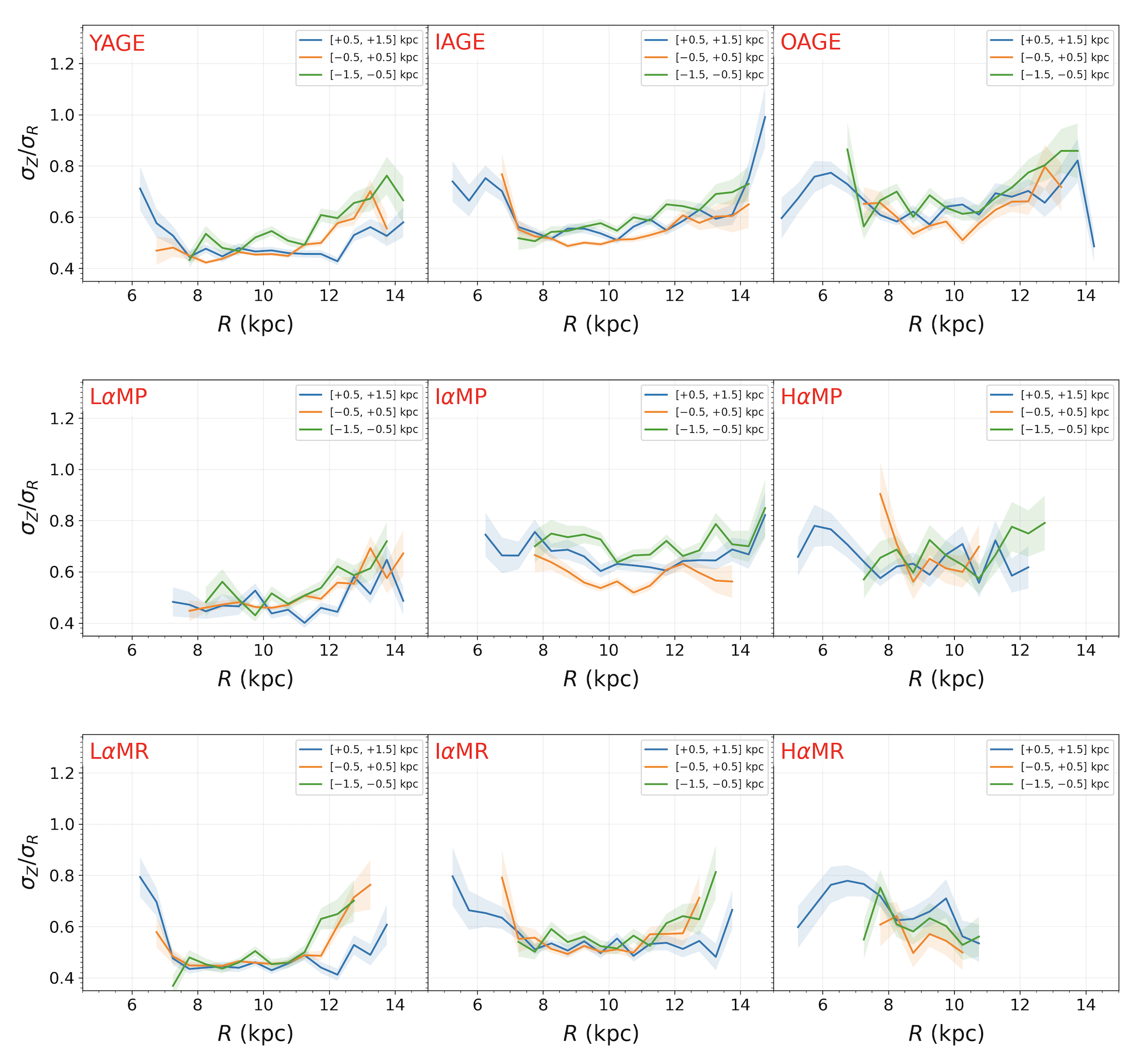}
}

\caption{Ratio of $\sigma_{Z}$/$\sigma_{R}$ of mono-age and mono-[$\alpha$/Fe]-[Fe/H] populations as a function of $R$ and $Z$.
The number of stars in each bin is required to be greater than 50 and the width of the bin is set to 0.5\,kpc.}
\end{figure*}
%%\label{Fig.D1}

\begin{figure*}[t]
\centering
\subfigure{
\includegraphics[width=16.5cm]{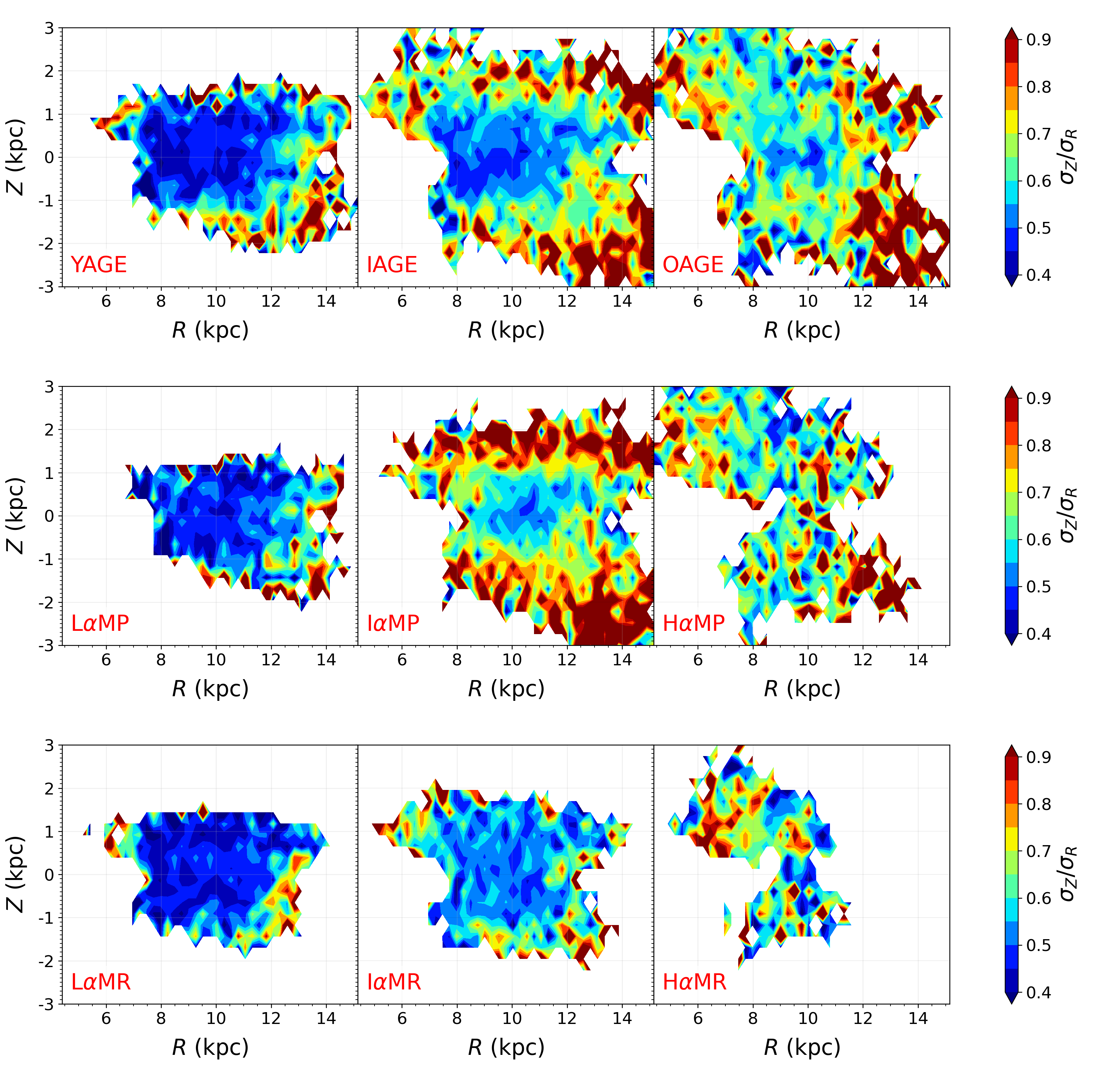}
}
\caption{Ratio of $\sigma_{Z}$/$\sigma_{R}$ of mono-age and mono-[$\alpha$/Fe]-[Fe/H] populations in $R$-$Z$ plane.
There are at least 5 stars in each bin, and both $R$ and $Z$ axes are spaced by 0.2\,kpc.}
\end{figure*}
%%\label{Fig.D2}

\begin{figure*}[t]
\centering

\subfigure{
\includegraphics[width=16.5cm]{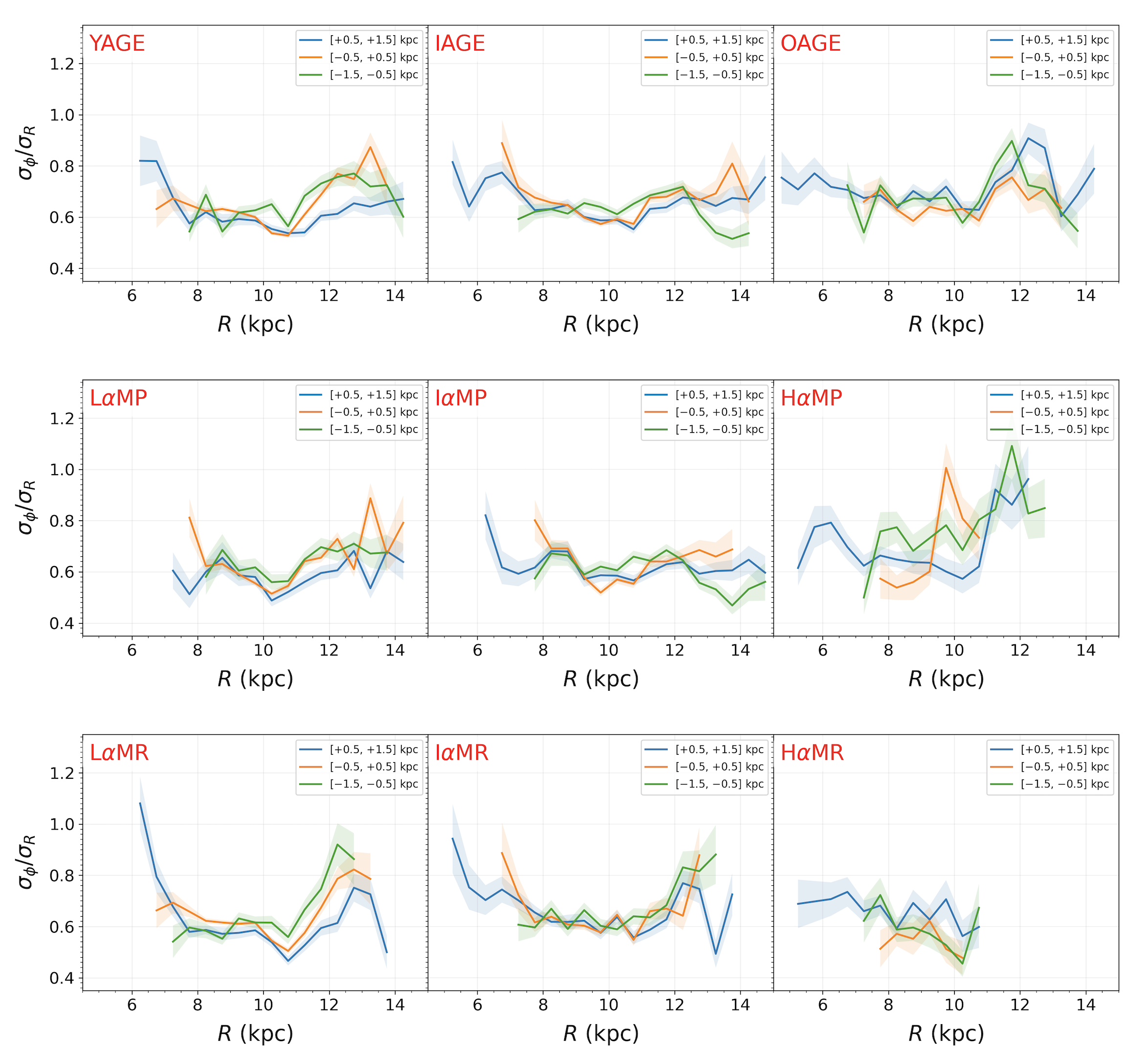}
}

\caption{Similar to the Appendix Fig.\,D1 but for the $\sigma_{\phi}$/$\sigma_{R}$.}
\end{figure*}
%%\label{Fig.D3}

\begin{figure*}[t]
\centering
\subfigure{
\includegraphics[width=16.5cm]{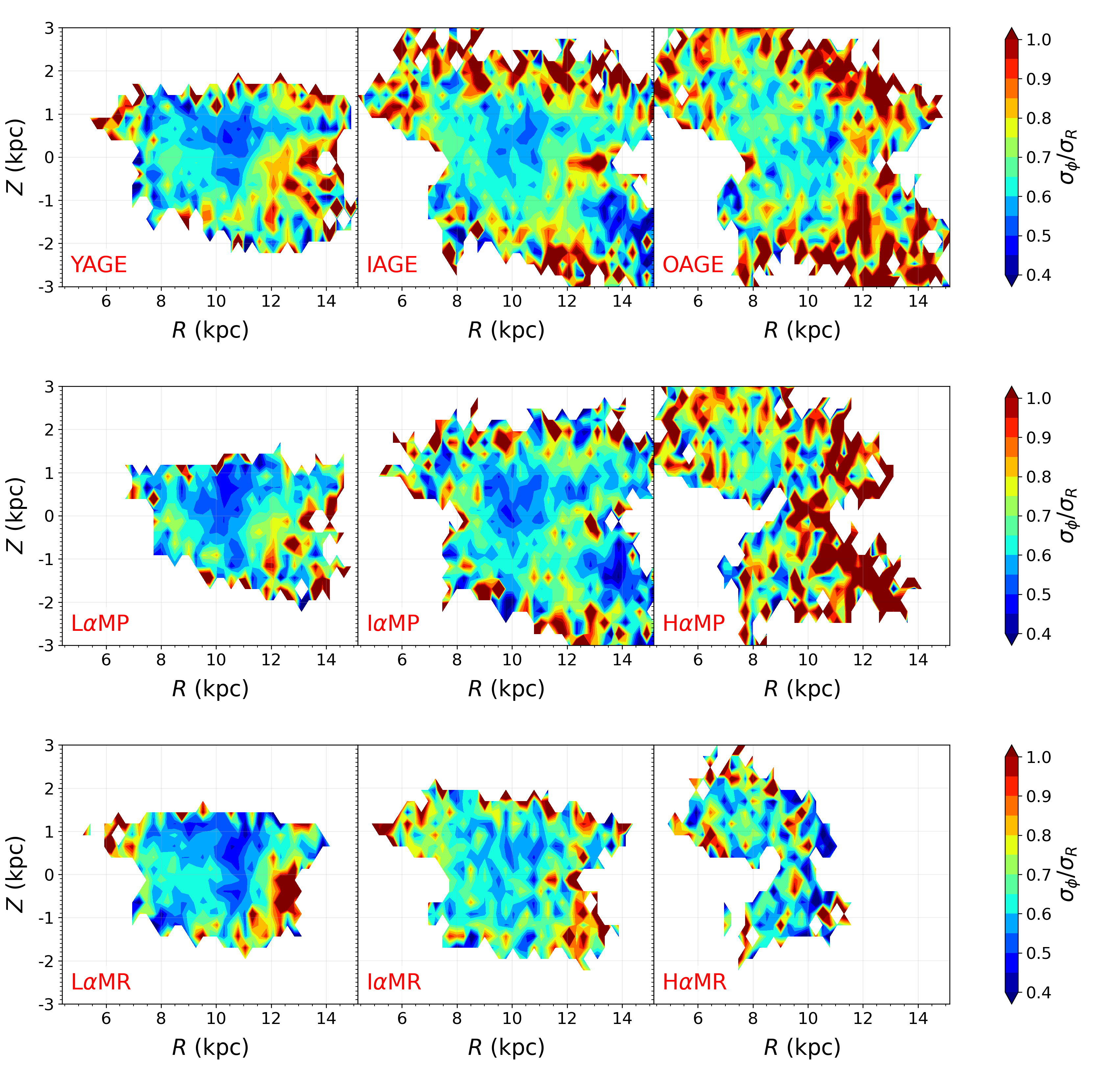}
}
\caption{Similar to the Appendix Fig.\,D2 but for the $\sigma_{\phi}$/$\sigma_{R}$.}
\end{figure*}
%%\label{Fig.D4}


\begin{thebibliography}{}
\bibitem[Abadi et al.(2003)]{Abadi2003} Abadi, M. G., Navarro, J. F., Steinmetz, M., Eke, V. R.\ 2003, \apj, 597, 21

\bibitem[Anguiano et al.(2018)]{Anguiano2018} Anguiano, B., Majewski, S. R., Freeman, K. C. et al.\ 2018, \mnras, 474, 854 

\bibitem[Antoja et al.(2018)]{Antoja2018} Antoja, T., Helmi, A., Romero-G{\'o}mez, M., et al.\ 2018, \nat, 561, 360
\bibitem[Aumer \& Binney(2009)]{Aumer2009} Aumer, M., Binney J. J.\ 2009, \mnras, 397, 1286
\bibitem[Aumer \& White (2013)]{Aumer2013} Aumer, M., White, S. D. M.\ 2013, \mnras, 428, 1055

\bibitem[Barbanis \& Woltjer(1967)]{Barbanis1967} Barbanis, B., Woltjer, L.\ 1967, \apj, 150, 461
\bibitem[Belokurov et al.(2018)]{Belokurov2018} Belokurov, V., Erkal, D., Evans, N. W., et al.\ 2018, \mnras, 478, 611

\bibitem[Bensby et al.(2005)]{Bensby2005} Bensby, T., Feltzing, S., Lundstr{\"o}m, I., Ilyin, I.\ 2005, \aap, 433, 185

\bibitem[Bensby, Feltzing \& Oey(2014)]{Bensby2014} Bensby, T., Feltzing, S., \& Oey, M. S.\ 2014, \aap, 562, A71

\bibitem[Bergemann et al.(2014)]{Bergemann2014} Bergemann, M., Ruchti, G. R., Serenelli, A., et al.\ 2014, \aap, 565, A89

\bibitem[Binney \& Tremaine(1987)]{Binney1987} Binney, J., \& Tremaine, S.\ 1987, Galactic Dynamics (Princeton, NJ: Princeton Univ. Press), 433

\bibitem[Binney \& Tremaine(2008)]{Binney2008} Binney, J., \& Tremaine, S.\ 2008, Galactic Dynamics (2nd ed.; Princeton, NJ: Princeton Univ. Press)

\bibitem[Bissantz et al. (2003)]{Bissantz2003} Bissantz, N., Englmaier, P., \& Gerhard, O.\ 2003, \mnras, 340, 949

\bibitem[Bland-Hawthorn \& Gerhard(2016)]{Bland-Hawthorn2016} Bland-Hawthorn, J., Gerhard, O.\ 2016, \araa, 54, 529

\bibitem[Bournaud et al.(2009)]{Bournaud2009} Bournaud, F., Elmegreen, B. G., \& Martig, M.\ 2009, \apj, 707, L1

\bibitem[Bovy et al.(2016)]{Bovy2016} Bovy, J., Rix, H. W., Schlafly, E. F., et al.\ 2016, \apj, 823, 30

\bibitem[Brook et al.(2004)]{Brook2004} Brook, C. B., Kawata, D., Gibson, B. K., \& Freeman, K. C.\ 2004, \apj, 612,  894

\bibitem[Brook et al.(2005)]{Brook2005} Brook, C. B., Gibson, B. K., Martel, H., \& Kawata, D.\ 2005, \apj, 630, 298

\bibitem[Brook et al.(2007)]{Brook2007} Brook, C., Richard, S., Kawata, D., Martel, H., \& Gibson, B. K.\ 2007, \apj, 658, 60

\bibitem[Brook et al.(2012)]{Brook2012} Brook, C. B., Stinson, G. S., Gibson, B. K., et al.\ 2012, \mnras, 426, 690

\bibitem[Carlin et al.(2013)]{Carlin2013} Carlin, J. L., DeLaunay, J., Newberg, H. J., et al.\ 2013, \apj, 777, L5

\bibitem[Carlin et al.(2014)]{Carlin2014} Carlin, J. L., DeLaunay, J., Newberg, H. J., et al.\ 2014, \apj, 784, L46

\bibitem[Carrillo et al.(2019)]{Carrillo2019} Carrillo, I., Minchev, I., Steinmetz, M., et al.\ 2019, \mnras, 490 797

\bibitem[Casagrande et al.(2011)]{Casagrande2011} Casagrande, L., Sch{\"o}nrich, R., Asplund, M., et al.\ 2011, \aap, 530, A138


\bibitem[Chen et al.(2019)]{Chen2019} Chen, B. Q., Huang, Y., Yuan, H. B., et al.\ 2019, \mnras, 483, 4277

\bibitem[Chiappini(2009)]{Chiappini2009} Chiappini, C. 2009, in IAU Symp. 254, The Galaxy Disk in Cosmological Context, ed. Andersen, J., Bland-Hawthorn, J., \& Nordstr{\"o}m, B., (Cambridge: Cambridge Univ. Press), 191

\bibitem[Combes (2014)]{Combes2014} Combes, F.\ 2014, in Structure and Dynamics of Disk Galaxies, eds. M. S. Seigar,
\& P. Treuthardt, ASP Conf. Ser., 480, 211

\bibitem[Cui et al.(2012)]{Cui2012} Cui, X. Q., Zhao, Y. H., Chu, Y. Q., et al.\ 2012,\ RAA, 12, 1197

\bibitem[Deason et al.(2018)]{Deason2018} Deason, A. J., Belokurov, V., Koposov, S. E., et al.\ 2018, \apj, 862, 1D

\bibitem[de Boer \& Weber (2011)]{de Boer2011} de Boer, W., Weber, M.,\ 2011, J. Cosmol. Astropart. Phys., 4, 002

\bibitem[Dehnen \& Binney(1998)]{Dehnen1998} Dehnen, W., \& Binney, J. J.\ 1998, \mnras, 298, 387

\bibitem[Deng et al.(2012)]{Deng2012} Deng, L. C., Newberg, H. J., Liu, C., et al.\ 2012,\ RAA, 12, 735
\bibitem[Duffy \& Sikivie(2003)]{Duffy2008} Duffy, L. D., Sikivie, P.\ 2008, Phys. Rev. D, 78, 063508

\bibitem[Evans \& Collett(1993)]{Evans1993} Evans, N. W., \& Collett, J. L.\ 1993, \mnras, 264, 353

\bibitem[Fuhrmann(2011)]{Fuhrmann2011} Fuhrmann, K.\ 2011, \mnras, 414, 2893

\bibitem[Foreman-Mackey et al.(2013)]{Foreman-Mackey2013} Foreman-Mackey, D., Hogg, D. W., Lang, D., et al.\ 2013, PASP, 125, 306.

\bibitem[Gaia Collaboration et al.(2016)]{Gaia Collaboration2016} Gaia Collaboration, Brown, A. G. A., Vallenari, A., Prusti, T., et al.\ 2016, \aap, 595, A2

\bibitem[Gaia Collaboration et al.(2018)]{Gaia Collaboration2018} Gaia Collaboration, Katz, D., Antoja, T., Romero-G{\'o}mez, M., et al.\ 2018, \aap, 616, A11
\bibitem[Gaia Collaboration et al.(2020)]{Gaia Collaboration2020} Gaia Collaboration, Brown, A. G. A., Vallenari, A., Prusti, T., et al.\ 2021, \aap, 649, A1

\bibitem[G{\'o}mez et al.(2012a)]{Gomez2012a} G{\'o}mez, F. A., Minchev, I., Villalobos, A., O'Shea, B. W., \& Williams, M. E. K. 2012, \mnras, 419, 2163

\bibitem[G{\'o}mez et al.(2012b)]{Gomez2012b} G{\'o}mez, F. A., Minchev, I., O'Shea, B. W., et al., 2012, \mnras, 423, 3727

\bibitem[Grand et al.(2016)]{Grand2016} Grand, R. J. J., Springel, V., G{\'o}mez, F. A., et al.\ 2016, \mnras, 459, 199

\bibitem[Guiglion et al.(2015)]{Guiglion2015} Guiglion, G., Recio-Blanco, A., de Laverny, P., et al.\ 2015, \aap, 583, A91

\bibitem[Han et al.(2020)]{Han2020} Han, D. R., Lee, Y. S., Kim, Y. K., et al.\ 2020, \apj, 896, 14H

\bibitem[H{\"a}nninen \& Flynn(2002)]{Hanninen2002} H{\"a}nninen, J., Flynn, C.\ 2002, \mnras, 337, 731

\bibitem[Hayden et al.(2015)]{Hayden2015} Hayden, M.~R., Bovy, J., Holtzman, J.~A., et al.\ 2015, \apj, 808, 132

\bibitem[Hayden et al.(2018)]{Hayden2018} Hayden, M. R., Recio-Blanco, A., de Laverny, P., et al.\ 2018, \aap, 609, A79

\bibitem[Hayden et al.(2020)]{Hayden2020} Hayden, M. R., Bland-Hawthorn, J., Sharma, S., et al.\ 2020, \mnras, 493, 2952

\bibitem[Haywood et al.(2013)]{Haywood2013} Haywood, M., Di Matteo, P., Lehnert, M. D., et al.\ 2013, \aap, 560, A109

\bibitem[Helmi et al.(2018)]{Helmi2018} Helmi, A., Babusiaux, C., Koppelman, H. H., et al.\ 2018, \nat, 563, 85

\bibitem[Holmberg \& Nordstr{\"o}m \& Andersen(2009)]{Holmberg2009} Holmberg, J., Nordstr{\"o}m, B., Andersen, J.\ 2009, \aap, 501, 941
\bibitem[House et al. (2011)]{House2011} House, E. L.; Brook, C. B.; Gibson, B. K., et al.\ 2011, \mnras, 415, 2652

\bibitem[Huang et al.(2015)]{Huang2015} Huang, Y., Liu, X. W., Yuan, H. B., et al.\ 2015, \mnras, 449, 162
\bibitem[Huang et al.(2016)]{Huang2016} Huang, Y., Liu, X. W., Yuan, H. B., et al.\ 2016, \mnras, 463, 2623
\bibitem[Huang et al.(2018a)]{Huang2018a} Huang, Y., Liu, X.-W., Chen, B.-Q., et al.\ 2018a, \aj, 156, 90

\bibitem[Huang et al.(2018b)]{Huang2018b} Huang, Y., Sch{\"o}nrich, R., Liu, X. W., et al.,\ 2018b, \apj, 864, 129

\bibitem[Huang et al.(2020)]{Huang2020} Huang, Y., Sch{\"o}nrich, R., Zhang, H. W., et al.\ 2020, ApJS, 249, 29

\bibitem[Hunter \& Toomre(1969)]{Hunter1969} Hunter, C., Toomre, A.\ 1969, \apj, 155, 747

\bibitem[Hunt et al.(2018)]{Hunt2018} Hunt, J. A. S., Hong, J., Bovy, J., Kawata, D., Grand, R. J. J.\ 2018, \mnras, 481, 3794

\bibitem[Jenkins \& Binney(1990)]{Jenkins1990} Jenkins, A., Binney, J.\ 1990, \mnras, 245, 305

\bibitem[Jenkins(1992)]{Jenkins1992} Jenkins, A.\ 1992, \mnras, 257, 620
\bibitem[Johnson \& Soderblom(1987)]{Johnson1987} Johnson, D. R. H., Soderblom, D. R.\ 1987, \aj, 93, 864

\bibitem[Kazantzidis et al.(2008)]{Kazantzidis2008} Kazantzidis, S., Bullock, J. S., Zentner, A. R., et al.\ 2008, \apj, 688, 254

\bibitem[Khoperskov \& Bertin(2017)]{Khoperskov2017} Khoperskov, S., Bertin, G.\ 2017, \aap, 597, A103

\bibitem[Kruijssen et al.(2019)]{Kruijssen2019} Kruijssen, J. M. D., Pfeffer, J. L., Reina-Campos, M., et al.\ 2019, \mnras, 486, 3180

\bibitem[Kuijken \& Tremaine(1991)]{Kuijken1991} Kuijken, K., \& Tremaine, S.\ 1991, in Dynamics of Disk Galaxies, ed. B. Sundelius (G{\"o}teborg: G{\"o}teborg Univ. Press), 71

\bibitem[Lacey(1984)]{Lacey1984} Lacey, C. G.\ 1984, \mnras, 208, 687

\bibitem[Lagarde et al.(2021)]{Lagarde2021} Lagarde, N., Reyl{\'e}, C., Chiappini, C., et al.\ 2021, \aap, 654, A13


\bibitem[Laporte et al.(2018)]{Laporte2018} Laporte, C. F. P., G{\'o}mez, F. A., Besla, G., Johnston, K. V., \& Garavito-Camargo, N.\ 2018, \mnras, 473, 1218


\bibitem[Lee et al.(2011)]{Lee2011} Lee, Y. S., Beers, T. C., An, D., et al.\ 2011, \apj, 738, 187

\bibitem[Li et al.(2020)]{Li2020} Li, X. Y., Huang, Y., Chen, B. Q., et al.\ 2020, \apj, 901, 56

\bibitem[Liu et al.(2014)]{Liu2014} Liu, X. W., Yuan, H. B., Huo, Z. Y., et al.\ 2014, in IAU Symp. 298, Setting the Scene for Gaia and LAMOST (Cambridge: Cambridge Univ. Press), 310

\bibitem[Mackereth et al.(2017)]{Mackereth2017} Mackereth, J. T., Bovy, J., Schiavon, R. P., et al.\ 2017, \mnras, 471, 3057

\bibitem[Mackereth et al.(2019a)]{Mackereth2019a} Mackereth, J. T., Schiavon, R. P. , Pfeffer, J., et al.\ 2019, \mnras, 482, 3426

\bibitem[Mackereth et al.(2019b)]{Mackereth2019b} Mackereth, J. T., Bovy, J., Leung, H. W., et al.\ 2019, \mnras, 489, 176 

\bibitem[Martig et al.(2014)]{Martig2014} Martig, M., Minchev, I., Flynn, C.\ 2014, \mnras, 443, 2452


\bibitem[Minchev \& Famaey(2010)]{Minchev2010} Minchev, I., Famaey, B.\ 2010, \apj, 722, 112

\bibitem[Minchev et al.(2012)]{Minchev2012} Minchev, I., Famaey, B., Quillen, A. C., et al.\ 2012, \aap, 548, A127
\bibitem[Minchev, Chiappini \& Martig (2013)]{Minchev2013} Minchev, I., Chiappini, C., \& Martig, M.\ 2013, \aap, 558, A9
\bibitem[Minchev et al.(2014)]{Minchev2014} Minchev, I., Chiappini, C., Martig, M., et al.\ 2014, \apjl, 781, L20
\bibitem[Minchev et al.(2015)]{Minchev2015} Minchev, I., Martig, M., Streich, D., et al.\ 2015, \apj, 804, L9
\bibitem[Minchev et al. (2018)]{Minchev2018} Minchev, I., Anders, F., Recio-Blanco, A., et al.\ 2018, \mnras, 481, 1645

\bibitem[Nidever et al.(2014)]{Nidever2014} Nidever, D.~L., Bovy, J., Bird, J.~C., et al.\ 2014, \apj, 796, 38

\bibitem[Nordstr{\"o}m et al.(2004)]{Nordstrom2004} Nordstr{\"o}m, B., Mayor, M., Andersen, J., et al.\ 2004, \aap, 418, 989 

\bibitem[Matteucci(2001)]{Matteucci2001} Matteucci, F. 2001, Astrophysics and Space Science Library, Vol. 253, The Chemical Evolution of the Galaxy (Dordrecht: Kluwer Academic Publishers)

\bibitem[Matteucci(2012)]{Matteucci2012} Matteucci, F.,\ 2012,\ Chemical Evolution of Galaxies (Berlin: Springer)

\bibitem[Pagel(2009)]{Pagel2009} Pagel, B. E. J. 2009, Nucleosynthesis and Chemical Evolution of Galaxies (Cambridge, UK: Cambridge University Press)

\bibitem[Perryman et al.(2001)]{Perryman2001} Perryman, M. A. C., de Boer, K. S., Gilmore, G., et al.\ 2001, \aap, 369, 339
\bibitem[Poggio et al.(2018)]{Poggio2018} Poggio, E., Drimmel, R., Lattanzi, M. G., et al.\ 2018, \mnras, 481, L21

\bibitem[Queiroz et al.(2020)]{Queiroz2020} Queiroz, A.~B.~A., Anders, F., Chiappini, C., et al.\ 2020, \aap, 638, A76

\bibitem[Quinn et al.(1993)]{Quinn1993} Quinn, P. J., Hernquist, L., \& Fullagar, D. P.\ 1993, \apj, 403, 74

\bibitem[Quillen \& Garnett(2001)]{Quillen2001} Quillen, A. C., \& Garnett, D. R. 2001, in ASP Conf. Ser. 230, Galaxy Disks and Disk Galaxies, ed. J. G. Funes \& E. M. Corsini (San Francisco, CA: ASP), 87

\bibitem[Recio-Blanco et al.(2014)]{Recio-Blanco2014} Recio-Blanco, A., de Laverny, P., Kordopatis, G., et al.\ 2014, \aap, 567, A5

\bibitem[Reid \& Brunthaler(2004)]{Reid2004} Reid, M. J., \& Brunthaler, A.\ 2004, \apj, 616, 872 
\bibitem[Reid et al.(2014)]{Reid2014} Reid, M. J., Menten, K. M., Brunthaler, A., et al.\ 2014, \apj, 783, 130
\bibitem[Rix \& Bovy(2013)]{Rix2013} Rix, H. W., Bovy, J.\ 2013, \aapr, 21, 61
\bibitem[Roman(1950a)]{Roman1950a} Roman, N. G.\ 1950a, \aj, 55, 182 
\bibitem[Roman(1950b)]{Roman1950b} Roman, N. G.\ 1950b, \apj, 112, 554
\bibitem[Ro{\v s}kar et al. (2010)]{Roskar2010} Ro{\v s}kar, R., Debattista, V. P., Brooks, A. M., et al.\ 2010, \mnras, 408, 783

\bibitem[Sarkar \& Jog(2018)]{Sarkar2018} Sarkar, S. \& Jog, C.~J.\ 2018, \aap, 617, A142.


\bibitem[Sanders(2018)]{Sanders2018} Sanders, J. L., Das, P.\ 2018, \mnras, 481, 4093

\bibitem[Sch{\"o}nrich \& Binney(2009a)]{Schonrich2009a} Sch{\"o}nrich, R., Binney, J.\ 2009a, \mnras, 396, 203 
\bibitem[Sch{\"o}nrich \& Binney(2009b)]{Schonrich2009b} Sch{\"o}nrich, R., Binney, J.\ 2009b, \mnras, 399, 1145
\bibitem[Sch{\"o}nrich et al.(2010)]{Schonrich2010} Sch{\"o}nrich, R., Binney, J., \& Dehnen, W.\ 2010, \mnras, 403, 1829
\bibitem[Sch{\"o}nrich(2012)]{Schonrich2012} Sch{\"o}nrich, R.\ 2012, \mnras, 427, 274

\bibitem[Sch{\"o}nrich \& Dehnen(2018)]{Schonrich2018} Sch{\"o}nrich, R., \& Dehnen, W.\ 2018, \mnras, 478, 3809

\bibitem[Seabroke \& Gilmore(2007)]{Seabroke2007} Seabroke, G. M., Gilmore, G.\ 2007, \mnras, 380, 1348 
\bibitem[Sharma et al.(2012)]{Sharma2012} Sharma, S., Steinmetz, M., \& Bland-Hawthorn, J.\ 2012, \apj, 750, 107
\bibitem[Sharma et al.(2014)]{Sharma2014} Sharma, S., Bland-Hawthorn, J., Binney, J., et al.\ 2014, \apj, 793, 51
\bibitem[Sharma et al.(2021)]{Sharma2021} Sharma, S., Hayden, M. R., Bland-Hawthorn, J., et al.\ 2021, \mnras, 506, 1761
\bibitem[Siebert et al.(2011)]{Siebert2011} Siebert, A., Famaey, B., Minchev, I., et al.\ 2011, \mnras, 412, 2026
\bibitem[Sikivie (2003)]{Sikivie2003} Sikivie, P.,\ 2003, Phys. Lett. B, 567, 1

\bibitem[Sofue et al. (2009)]{Sofue2009} Sofue, Y., Honma, M., Omodaka, T.,\ 2009, PASJ, 61, 227

\bibitem[Spina et al.(2021)]{Spina2021} Spina, L., Ting, Y. S., De Silva, G. M., et al.\ 2021, \mnras, 503, 3279
\bibitem[Spitzer \& Schwarzschild(1951)]{Spitzer1951} Spitzer Lyman J., Schwarzschild M.\ 1951, \apj, 114, 385
\bibitem[Spitzer \& Schwarzschild(1953)]{Spitzer1953} Spitzer Lyman J., Schwarzschild M.\ 1953, \apj, 118, 106

\bibitem[Str{\"o}mberg(1946)]{Stromberg1946} Str{\"o}mberg, G.\ 1946, \apj, 104, 12
\bibitem[Sun et al.(2015)]{Sun2015} Sun, N. C., Liu, X. W., Huang, Y., et al.\ 2015, RAA, 15, 1342
\bibitem[Sun et al.(2020)]{Sun2020} Sun, W. X., Huang, Y., Wang, H. F., et al.\ 2020, \apj, 903, 12
\bibitem[Sun et al.(2023)]{Sun2023} Sun, W. X., Shen, H., Liu, X. W.\ 2023, \apj, 952, 163
\bibitem[Trick et al.(2019)]{Trick2019} Trick, W. H., Coronado, J., Rix, H. W.\ 2019, \mnras, 484, 3291
\bibitem[Vera-Ciro et al.(2014)]{Vera-Ciro2014} Vera-Ciro, C., D'Onghia, E., Navarro, J., Abadi, M.\ 2014, \apj, 794, 173

\bibitem[Villalobos \& Helmi(2008)]{Villalobos2008} Villalobos, {\'A}., \& Helmi, A.\ 2008, \mnras, 391, 1806 

\bibitem[Wang et al.(2019a)]{Wangc2019} Wang, C., Huang, Y., Yuan, H. -B., et al.\ 2019, \apj, 877, L7

\bibitem[Wang et al.(2019)]{Wang2019} Wang, H.-F., Carlin, J. L., Huang, Y., et al.\ 2019, \apj, 884, 135.

\bibitem[Widrow et al.(2012)]{Widrow2012} Widrow, L. M., Gardner, S., Yanny, B., Dodelson, S., Chen, H. Y.\ 2012, \apj, 750, 41

\bibitem[Wielen(1977)]{Wielen1977} Wielen, R.\ 1977, \aap, 60, 263
\bibitem[Williams et al.(2013)]{Williams2013} Williams, M. E. K., Williams, M., Binney, J., et al.\ 2013, \mnras, 436, 101
\bibitem[Wisnioski et al.(2015)]{Wisnioski2015} Wisnioski, E., F{\"o}rster Schreiber, N. M., Wuyts, S., et al.\ 2015, \apj, 799, 209
\bibitem[Xu et al.(2018)]{Xu2018} Xu, Y., Bian, S. B., Reid, M. J., et al.\ 2018, \aap, 616, L15
\bibitem[Yan et al.(2019)]{Yan2019} Yan, Y., Du, C., Liu, S., et al.\ 2019, \apj, 880, 36

\bibitem[Yoachim \& Dalcanton(2006)]{Yoachim2006} Yoachim, P., Dalcanton, J. J.\ 2006, \aj, 131, 226
\bibitem[York et al.(2000)]{York2000}York, D. G., Adelman, J., Anderson, J. E. J., et al.\ 2000, \aj, 120, 1579
\bibitem[Yu \& Liu(2018)]{Yu2018} Yu, J. C., Liu, C.\ 2018, \mnras, 475, 1093
\bibitem[Yuan et al.(2015)]{Yuan2015} Yuan, H. B., Liu, X. W., Huo, Z. Y., et al.\ 2015, \mnras, 448, 855


\end{thebibliography}
\end{document}